\documentclass[prl,amsmath,amssymb,twocolumn, showpacs, superscriptaddress,10pt]{revtex4-1}

\usepackage{amsmath}
\usepackage{hyperref}
\usepackage{graphicx}
\usepackage{amsfonts}
\usepackage{amsthm}
\usepackage{cases}
\usepackage{mathtools,amssymb}

\usepackage{color}
\definecolor{Blue}{rgb}{0.00, 0.00, 1.00}
\definecolor{Red}{rgb}{1.00, 0.00, 0.00}

\newcommand{\nn}{\nonumber}
\newcommand{\be}{\begin{equation}}
\newcommand{\ee}{\end{equation}}
\newcommand{\bea}{\begin{eqnarray}}
\newcommand{\eea}{\end{eqnarray}}

\newcommand{\Ai}{{\rm Ai}}
\newcommand{\Tr}{{\rm Tr}}

\newcommand{\beq}{\begin{equation}}
\newcommand{\eeq}{\end{equation}}
\newcommand{\beqn}{\begin{eqnarray}}
\newcommand{\eeqn}{\end{eqnarray}}

\begin{document}

\title{Maximum of Airy plus Brownian processes, and persistent correlations in KPZ growth}

\title{Maximum of an Airy process plus Brownian motion and memory in KPZ growth}

\author{Pierre Le Doussal}
\affiliation{CNRS-Laboratoire de Physique Th\'eorique de l'Ecole Normale Sup\'erieure, 24 rue Lhomond, 75231 Paris Cedex, France}

\date{\today}

\begin{abstract}
We obtain several exact results for universal distributions involving the 
maximum of the Airy$_2$ process minus a parabola and plus
a Brownian motion, with applications to the 1D Kardar-Parisi-Zhang (KPZ) 
stochastic growth universality class. This allows to obtain (i) the universal limit, for large time separation,
of the two-time height correlation for droplet initial conditions, e.g. 
$C_{\infty} = \lim_{t_2/t_1 \to +\infty} \overline{h(t_1) h(t_2)}^c/\overline{h(t_1)^2}^c$,
with $C_{\infty} \approx 0.623$, as well as conditional moments, which quantify ergodicity breaking in the time evolution;
(ii) in the same limit, the distribution of the midpoint position $x(t_1)$ of a directed polymer of length $t_2$, 
and (iii) the height distribution in stationary KPZ with a step. These results are derived
from the replica Bethe ansatz
for the KPZ continuum equation, with
a "decoupling assumption" in the large time limit. 
They agree and confirm, whenever they can be compared, with (i) our recent tail results for two-time KPZ with de Nardis \cite{deNardisPLD2timeLong},
checked in experiments
with Takeuchi \cite{deNardisPLDTakeuchi}, (ii) a recent result of Maes and Thiery 
\cite{MaesThiery} on midpoint position. 
\end{abstract}

\pacs{05.40.-a, 02.10.Yn, 02.50.-r}
\pacs{72.20.-i, 71.23.An, 71.23.-k}


\maketitle

Stochastic processes, such as the Brownian motion, are useful unifying mathematical
tools to describe universal behavior of complex systems. In recent years, the
Airy$_2$ process, introduced in \cite{PraSpo02}, appeared in several contexts in physics
and mathematics. Its simplest definition (see e.g. \cite{ps-npoint}) 
involves the Dyson Brownian motion (DBM) \cite{Dys62}: consider a large Hermitian random matrix $H$
whose independent entries (both real and imaginary parts) perform independent 
stationary Orstein-Ulhenbeck processes (i.e. Brownian motions 
equilibrated in an harmonic well).
The Airy$_2$ process describes the evolution 
of the largest eigenvalue of $H$ (centered and scaled). 
The Airy$_2$ process appears as a limit process 
in directed last passage percolation \cite{Joh03}, non-intersecting Brownian bridges 
\cite{TW07,PraSpo02, Joh03, watermelon_us,KIK08,FMS11,SMCF13,CH11}, 
random tilings \cite{Joh05}, 
interacting particle transport in 1D \cite{Ferrari,KrugReview}, quantum dynamics of fermions
\cite{EislerDynamics,Dubail1,UsAiryPeriodic}
and stochastic growth models, either discrete \cite{PraSpo02,spohn_oup}
or the continuum 1D Kardar-Parisi-Zhang (KPZ) equation 
\cite{KPZ,CH16} (for a review see \cite{QR14,Baik_book,Cor12}). In fact the Airy$_2$ process is a hallmark of the very broad 1D-KPZ universality class, 
which arises in all these models.

Models in the 1D-KPZ class usually allow for the definition of a height field $h(x,t)$,
which undergoes stochastic growth. The prominent example is the continuum
KPZ equation \cite{KPZ}, where $h(x,t)$ is an interface height at point $x \in \mathbb{R}$, 
evolving as a function of time $t$ as
\be \label{kpzeq}
\partial_t h(x,t) = \nu \partial_x^2 h(x,t) + \frac{\lambda_0}{2}  (\partial_x h(x,t))^2 + \sqrt{D} ~ \xi(x,t)
\ee
driven by unit white noise $\overline{\xi(x,t) \xi(x',t')}=\delta(x-x') \delta(t-t')$. For the curved (i.e. droplet) initial condition (IC)
it is known (in some cases proved) that it converges at large time $t \to +\infty$ (rescaled and centered)
to ~\cite{PraSpo02, Joh03, Cor12,KPZFixedPoint2} 
\be
(\Gamma t)^{-\frac{1}{3}} (h_{\rm drop}(x,t) - v_{\infty} t ) \simeq {\cal A}_2(\hat x) - \hat x^2
\, , \, \hat x= A \frac{x}{2 t^{\frac{2}{3}}}  \label{conv1}
\ee
where ${\cal A}_2(\hat x)$ is the Airy$_2$ process, as an identity
between processes (i.e. as $\hat x$ is varied). Since ${\cal A}_2$ is stationary 
(statistical translational and reflection invariant in $\hat x$) the $- \hat x^2$ term
embodies the mean parabolic profile. We use units 
such that the non-universal constants $\Gamma=A=1$, i.e. $\lambda_0=D=2$ and $\nu=1$ for the KPZ equation \eqref{kpzeq}, and set $v_{\infty}=0$ (upon a shift of $h$). 
The equilibrium measure of the DBM being the
Gaussian unitary ensemble (GUE) measure for $H$, the fluctuations
of the Airy$_2$ process, hence of the KPZ height from \eqref{conv1},
at any given point, e.g. $x=0$, is the Tracy-Widom (TW) distribution for the largest eigenvalue 
of a GUE random matrix \cite{TW1994}. Its cumulative distribution function (CDF) is explicitly
known as a Fredholm determinant
\be
{\rm Prob}( {\cal A}_2(0) < \sigma) = F_2(\sigma) = {\rm Det}(I - P_\sigma K_{\Ai}) 
\ee
where $K_{\Ai}(u,v)=\int_0^{+\infty} dy \, \Ai(y+u) \Ai(y+v)$ is the Airy kernel, 
$P_\sigma$ being here 
the projector on $[\sigma,+\infty[$. Furthermore, from properties of the DBM,
the Airy$_2$ process is determinantal, i.e. any $p$-point joint CDF (JCDF) of
${\cal A}_2(\hat x)$ can be written as $p \times p$ matrix Fredholm determinants,
in term of an extended Airy kernel \cite{PraSpo02,QR14}.

Although much studied, and fully characterized by its determinantal structure, important
open questions remain about the Airy process, with applications to the 1D KPZ class. First, for more general
initial conditions $h(x,t=0)$, the value at a given point, e.g. $x=0$, is obtained from the variational problem \cite{QuastelRemenikParabola}
\bea
t^{-1/3} h(0,t)  \simeq \max_{\hat y} \left( {\cal A}_2(\hat y) - \hat y^2 + {\sf h}_0(\hat y) \right)
\label{hgen}
\eea 
when a limit exists for the rescaled IC ${\sf h}_0(\hat y) \simeq t^{-1/3} h(2 t^{2/3} \hat y,0)$.
Droplet subclass IC's correspond to ${\sf h}_0(0)=0$ and ${\sf h}_0(\hat y)= - \infty$ for $\hat y \neq 0$, recovering \eqref{conv1}, while flat subclass IC's correspond to ${\sf h}_0(\hat y)=0$. The CDF of $h(0,t)$ and of the {\rm argmax} in \eqref{hgen} (i.e. the endpoint distribution a directed polymer - see below) for flat IC, and 
and other results such as intermediate classes of IC, 
have been obtained from exact solutions of models in the KPZ class at large $t$, or
from powerful methods directly on the Airy process which allow to treat a large class of ${\sf h}_0$ \cite{QR14,QuastelRemenikParabola,KPZFixedPoint2}. 
The latter, however, do not readily extend to {\it random initial conditions}, such as the Brownian IC, 
related to the important stationary KPZ subclass. 

The aim of this Letter is to study some properties of the optimization problem
\be
\max_{\hat y} \left( {\cal A}_2(\hat y) - \hat y^2 + \sqrt{2} B(\hat y) \right) \label{maxgen0} 
\ee
where $B(\hat y)$ is the two-sided unit Brownian motion. Eq. \eqref{maxgen0} defines ${\cal A}_{\rm stat}(0)$, the Airy process associated to stationary KPZ equation, at $\hat x=0$. We first describe our three main results and 
applications, then give explicit formula, and finally sketch the replica Bethe ansatz
method.

\begin{figure}[h]
\centering
\includegraphics[scale=0.2]{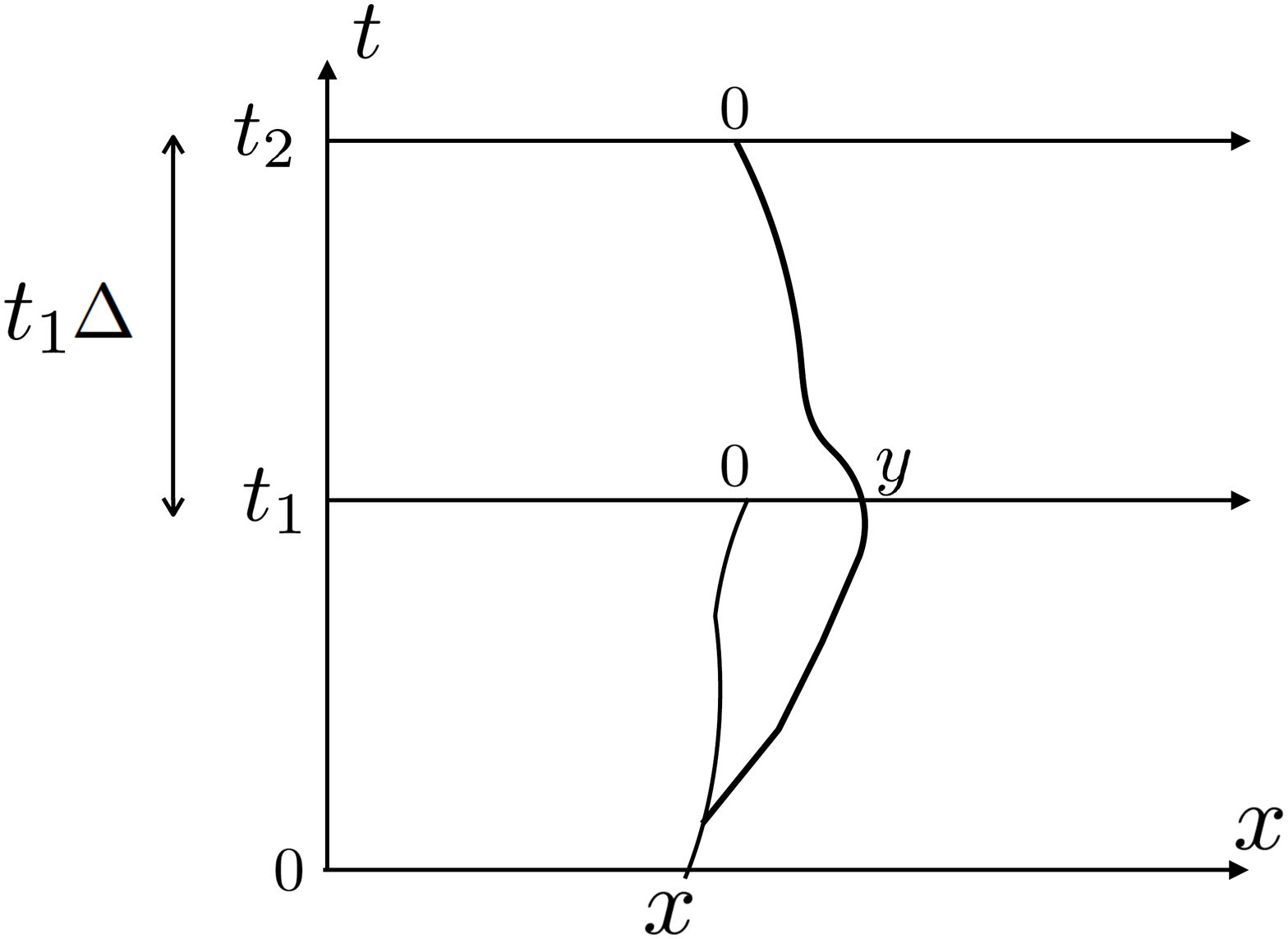}
\caption{Two directed polymers in a random potential with fixed endpoints, starting at $(x,0)$, ending respectively 
at $(0,t_1)$ and $(0,t_2)$. Minus their free energies maps to KPZ heights $h(0,t_1)$
and $h(0,t_2)$ with droplet initial condition centered at $x$. The PDF of the midpoint $y$,
and the correlations between $h(0,t_1)$ and $h(0,t_1)$ are obtained here exactly in the
limit $t_1, \Delta=\frac{t_2-t_1}{t_1} \to +\infty$.}
\label{fig:DP}
\end{figure}

{\bf Our first result} is the probability distribution function (PDF) of the position $\hat y_m$ of the maximum in \eqref{maxgen0}, i.e.
\bea
\hat y_m={\rm argmax}_{\hat y \in \mathbb{R}} \left( {\cal A}_2(\hat y) - \hat y^2 + \sqrt{2} B(\hat y) \right) \label{maxgen1} 
\eea
This distribution arises in the {\it midpoint probability of a directed polymer} (DP) in the white noise $d=1+1$ random 
potential $\xi$. Recall that the partition sum $Z(x,t|y,0)$ of continuum directed paths 
from $(y,0)$ to $(x,t)$ defined as
\be \label{zdef} 
Z(x,t|y,0) := \int_{x(0)=y}^{x(t)=x}  Dx e^{-  \int_0^t d\tau [ \frac{1}{4}  (\frac{d x}{d\tau})^2  - \sqrt{2} ~ \xi(x(\tau),\tau) ]}
\ee
equals $e^{h(x,t)}$ where $h(x,t)$ is the solution of \eqref{kpzeq} with droplet initial condition (centered at $y$).  
Consider a DP from $(0,0)$ to $(0,t_2)$ and ask about the PDF, $P_{t_1,t_2}(y)$, of the position $x(t_1)=y$ at
intermediate time $t_1$, see Fig.\ref{fig:DP}. In the limit of large times $t_1,t_2$, with $\hat y = y/(2 t_1^{2/3})$,
\be
\overline{P_{t_1,t_2}(y)} dy = \overline{\frac{Z(0,t_2|y,t_1)  Z(y,t_1|0,0)}{Z(0,t_2|0,0)}} dy \to P_\Delta(\hat y) d\hat y
\ee
One finds (see below) that as $\Delta=\frac{t_2-t_1}{t_1} \to +\infty$, 
$P_{t_1,t_2}(y)$ concentrates on $\hat y=\hat y_m$ defined in \eqref{maxgen1}, hence
\be
{\cal P}(\hat y) d\hat y := P_{+\infty}(\hat y) d\hat y = {\rm Prob}( \hat y_m \in [\hat y, \hat y + dy[ ) 
\ee
Here we calculate this distribution, which also arises in the study of the coalescence of optimal
paths \cite{Pimentel}.

{\bf Our second result} is the following joint CDF 
\bea
&& G(\sigma_1,\sigma_2) := \label{JCDF1} \\
&&{\rm Prob}\big( {\cal A}_2(0)  < \sigma_1 ,  ~ \max_{\hat y \in \mathbb R} ( {\cal A}_2(\hat y) - \hat y^2 + \sqrt{2} B(\hat y) ) < \sigma_2 
\big) 
\nn
\eea 
It is important in the study of the so-called {\it persistence of correlations in the two-time KPZ problem for droplet initial conditions}, which exhibits an interesting memory effect, also called ergodicity breaking \cite{Takeuchi,TakeuchiHHLReview,deNardisPLDTakeuchi,deNardisPLD2timeLong,FerrariSpohn2times}.
Indeed, consider the rescaled
heights at $t_1$ and at $t_2>t_1$, in the limit where both times are large, with $\Delta=(t_2-t_1)/t_1$ fixed
\bea
&& t_1^{-1/3} h(0,t_1)  \simeq {\cal A}_2(0)  \label{h1h2} \\
&& t_1^{-1/3} h(0,t_2) \simeq \max_{\hat y \in \mathbb R} \big( {\cal A}_2(\hat y) - \hat y^2 + \Delta^{\frac{1}{3}} (\tilde {\cal A}_2(\frac{\hat y}{\Delta^{\frac{2}{3}}}) - \frac{\hat y^2}{\Delta^{\frac{4}{3}}}) \big) \nn \\
&& \!\! \mathrel{\substack{\\ \simeq \\ \Delta \to +\infty}} 
\Delta^{\frac{1}{3}} \tilde {\cal A}_2(0) +  \max_{\hat y \in \mathbb R} \big( {\cal A}_2(\hat y) - \hat y^2
+ \sqrt{2} B(\hat y) \big) + O(\frac{1}{\Delta^{\frac{1}{3}}}) \nn 
\eea
where ${\cal A}_2$ and $\tilde {\cal A}_2$ denote two independent Airy processes.
The second line expresses that the height at $t_2$ is the sum of a first contribution
from the time interval $[0,t_1]$ and the second from $[t_1,t_2]$ which, for 
a fixed intermediate point $y$, are independent, see Fig.\ref{fig:DP}. 
Optimization over $\hat y$ correlates
them. Obtaining the resulting joint PDF (JPDF) of the two rescaled heights
for arbitrary $\Delta$ is a difficult problem
\cite{dotsenko2times1,dotsenko2times2,dotsenko2times3,Johansson2times,FerrariSpohn2times,deNardisPLD2timeLong,deNardisPLDTakeuchi,CorwininPrep}. In the 
limit of well separated times, i.e. large $\Delta$, 
using that the Airy process $\tilde {\cal A}_2$ is locally a Brownian,
one obtains the third line in \eqref{h1h2}, where $B$ and ${\cal A}_2$ are independent
processes \cite{Haag}. As is clear from \eqref{h1h2}, $h(0,t_1) \sim t_1^{\frac{1}{3}}$, $h(0,t_2)\sim t_2^{\frac{1}{3}}$ are quite different
in magnitude (for large $t_2/t_1$), but remain correlated by the $O(t_1^{1/3})$ term. To measure this
memory effect one usually define the dimensionless ratio
of the two covariances
\bea \label{defC} 
&& C(t_1,t_2) = \frac{\overline{ h(0,t_1) h(0,t_2)}^c}{\overline{h(0,t_1)^2}^c} \simeq C_\Delta
\eea 
which, at large times $t_1,t_2 \to +\infty$, becomes a universal function $C_\Delta$ of $\Delta$.
From \eqref{h1h2} the latter has a finite limit
\bea \label{Cinfty1} 
&& C_{\infty} = \frac{ \int d \sigma_1 d \sigma_2 \sigma_1 \sigma_2 p(\sigma_1,\sigma_2) }{\kappa_2^{\mbox{\tiny{GUE}}}}
\, , \,  
\eea 
where $p(\sigma_1,\sigma_2) = \partial_{\sigma_1} \partial_{\sigma_2} G(\sigma_1,\sigma_2)$ 
is the JPDF associated to \eqref{JCDF1}, obtained here exactly
(here and below $\kappa_p^{\mbox{\tiny{GUE}}}$ is the $p$-th cumulant of the GUE-TW 
distribution). 

Finally, {\bf our third main result} is the CDF for the height $h(x,t)$ 
for an {\it IC equal to a Brownian plus a step}, corresponding to a rescaled IC
${\sf h}_0(\hat y) = \sqrt{2} B(\hat y) - \hat H {\rm sgn}(\hat y)$. It will be
relevant for any KPZ system where two half spaces, each stationary, are put
in contact at $t=0$, with a mistmatch in height $2 \hat H t^{1/3}$.

Before displaying our explicit formula, it is important to recall some known results about the
stationary KPZ IC subclass, which corresponds to ${\sf h}_0(\hat y)= \sqrt{2} B(\hat y)$
where $B(x)$ is a two sided unit Brownian $\langle dB(x)^2 \rangle=dx$ with $B(0)=0$.
It is realized, e.g. by the solution $h_{\rm stat}(x,t)$ of the KPZ equation \eqref{kpzeq} with a
unit two sided Brownian initial condition $h(x,t=0)=B_0(x)$ \cite{footnote3} as
\bea
&& t^{-\frac{1}{3}} h_{\rm stat}(x,t)  \simeq {\cal A}_{\rm stat}(\hat x)  \\
&& = \max_{\hat y} \left( {\cal A}_2(\hat y-\hat x) - (\hat y-\hat x)^2 + \sqrt{2} B(\hat y) \right)
\label{hstat}
\eea
where here the last equality is only at fixed $\hat x$ (not as a process). 
Let us now define the two functions
\bea
&& {\cal B}_w(v) = e^{\frac{1}{3} w^3 - v w} - \int_0^{+\infty} dy \Ai(v+y) e^{w y} \label{defBL} \\
&& {\cal L}_{\hat x}(\sigma) =  \sigma - 1 - \hat x^2 
+  \int_{\sigma}^{+\infty} dv (1- {\cal B}_{\hat x}(v) {\cal B}_{-\hat x}(v)) \nn
\eea
It is known that the one-point CDF of the ${\cal A}_{\rm stat}$ process is given by
the extended Baik-Rains (EBR) distribution
\cite{png,BaikFerrariPeche2010,BaikLiuAiry,QR14,CorwinLiuWang,SasamotoStationary,BCFV,FerrariSpohnStationary2006}, which has the following exact expression
\be
 {\rm Prob}({\cal A}_{\rm stat}(\hat x) < \sigma) =: F_0(\sigma-\hat x^2,\hat x) = 
 \partial_\sigma (F_2(\sigma) Y_{\hat x}(\sigma)) \label{EBR1}
 \ee
 in terms of the auxiliary function
 \be \label{defYx} 
  Y_{\hat x}(\sigma) := 1+ {\cal L}_{\hat x}(\sigma) -  
{\rm Tr} [ P_\sigma K_{\Ai} (I- P_\sigma K_{\Ai})^{-1} P_\sigma {\cal B}_{-\hat x} {\cal B}_{\hat x}^T]  
\ee
where we denote $A B^T$ the projector $A B^T(u,v)=A(u) B(v)$. For 
$\hat x=0$, the function $F_0(\sigma,0) = F_0(\sigma) =
 \partial_\sigma (F_2(\sigma) Y_{0}(\sigma))$ is the CDF of the
 standard Baik-Rains (BR) distribution $F_0$.\\

{\bf The PDF of {\rm argmax}.} We now display our first result. Let
$H(\hat x) = {\rm Prob}(\hat y_m >  \hat x)$ the CDF of the position $\hat y_m$ of 
the maximum defined by \eqref{maxgen1}. Our method, detailed below, gives
\bea \label{resH1} 
H(- \hat x) 
&=& \int  d\sigma F_2(\sigma) \bigg(  Y_{\hat x}(\sigma) \\
& \times & \Tr [(I- P_\sigma K_{\Ai})^{-1} P_\sigma  (\Ai' + \hat x \Ai) \Ai^T] \nn
\\
& + &(\Tr[(I- P_\sigma K_{\Ai} )^{-1} P_\sigma \Ai {\cal B}_{\hat x}^T] - 1)  \nn \\
 &\times& \Tr[ (I- P_\sigma K_{\Ai} )^{-1} 
P_\sigma (\Ai' + \hat x \Ai) {\cal B}_{-\hat x}^T] \nn
 \bigg)
\eea 
where $\Ai'$ is the derivative of the Airy function.
Interestingly, in a recent work, Maes and Thiery \cite{MaesThiery} noted that 
the distribution of {\rm argmax} $\hat y_m$ can be related,
using the Fluctuation-Dissipation theorems (FDT), to the 
variance of the height in stationary KPZ, defined as
\be \label{defg1} 
g(\hat x) = \langle \sigma^2 \rangle_{F_0,\hat x} - \langle \sigma \rangle^2_{F_0,\hat x} 
\ee
where $\langle O(\sigma) \rangle_{F_0,\hat x} = \int d\sigma O(\sigma) \partial_\sigma F_0(\sigma-\hat x^2,\hat x)$
denotes an average over the extended Baik Rains distribution, which is an even function of $\hat x$.
Note that the second 
term is simply $-\langle \sigma \rangle^2_{F_0,\hat x} =- \hat x^4$. As a consequence of \cite{MaesThiery},
the scaled PDF of the midpoint probability is predicted as
\be \label{rescalP} 
{\cal P}(\hat y)= -H'(\hat y) = f_{\rm KPZ}(\hat y) \quad , \quad \!\!\! f_{\rm KPZ}(\hat y):=\frac{1}{4} g''(\hat y)
\ee
where the notation $f_{\rm KPZ}(\hat y)$ for the second derivative of $g$ in \eqref{defg1} was 
introduced in the context of the PNG and TASEP models \cite{PrahoferSpohn2004}.

It is thus important to check whether our result \eqref{resH1}, obtained through an
independent and completely different route, agrees with this prediction.
Using the identities
\cite{SM}
\bea \label{ident1} 
& \partial_\sigma Y_{\hat x}(\sigma) =& (\Tr[ (I- P_\sigma K_{\Ai} )^{-1} P_\sigma \Ai {\cal B}_{\hat x}^T] - 1)  \\
&& \times  (\Tr[(I- P_\sigma K_{\Ai} )^{-1} P_\sigma \Ai {\cal B}_{- \hat x}^T] - 1) \nn \\
& \partial_\sigma F_2(\sigma) =& {\rm Tr} [P_{\sigma}  (I- P_{\sigma} K_{\Ai})^{-1} P_\sigma \Ai \Ai^T ] \nn
\eea 
a few algebraic manipulations \cite{SM} show that \eqref{resH1} can indeed be rewritten as
\bea
H(\hat x) = \frac{1}{2} - \frac{1}{4} g'(\hat x)  \label{Hg}
\eea 
where, we recall, $g'(\hat x)$ is odd, and $g'(\pm \infty)=\pm 2$.
Hence our result \eqref{resH1} provides an equivalent, though
different form for the midpoint probability ${\cal P}(\hat y)=H(\hat y)$.
This provides a test of our method (the decoupling assumption, see
below) and of the FDT for the KPZ problem. Note that the result of
\cite{MaesThiery} extends to finite time, while our method deals with
large times. \\

{\bf Joint PDF of Airy and Airy minus parabola plus Brownian and persistent KPZ two time
correlations.} We now give our result for the JCDF \eqref{JCDF1}. We find, for $\sigma_1 \leq \sigma_2$
\bea
&& G(\sigma_1,\sigma_2) \label{ResF} \\
&& = F_2(\sigma_1) Y_{0}(\sigma_1) 
{\rm Tr} [ (I- P_{\sigma_1} K_{\Ai})^{-1} P_{\sigma_1}  \Ai_{\sigma_2-\sigma_1} \Ai_{\sigma_2-\sigma_1}^T ] \nn \\
&& + F_2(\sigma_1) ({\rm Tr}[(I- P_{\sigma_1} K_{\Ai} )^{-1} P_{\sigma_1}  \Ai_{\sigma_2-\sigma_1} {\cal B}_{0}^T] - 1)^2 \nn
\eea
where $\Ai_{\sigma}(u)=\Ai(u+\sigma)$ and $G(\sigma_1,\sigma_2)=F_0(\sigma_2)$ for $\sigma_1 \geq \sigma_2$.
An extended result for $\hat x \neq 0$ is displayed in \cite{SM}. It is easy to check \cite{SM} the continuity, $G(\sigma,\sigma)=F_0(\sigma)$ using the identities \eqref{ident1}. It is also easy to see that the marginal CDF of $\sigma_1$, $G(\sigma_1,+\infty)=F_2(\sigma_1)$ is the
GUE-TW and the marginal CDF of $\sigma_2$, $G(+\infty,\sigma_2)=F_0(\sigma_2)$ is the BR distribution.

We now apply this result to the large time separation
limit of the two-time correlation in the 1D KPZ class. Using integration by parts one obtains \cite{SM}
\be
\langle (\sigma_2-\sigma_1)^2 \rangle = 2  \int_{-\infty}^{+\infty} d\sigma_2 \int_{-\infty}^{\sigma_2}  d\sigma_1  (F_2(\sigma_1) -G(\sigma_1,\sigma_2)) \label{var1} 
\ee 
where here and below $\langle .. \rangle$ denotes averages w.r.t. $p(\sigma_1,\sigma_2) = \partial_{\sigma_1} \partial_{\sigma_2} G(\sigma_1,\sigma_2)$, the associated JPDF. This allows us to
compute the two-time persistent dimensionless covariance ratio \eqref{Cinfty1} as
\bea
&& C_{\infty} = \frac{\langle \sigma_2 \sigma_1 \rangle^c}{\langle \sigma_1^2 \rangle} 
= \frac{\langle \sigma_2 \sigma_1 \rangle}{\kappa_2^{\mbox{\tiny{GUE}}}}
= \frac{\langle \sigma_1^2 \rangle + \langle \sigma_2^2 \rangle - \langle (\sigma_2-\sigma_1)^2 \rangle}{2 \kappa_2^{\mbox{\tiny{GUE}} }} \nn \\
&& = \frac{1}{2} + 
\frac{(\kappa_1^{\mbox{\tiny{GUE}}})^2 + 
\kappa_2^{\mbox{\tiny{BR}}}}
{2 \kappa_2^{\mbox{\tiny{GUE}} }} - \frac{\langle (\sigma_2-\sigma_1)^2 \rangle}{2 \kappa_2^{\mbox{\tiny{GUE}} }} 
\label{exact1} \\
&& = 3.13598 - \frac{\langle (\sigma_2-\sigma_1)^2 \rangle}{2 \kappa_2^{\mbox{\tiny{GUE}} }}
\approx 0.6225 \pm 0.0015 \label{resultCinfty}
\eea
using the known GUE-TW and BR cumulants, i.e.
$\langle \sigma_2 \rangle= \kappa_1^{\mbox{\tiny{BR}}}=0$,
$\kappa_1^{\mbox{\tiny{GUE}}} = -1.7710868$, 
$\kappa_2^{\mbox{\tiny{GUE}}} = 0.81319$ 
and $\kappa_2^{\mbox{\tiny{BR}}} = 1.15039$, 
and evaluated \eqref{var1} numerically (see Sec IV. 6. in \cite{SM}).
Eq. \eqref{resultCinfty} compares quite well with recent numerical simulations and 
experiments \cite{Y,X}.

Let us recall our recent study \cite{deNardisPLD2timeLong,deNardisPLDTakeuchi}
and reexamine the observables defined there, using our new exact results.
There we defined 
the variables $h_1:= h(0,t_1)/t_1^{1/3}$ and the scaled height difference
\be
h := (h(0,t_2)-h(0,t_1))/(t_2-t_1)^{1/3} 
\ee
Defining the (unknown) exact JPDF $P_\Delta(\sigma_1,\sigma):=\lim_{t_1,t_2=t_1(1+\Delta) \to +\infty}
\overline{\delta(h_1-\sigma_1)  \delta(h-\sigma)}$,
we derived an approximation of it, denoted $P^{(1)}_\Delta(\sigma_1,\sigma)$, conjectured to be 
exact to leading order in large positive $\sigma_1$, for any fixed $\sigma$ and $\Delta$.
It was shown in \cite{deNardisPLDTakeuchi} to be good enough an approximation to fit 
experiments and numerics in a broad range of values 
$\sigma_1 > \langle \sigma_1 \rangle= \kappa_1^{\mbox{\tiny{GUE}}}$. It is thus of
great importance to check whether our present exact result, valid only
for large $\Delta$, but for any $\sigma_1,\sigma$, confirms
these predictions.

\begin{figure}[h]
\centering
\includegraphics[scale=0.5]{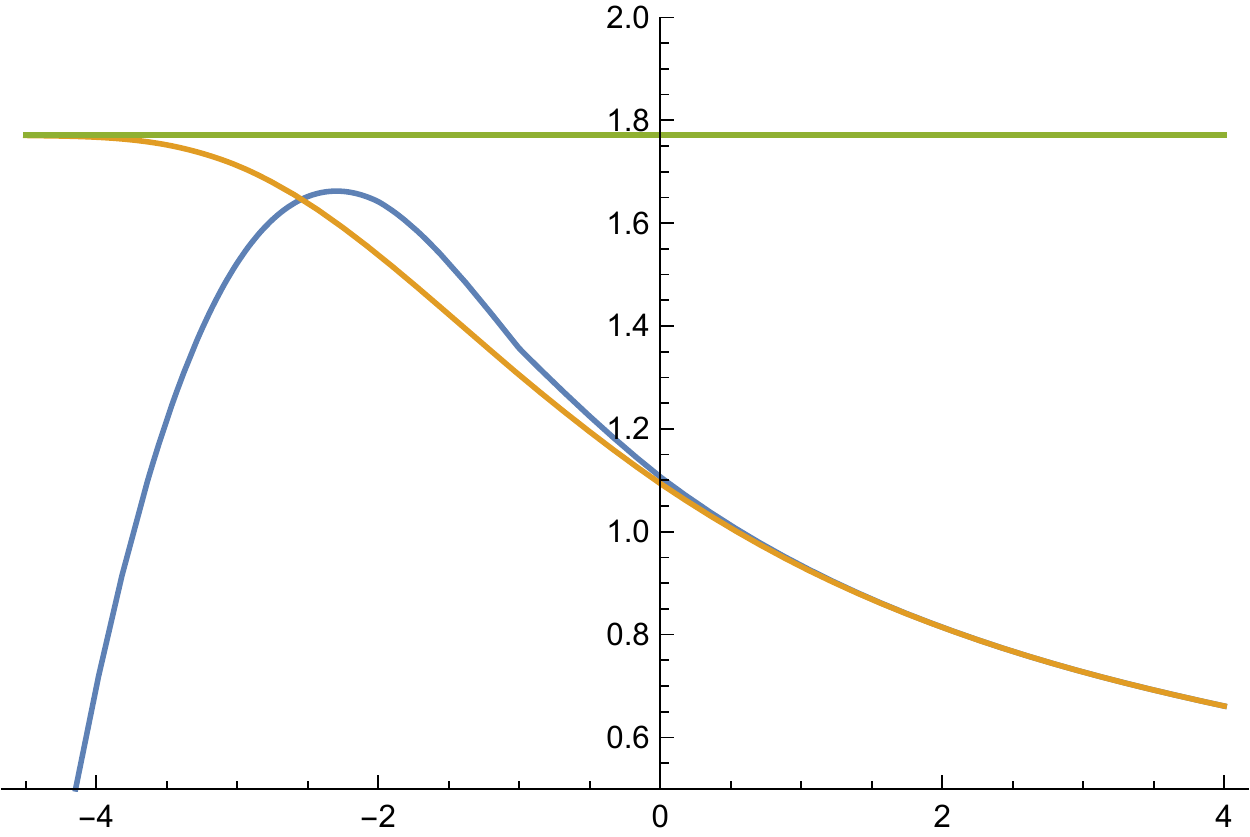}
\caption{Conditional average
$\langle \sigma_2-\sigma_1\rangle_{\sigma_1 > \sigma_{1c}}$ ($y$ axis) as
a function of $\sigma_{1c}$ ($x$ axis), which described the averaged scaled KPZ height difference $h$ at large $\Delta=\frac{t_2-t_1}{t_1}$, see \eqref{hhh}. 
(i) Orange: exact result obtained here.
(ii) Blue: prediction obtained in 
\cite{deNardisPLD2timeLong,deNardisPLDTakeuchi}, which becomes exact
for large positive $\sigma_{1c}$ (indistinguishable from (i) for 
$\sigma_{1c} >0$). The limit $\sigma_{1c} \to -\infty$ (unconditionned mean),
exact from \eqref{uncond} and to which our result (i) converges, is the orange horizontal line at $- \kappa_1^{\mbox{\tiny{GUE}}} =
1.77109$.}
\label{fig:condmean}
\end{figure}

At large times, the height difference, 
from \eqref{h1h2}, takes the form, up to $O(\Delta^{-2/3})$ terms 
\be \label{hh}
h \simeq \tilde {\cal A}_2(0) + \Delta^{-\frac{1}{3}}  (\sigma_2 - \sigma_1)
\ee
where we denote (with a slight abuse of notations) the two random variables
\be
\sigma_1= {\cal A}_2(0) \quad , \quad 
\sigma_2= \max_{\hat y \in \mathbb R} \big( {\cal A}_2(\hat y) - \hat y^2
+ \sqrt{2} B(\hat y) \big)
\ee
and we recall that $\tilde {\cal A}_2(0)$ is a GUE-TW random variable independent of
the $O(\Delta^{-1/3})$ term. The first consequence, averaging \eqref{hh},
is that 
\be \label{uncond} 
\overline{h} = \kappa_1^{\mbox{\tiny{GUE}}} (1 - \frac{1}{\Delta^{1/3}})  + O(\Delta^{-1})
\ee 
since $\langle \sigma_2 \rangle=\kappa_1^{\mbox{\tiny{BR}}}=0$ and
$\langle \sigma_1 \rangle = \kappa_1^{\mbox{\tiny{GUE}}}$, in agreement with
the general formula for $\overline{h}$ for any $\Delta$ (see (48) in 
\cite{deNardisPLD2timeLong}). Important quantities, introduced in \cite{deNardisPLD2timeLong,deNardisPLDTakeuchi},
are the {\it conditional averages} of $h$, either for a fixed value of $h_1=\sigma_1$,
$\overline{h}_{h_1=\sigma_1}$, or, for a value larger than some threshold $h_1=\sigma_1>\sigma_{1c}$,
$\overline{h}_{h_1=\sigma_1>\sigma_{1c}}$.
From the above one predicts
\be
\overline{h}_{h_1=\sigma_1} \simeq \kappa_1^{\mbox{\tiny{GUE}}} + \frac{1}{\Delta^{1/3}} 
\langle \sigma_2-\sigma_1 \rangle_{\sigma_1} + o(\frac{1}{\Delta^{1/3}})
\ee
where the conditional average w.r.t. $p$, denoted as
\be \label{condmom1} 
\langle \sigma_2-\sigma_1 \rangle_{\sigma_1} = \frac{1}{F_2'(\sigma_1)} 
\int_{-\infty}^{+\infty} d\sigma_2 (\sigma_2-\sigma_1) p(\sigma_1,\sigma_2)
\ee
can be calculated from \eqref{ResF}. For large positive $\sigma_1$ one shows from \eqref{ResF} that (see \cite{SM} where next order is also displayed)
\bea
&& p(\sigma_1,\sigma_2) \simeq  - 2 \partial_{\sigma_2}  K_{\Ai}(\sigma_1,\sigma_2)  - \Ai(\sigma_2)^2 
\eea
which leads to, for large positive $\sigma_1$ 
\bea
&& \langle \sigma_2-\sigma_1 \rangle_{\sigma_1} \simeq  R_{1/3}(\sigma_1) \label{firstmomcond} \\
&& R_{1/3}(\sigma_1) :=  \frac{[ \int_{\sigma_1}^{+\infty} dy \Ai(y) ]^2 - 
\int_{\sigma_1}^{+\infty} dy K_{\Ai}(y,y)}{K_{\Ai}(\sigma_1,\sigma_1)} 
\eea 
which is precisely the prediction obtained in \cite{deNardisPLD2timeLong}. 
This is encouraging evidence that the method of  \cite{deNardisPLD2timeLong} is good enough to
capture, as claimed there, the tail of the two-time JPDF. 
%
We can thus calculate the conditional averages beyond the large positive $\sigma_1$ 
regime. We show in Fig. \ref{fig:condmean} the leading order of 
\be \label{hhh} 
\overline{h}_{h_1>\sigma_{1c}} = \kappa_1^{\mbox{\tiny{GUE}}}  
+ \Delta^{-1/3} \langle \sigma_2-\sigma_1 \rangle_{\sigma_1 > \sigma_{1c}} 
+ O(\Delta^{-2/3}) 
\ee
evaluated numerically \cite{SM} from \eqref{ResF},
a quantity which can be measured accurately in experiments and numerics.

\begin{figure}[h]
\centering
\includegraphics[scale=0.6]{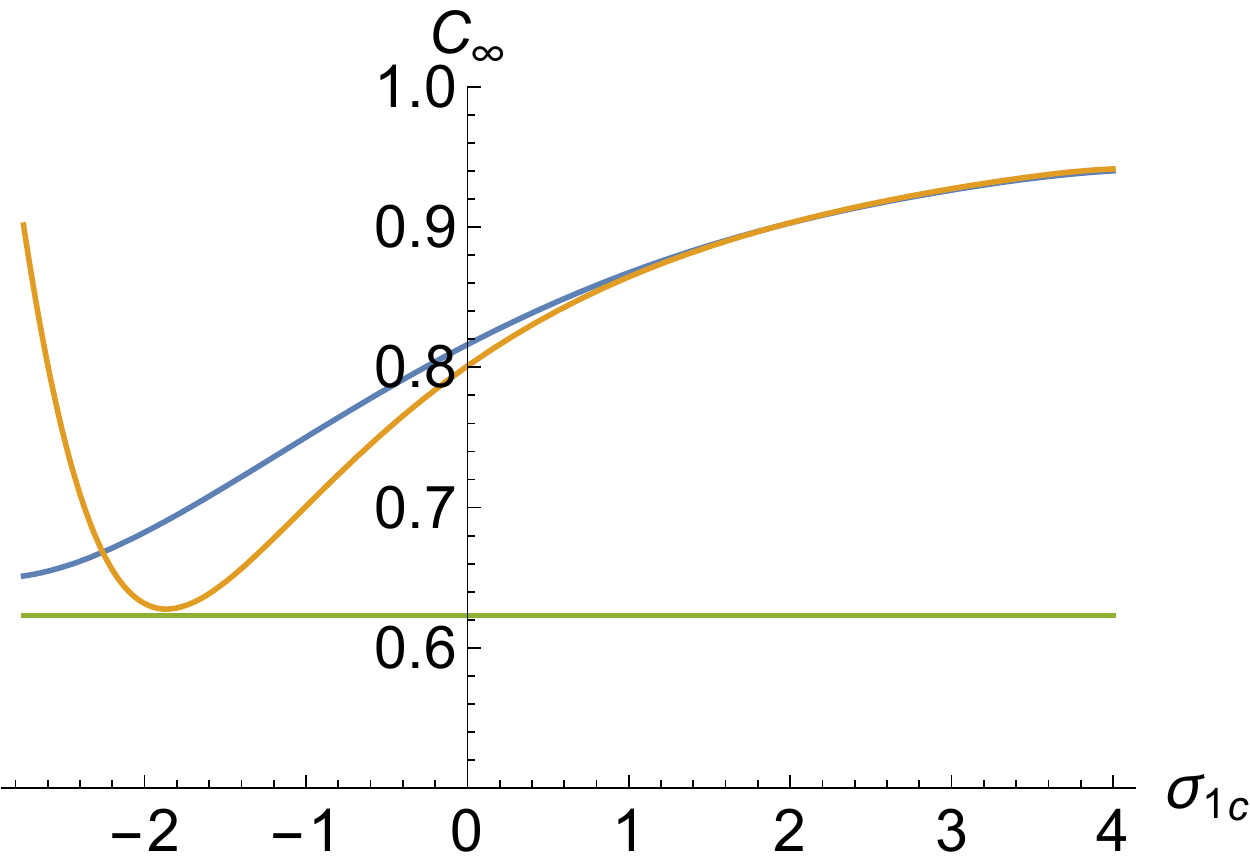}
\caption{Conditional covariance ratio $C_{\infty}(\sigma_{1c})$, Eq. \eqref{Cinfty2}. 
Orange: asymptotic prediction
for large positive $\sigma_{1c}$ from
\cite{deNardisPLD2timeLong,deNardisPLDTakeuchi}. Blue: exact result
from \eqref{ResF}, converging to \eqref{resultCinfty} (orange horizontal line)
for $\sigma_{1c}=-\infty$.}
\label{fig:C}
\end{figure}

Finally, the conditional covariance ratio was introduced and measured 
in \cite{deNardisPLD2timeLong,deNardisPLDTakeuchi}
\bea
&& C_{\Delta}(\sigma_{1c}) := \lim_{\Delta \to \infty} 
\frac{\overline{h_1 h_2}^c_{h_1> \sigma_{1c}} }{\overline{h_1^2}^c_{h_1> \sigma_{1c}}} 
\eea 
We obtain here its large $\Delta$ limit, 
\bea
&& C_{\infty}(\sigma_{1c}) = \frac{< \sigma_2 \sigma_1 >^c_{\sigma_1>\sigma_{1c}}}{<\sigma_1^2>^c_{\sigma_1>\sigma_{1c}}} \label{Cinfty2}
\eea
a function of $\sigma_{1c}$ which interpolates between $C_{\infty}(\sigma_{1c}=-\infty)= C_{\infty}$ the unconditionned 
two-time covariance ratio obtained in \eqref{resultCinfty}
and $C_{\infty}(\sigma_{1c}=+\infty)=1$. It is evaluated numerically 
\cite{SM} from \eqref{ResF} and plotted in Fig. \ref{fig:C}.

We also explored the case where the longer polymer in Fig.\ref{fig:DP}
is constrained to pass to the right of $0$, $y>0$: 
we find $C_{\infty} \approx 0.6925$, i.e. that the correlations are increased.

{\bf Stationary KPZ in presence of a step}. Finally, the heigth $h(x,t)$ in the KPZ class with a step at $x=0$ in the initial condition,
and independent Brownian initial condition on each side, takes the scaling
form at large $t$
\bea
&& t^{- \frac{1}{3}} h(x=2 t^{\frac{2}{3}} \hat x,t) \approx \hat h(\hat x) 
\\
&& := \max_{\hat y} \left( {\cal A}_2(\hat x-\hat y) - (\hat x-\hat y)^2 + \sqrt{2} B(y) - \hat H {\rm sgn}(\hat y) \right) \nn
\eea
Defining $G_{\hat H}(\sigma_L) ={\rm Prob}( \hat h(\hat x)- \hat H + \hat x^2  < \sigma_L)$,
we obtain
\bea \label{GH}
&& G_{\hat H}(\sigma_L) = F_2(\sigma_L) \nn \\
&& \times \bigg(  Y_{\hat x}(\sigma_L) e^{2 \hat x \hat H} 
\Tr [(I- P_{\sigma_L} K_{\Ai})^{-1} P_{\sigma_L}  \Ai_{2 \hat H} \Ai^T ] \nn \\
&& + (\Tr[(I- P_{\sigma_L} K_{\Ai} )^{-1} P_{\sigma_L}  \Ai {\cal B}^T_{\hat x}] - 1)
  \\
&& \times 
(e^{2 \hat x \hat H} \Tr[(I- P_{\sigma_L} K_{\Ai} )^{-1} \Ai_{2 \hat H} {\cal B}^T_{-\hat x}] - 1) \nn
\eea 

\medskip

{\bf Method}. The method is based on the replica Bethe ansatz, 
which led to exact solutions for one-time observables
for various initial conditions
\cite{we,dotsenko,we-flat,SasamotoStationary,dotsenkoGOE,ps-2point,ps-npoint,dotsenko2pt,Spohn2ptnew,dotsenkoEndpoint,PLDCrossoverDropFlat,KPZFixedPoint}.
Since it is an extension of the calculation in
\cite{PLDCrossover17} we only sketch the idea, with details in Sec. II-III of \cite{SM}.
We define, jointly in the same noise realization, 
$h_1(x,t) = \ln Z(x,t|0,0)$ and $h_{L,R}(x,t)$ the solutions of the KPZ equation
with IC's respectively droplet (1), and Brownian on $y<0$ (L) and on $y>0$ (R).
We define the joint Laplace transform 
\bea \label{genfunctLR0} 
\hat g_t(\sigma_1,\sigma_L,\sigma_R;\hat x) 
= \overline{e^{- \sum_{b=1,L,R} u_b Z_b(x,t) }}
\eea
with $u_b=e^{- t^{1/3} (\sigma_b-\hat x^2) }$. Its large time limit gives 
\bea \label{gcontain}
&& \hat g_{\infty} (\sigma_1,\sigma_L,\sigma_R;\hat x) = G_{\hat x}(\sigma_1,\sigma_L,\sigma_R;\hat x) :=
\\
&& {\rm Prob}( {\cal A}_2(-\hat x)  < \sigma_1 , \hat h_L(\hat x) + \hat x^2 < \sigma_L ,  
\hat h_R(\hat x) + \hat x^2 < \sigma_R ) \nn
\eea
with
\be
\hat h_{L,R}(\hat x)= \max_{\hat y<0,\hat y>0}( {\cal A}_2(\hat y-\hat x) - (\hat y - \hat x)^2 + \sqrt{2} 
B(\hat y) )
\ee
containing all desired information (and more). To compute $\hat g_t$ we
expand the exponential in \eqref{genfunctLR0}, and write the
joint moments as quantum mechanical expectations
\be \label{expect} 
\overline{\prod_{b=1,L,R} Z_b(x,t)^{n_b}} =\langle x,..x|e^{- H_n t}|n_L,n_1,n_R\rangle
\ee
where 
\be
H_n = -\sum_{\alpha=1}^n \frac{\partial^2}{\partial {x_\alpha^2}}  - 2  \sum_{1 \leq  \alpha< \beta \leq n} \delta(x_\alpha - x_\beta) 
\label{LL}
\ee
is the Hamiltonian of the attractive 
Lieb Liniger $\delta$-Bose gas model \cite{ll}, and $|n_L,n_1,n_R\rangle$ is a state with
$n_1$ bosons at $y=0$, and $n_{L,R}$ in $y<0$ and $y>0$ respectively, with $n=n_1+n_L+n_R$.
One inserts in \eqref{expect} the known Bethe Ansatz eigenstates, each consisting of $1 \leq n_s \leq n$ strings (bound states) of $m_j \geq 1$ bosons
with rapidities $\lambda_{j, a}=k_j - \frac{i}2(m_j+1-2a)$, $a = 1,...,m_j$, and $\sum_{j=1}^{n_s} m_j=n$.
For the Brownian (and flat) IC the overlap of $|n_L,n_1,n_R\rangle$ with any Bethe state can be expressed explicitly, although as a complicated sum over products of Gamma functions, extending as in \cite{PLDCrossover17} the combinatoric method introduced in \cite{dotsenkoGOE}. It can be simplified in the large $t$ limit through the {\it decoupling assumption} (which sets all inter-string double products to unity) as done in
\cite{dotsenkoGOE,ps-2point,ps-npoint,dotsenko2pt,Spohn2ptnew,dotsenkoEndpoint,PLDCrossoverDropFlat,KPZFixedPoint}.
Summing over the eigenstates becomes possible 
and leads to a Fredholm determinant formula for $\hat g_{\infty} $
(\ref{final10}-\ref{K12}), and (\ref{final10w}-\ref{Khat1}) in \cite{SM}. For regularization the calculation includes finite drifts $w_{L,R}>0$ in the Brownian, and the 
(delicate) limit $w_{L,R}=0^+$ converts the Fredholm determinant into a final expression \eqref{mainres} in \cite{SM} for
$G_{\hat x}(\sigma_1,\sigma_L,\sigma_R;\hat x)$. Specializing to $\sigma_L=\sigma_R=\sigma_2$ gives the 
result \eqref{resFx} in \cite{SM} for the JCDF $G_{\hat x}(\sigma_1, \sigma_2)$ for general $\hat x$, which reduces to \eqref{ResF} for $G=G_{0}$ for $\hat x=0$. Specializing instead to $\sigma_1=+\infty$, one obtains (i) the result \eqref{resH1} for the CDF of the argmax of Airy minus parabola plus Brownian and \eqref{rescalP} for the limiting midpoint
DP probability, from
\be
H(- \hat x) = \int_{-\infty}^{+\infty} d\sigma_R [\partial_{\sigma_R} 
\hat g_{\infty} (+\infty,\sigma_L,\sigma_R; \hat x) )|_{\sigma_L=\sigma_R^-} \nn 
\ee
(ii) the CDF of the KPZ height in presence of a step IC: setting $\sigma_R=\sigma_L+\hat H$ one obtains $G_{\hat H}(\sigma_L)=
\hat g_{\infty} (+\infty,\sigma_L,\sigma_R;\hat x)$ leading to the result \eqref{GH}.

In conclusion, from a replica Bethe ansatz calculation, using a decoupling assumption, we 
obtained several distributions involving the maximum of the Airy process minus parabola plus Brownian. This leads to exact universal results for two-time KPZ in the large time separation limit $\Delta=\frac{t_2-t_1}{t_1}>>1$, which correctly match, and nicely complement, our recent tail 
results \cite{deNardisPLD2timeLong,deNardisPLDTakeuchi} for any $\Delta$, putting both
methods on firmer ground. Taken together
they should also lead to further accurate comparisons with experiments and numerics
in the universal large time limit, and allow to test other observables, e.g. the effect of the endpoint position $\hat x \neq 0$
as predicted here and in \cite{deNardisPLD2timeLong}.

We thank J. de Nardis, K. Takeuchi and T. Thiery for stimulating discussions
and collaborations, 
and A. Borodin, I. Corwin, T. Halpin-Healy, S. Majumdar, J. Quastel, G. Schehr for interesting
remarks.

{}

\newpage

.

\begin{widetext} 

\bigskip

\bigskip

\begin{large}
\begin{center}

SUPPLEMENTARY MATERIAL

\end{center}
\end{large}

\bigskip

We give here the details of the calculations and their applications, as
described in the main text of the Letter.
Calculations being performed in a slightly more general framework, further explicit results
are being displayed (e.g. for $\hat x \neq 0$, triple JCDF's, with and without Brownian),
which are then specialized to the cases considered in the main text. 

\bigskip

\tableofcontents

\section{I. Continuum KPZ equation, directed polymer and Airy process}

We start from the 1D KPZ equation \eqref{kpzeq} and work in the units $x^* = \frac{(2 \nu)^3}{D \lambda_0^2}$, $t^* = \frac{ 2 (2 \nu)^5}{D^2 \lambda_0^4}$ and $h^* = \frac{2 \nu}{\lambda_0}$, such that it reads \bea \label{kpzeq2}
&& \partial_t h(x,t) =  \partial_x^2 h(x,t) + (\partial_x h(x,t))^2 + \sqrt{2} ~ \eta(x,t)  
\eea
where $\eta$ is a unit white noise $\overline{\eta(x,t) \eta(x',t')}=\delta(x-x') \delta(t-t')$. 
The KPZ equation is mapped, via the Cole-Hopf transformation, to the 
continuous directed polymer (DP) in a quenched random potential, such
that $h(x,t)= \ln Z(x,t)$ is (minus) the free energy of the DP of length $t$ with one fixed endpoint 
at $x$. For an arbitrary initial condition, the solution at time $t$ can be written as:
\be \label{ch}
e^{h(x,t)} = Z(x,t) := \int dy Z_\eta(x,t|y,0) Z_0(y) \quad , \quad Z_0(y) = e^{h(y,t=0)}.
\ee
where here and below we denote $\int dy = \int_{-\infty}^{+\infty} dy$.
Here $Z_\eta(x,t|y,0)$ is the partition function of the continuum directed polymer in the random potential
$- \sqrt{2} ~ \eta(x,t)$ with fixed endpoints at $(x,t)$ and $(y,0)$:
\be \label{zdef} 
Z_\eta(x,t|y,0) = \int_{x(0)=y}^{x(t)=x}  Dx e^{-  \int_0^t d\tau [ \frac{1}{4}  (\frac{d x}{d\tau})^2  - \sqrt{2} ~ \eta(x(\tau),\tau) ]}
\ee
which is the solution of the (multiplicative) stochastic heat equation (SHE):
\bea \label{dp1} 
\partial_t Z = \nabla^2 Z + \sqrt{2} ~ \eta Z
\eea 
with Ito convention and initial condition $Z_\eta(x,t=0|y,0)= \delta(x-y)$. Equivalently, $Z(x,t)$ is the solution of
(\ref{dp1}) with initial conditions $Z(x,t=0)=e^{h(x,t=0)}=Z_0(x)$. We will adopt the notation (for the 
solution of the droplet initial condition started in $y$):
\bea \label{drop0} 
h_\eta(x,t|y,0) = \ln Z_\eta(x,t|y,0) 
\eea 
although it is somewhat improper (it requires a short time regularization, irrelevant here). 
We will most often omit the "environment" index $\eta$.
Here and below overbars denote averages over the white noise $\eta$.

Note that the time reversed path sees the time reversed random potential $\tilde \eta$, which
has the same distribution as $\eta$, hence in law
\bea
Z_\eta(x,t|y,0) = Z_{\tilde \eta}(y,t|x,0) \equiv_{\rm in law} Z_\eta(y,t|x,0)
\eea 
a property used extensively below. 

In order to translate our results below in terms of Airy processes, it is
useful to recall that the following convergence to the Airy$_2$ process, ${\cal A}_2(\hat x)$,
is expected at large time 
\cite{PraSpo02,Ferrari,KPZFixedPoint2,CH16}
\bea
&& h(x,t|y,0) \simeq  
t^{1/3} (  {\cal A}_2(\hat x-\hat y) - (\hat x-\hat y)^2 ) + o(t^{1/3}) \quad , \quad \hat x = \frac{x}{2 t^{2/3}} \quad , \quad \hat y = \frac{y}{2 t^{2/3}} \label{Ai1}
\eea 
where $h(x,t|y,0)$ is the droplet solution with arbitrary endpoints (\ref{drop0}). 
In terms of processes, this equivalence is only valid at either fixed $y$ or fixed
$x$. The process as $(x,y)$ are both varied is called the Airy sheet 
(see e.g. \cite{QR14,KPZFixedPoint,Pimentel}) and is
not yet fully characterized. 

The formula \eqref{ch} for a general initial condition then leads, in the large time limit, to a variational formula
in terms of an Airy process. Indeed the variations of $h(x,t|y,0)$ being $O(t^{1/3})$ the
integral in \eqref{ch} becomes dominated by the maximum value of the integrand, hence we can write,
in the sense of the one-point PDF at fixed $x$
\bea
&& h(x,t) \simeq  
t^{1/3} \max_{\hat y} ( {\cal A}_2(\hat x-\hat y) - (\hat x-\hat y)^2 + {\sf h}_0(\hat y)) + o(t^{1/3})  \label{Ai2}
\eea 
where we have defined the rescaled initial condition
\be
{\sf h}_0(\hat y) = t^{-1/3} h(2 t^{2/3} \hat y,0)  \label{Ai3}
\ee
This allows to classify the initial conditions, depending on whether ${\sf h}_0(\hat y)$ has a time independent
limit. When ${\sf h}_0(\hat y) \to 0$ one is in the flat IC class, with the GOE TW distribution $F_1$
for $h(x,t)$. 
The case of the Brownian initial condition is discussed below. Other intermediate classes have been
identified and studied \cite{QuastelRemenikParabola,KPZFixedPoint2}.

\section{II. Definition of the generating function and its physical content}

We define, in the same environment $\eta$, the three partition sums of all DP paths with one endpoint at $(x,t)$ and the second endpoint at $t=0$ either on the
negative $y<0$, or positive $y>0$ axis, or at $y=0$, together with their logarithms, as
\bea \label{ZZdef} 
&& Z_L(x,t) = \int_{y<0} Z(x,t|y,0) e^{a_L B_0(y) + w_L y} \quad , \quad h_L(x,t)= \ln Z_L(x,t) \\
&& Z_R(x,t) = \int_{y>0} Z(x,t|y,0) e^{a_R B_0(y)-  w_R y} \quad , \quad h_R(x,t)= \ln Z_R(x,t)   \\
&& Z_c(x,t) = Z(x,t|0,0) \quad , \quad h_c(x,t)= \ln Z_c(x,t) 
\eea 
We are multiplying by an additional weight on the $y$ axis, corresponding to the cases which we can solve.
Each parameter $a_R$ and $a_L$ is chosen zero or unity. Here $B_0(y)$ is a two-sided unit centered Brownian, with $B_0(0)=0$, i.e. $\langle B_0(y) B_0(y') \rangle=\min(y,y') \theta(y) \theta(y') + \min(-y,-y') \theta(-y) \theta(-y')$ (it can also be written as the sum of two independent one-sided Brownians on respectively the negative and positive half line). 
The parameters $w_L,w_R$ (usually chosen positive) represent the drifts of the Brownians. 
The partition sums $Z_{L,R}(x,t)$ give the relative weights that a DP path with one endpoint at $(x,t)$
has it second endpoint at $t=0$ either on $(y<0,0)$ or $(y>0,0)$, in presence of the additional weights
$e^{a_{L,R} B_0(y) \pm w_{L,R} y}$. More precisely,
\bea  \label{proba1} 
p_\eta(x,t)=Z_R(x,t)/(Z_L(x,t) + Z_R(x,t))
\eea
is the probability, in a given sample, that the DP with one fixed endpoint at $(x,t)$ ends up at $y>0$ at $t=0$. Equivalently,
"reversing time", it is clear from \eqref{ch} that $h_{L,R}(x,t)$ are also the solutions of the KPZ equation with initial
conditions (Brownian, flat or wedge) on the corresponding half-line, i.e.
$h_L(y,0)=a_L B_{0}(y) + w_L y$ for $y<0$ and $h_L(y,0)=-\infty$ for $y>0$, and
$h_R(y,0)=a_R B_{0}(y) - w_R y$ for $y>0$ and $h_R(y,0)=-\infty$ for $y<0$, respectively.
In addition, $h_c(x,t)$ is the solution with the droplet IC centered at $0$. 
Both interpretations will be used below. \\

We will be interested in the {\it joint} distribution of these three partition sums. To this aim we
define the following generating function in terms of scaled parameters
\bea \label{genfunctLR} 
\hat g_t(\sigma_1,\sigma_L,\sigma_R;\hat x) 
:= \overline{\exp( -e^{- t^{1/3} (\sigma_1-\hat x^2) } Z_c(x,t) -e^{- t^{1/3} (\sigma_L -\hat x^2) } Z_L(x,t)  -e^{- t^{1/3} (\sigma_R -\hat x^2) } Z_R(x,t))} 
\eea 
where the average is implicitly over the noise $\eta$ and the Brownian IC. 
We obtain below a formula, Eqs. \eqref{final10}-\eqref{K12},
for this generating function in the limit of large $t$. In that limit it becomes 
equal to the following joint CDF 
\be
\hat g_{\infty} (\sigma_1,\sigma_L,\sigma_R;\hat x) 
= \lim_{t \to +\infty} {\rm Prob}\left( t^{-\frac{1}{3}} (h_c(x,t) + \frac{x^2}{4 t} ) < \sigma_1 , \,
 t^{-\frac{1}{3}} (h_L(x,t) + \frac{x^2}{4 t} ) < \sigma_L , \, t^{-\frac{1}{3}} (h_R(x,t) + \frac{x^2}{4 t} ) < \sigma_R \right)
\ee 

In the limit of large time, using \eqref{Ai1},\eqref{Ai2},\eqref{Ai3}, we can translate this equality in terms of Airy processes.
One finds
\bea
h_{L,R}(x,t) \simeq t^{1/3} \hat h_{L,R}(\hat x) \quad , \quad h_c(x,t) \simeq t^{1/3} ({\cal A}_2(-\hat x) - \hat x^2)
\eea 
where the 
random variables $\hat h_{L,R}(\hat x)$ are defined as maxima over the following sum of processes in $y$
\bea \label{hLRdef} 
&& \hat h_L(\hat x)= \max_{\hat y<0}( {\cal A}_2(\hat y-\hat x) - (\hat y - \hat x)^2 + 2 \hat w_L \hat y + a_L \sqrt{2} 
B(\hat y) ) \\
&& \hat h_R(\hat x)= \max_{\hat y>0}( {\cal A}_2(\hat y-\hat x) - (\hat y - \hat x)^2 -  2 \hat w_R \hat y + a_R 
\sqrt{2} B(\hat y) ) \nn
\eea  
at fixed $\hat x$. We used that $ t^{-1/3} B_0(2 t^{2/3} \hat y) \equiv \sqrt{2} B(\hat y)$ where $B(\hat y)$ 
is another unit two-sided Brownian in the variable $\hat y$, with $B(0)=0$. We also
reversed for convenience the sign of the argument of the Airy process, which is 
statistically symmetric. Hence we have (the result for this quantity being given in Eqs. \eqref{final10}-\eqref{K12})
\be \label{gcontain}
\hat g_{\infty} (\sigma_1,\sigma_L,\sigma_R;\hat x) 
= {\rm Prob}\left( {\cal A}_2(-\hat x)  < \sigma_1 , \hat h_L(\hat x) + \hat x^2 < \sigma_L ,  
\hat h_R(\hat x) + \hat x^2 < \sigma_R  \right)
\ee 

We now specialize to the cases of most interest.

\begin{itemize}

\item {\it Joint CDF of Airy$_2$ at a point and value of a maximum involving Airy$_2$.}

Let us set $\hat x=0$ and $\sigma_L=\sigma_R=\sigma_2$.
\bea \label{joint1} 
&& \hat g_{\infty} (\sigma_1,\sigma_2,\sigma_2;0) 
= {\rm Prob}\left( {\cal A}_2(0)  < \sigma_1 , \max( \hat h_L(0), \hat h_R(0))  < \sigma_2  \right) \\
&& 
 = {\rm Prob}\left( {\cal A}_2(0)  < \sigma_1 , 
 \max_{\hat y}( {\cal A}_2(\hat y) - \hat y^2 + V(\hat y) ) < \sigma_2  \right) \\
 && V(\hat y)=
 ( 2 \hat w_L \hat y + a_L \sqrt{2} B(\hat y) ) \theta(-\hat y) + 
 ( - 2 \hat w_R \hat y + a_R \sqrt{2} B(\hat y) ) \theta(\hat y)   \label{defV} 
\eea
Hence we obtain the JCDF of ${\cal A}_2(0)$ and  $\max_{\hat y}( {\cal A}_2(\hat y) - \hat y^2 + V(\hat y) )$
in the same environment (same realization of the Airy$_2$ process) for a family of potentials $V(\hat y)$.

An example of particular interest for the two-time KPZ problem in the large time separation limit, is the case $a_L=a_R=1$ in the
limit $\hat w_L=\hat w_R=0^+$ where one recovers the JPDF defined in the main text (the result for it being displayed in \eqref{ResF})
\bea
G(\sigma_1,\sigma_2)  &:=& {\rm Prob}\left( {\cal A}_2(0)  < \sigma_1 , 
 \max_{\hat y}( {\cal A}_2(\hat y) - \hat y^2 + \sqrt{2} B(\hat y) ) < \sigma_2  \right) \\
&=& \lim_{\hat w_L \to 0^+, \hat w_R \to 0^+} \hat g_{+\infty,a_{L,R}=1}(\sigma_1,\sigma_2,\sigma_2;0)  
\nn
\eea
where $B(\hat y)$ is a doubled sided unit Brownian (with $B(0)=0$.). It is
generalized to arbitrary $\hat x$ as
\bea  \label{Fx} 
G_{\hat x}(\sigma_1,\sigma_2)  &:=& {\rm Prob}\left( {\cal A}_2(-\hat x)  < \sigma_1 , 
 \max_{\hat y}( {\cal A}_2(\hat y-\hat x) - (\hat y-\hat x)^2 + \sqrt{2} B(\hat y) ) < \sigma_2 - \hat x^2 \right) \\
&=& \lim_{\hat w_L \to 0^+, \hat w_R \to 0^+} \hat g_{+\infty,a_{L,R}=1}(\sigma_1,\sigma_2,\sigma_2;\hat x) \nn
\eea
which applies to the two-time KPZ problem (equivalently to the two-DP problem) with 
a shift $\hat x$ in the droplet initial condition (equivalently in the DP endpoint). Its exact
expression for arbitrary $\hat x$ is given in \eqref{resFx}. 

\item {\it PDF or argmax of Airy$_2$ minus a parabola plus a Brownian, and extensions.}

Let us set $w_L=w_R=0^+$, we can rewrite, defining $\hat z=\hat y - \hat x$
\bea
&& \hat h_L(\hat x) = \max_{\hat z < - \hat x}( {\cal A}_2(\hat z) - \hat z^2  + a_L 
\sqrt{2} B(\hat x + \hat z) )  \\
&& \hat h_R(\hat x)  = \max_{\hat z> - \hat x}( {\cal A}_2(\hat z) - \hat z^2  + a_R 
\sqrt{2} B(\hat x + \hat z) ) 
\eea
which can also be written as
\bea
&& \hat h_L(\hat x) = \max_{\hat z < - \hat x}( {\cal A}_2(\hat z) - \hat z^2  + a_L 
\sqrt{2} \tilde B(\hat z) ) + a_L \sqrt{2}  B(\hat x)   \\
&& \hat h_R(\hat x)  = \max_{\hat z> - \hat x}( {\cal A}_2(\hat z) - \hat z^2  + a_R 
\sqrt{2} \tilde B(\hat z) ) + a_R \sqrt{2} B(\hat x)  
\eea
where we have defined
\bea
\tilde B(\hat z) := B( \hat z + \hat x) - B(\hat x)
\eea
which is also a unit two-sided Brownian with $\tilde B(0)=0$, which is, however, correlated to $B$.
Because of this correlation, there are only two main applications.

The first is for the PDF of argmax of Airy$_2$ minus a parabola plus a Brownian, which is equivalent to 
the PDF of the midpoint of a DP with Brownian initial conditions (see discussion in the main text
and Fig. \ref{fig:DP}).
Consider now $a_L=a_R=1$. The term $\sqrt{2} B(\hat x)$ cancels in the difference 
$\hat h_R(\hat x)-\hat h_L(\hat x)$, i.e. 
\bea
&& \hat h_R(\hat x)-\hat h_L(\hat x) = \max_{\hat z> - \hat x}( {\cal A}_2(\hat z) - \hat z^2 + \sqrt{2} \tilde B(\hat z)  )
- \max_{\hat z < - \hat x}( {\cal A}_2(\hat z) - \hat z^2 + \sqrt{2} \tilde B(\hat z)  ) 
\eea
Hence we have
\bea
&&  {\rm Prob}\left( {\cal A}_2(- \hat x)  < \sigma_1 ,  \hat h_R(\hat x)-\hat h_L(\hat x)  \in [\sigma,\sigma + d\sigma]   \right) \\
&& = \lim_{\hat w_L \to 0^+, \hat w_R \to 0^+}  \int_{-\infty}^{+\infty}  d\sigma_R d\sigma_L  ( \partial_{\sigma_L} \partial_{\sigma_R} \hat g_{+\infty, a_{L,R}=1}(\sigma_1,\sigma_L,\sigma_R; \hat x) ) \delta(\sigma_R-\sigma_L-\sigma)
d\sigma \nn
\eea 
An application is to the PDF of argmax of the following variational problem. Let us define
(we have now suppressed the tilde subscript on $B$)
\bea
\hat z_m={\rm argmax}_{\hat z \in \mathbb{R}} \left( {\cal A}_2(\hat z) - \hat z^2 + \sqrt{2} B(\hat z) \right) \label{maxgen} 
\eea
Clearly one has 
\bea
&& H(- \hat x) := {\rm Prob}(\hat z_m >  - \hat x) = {\rm Prob}( \hat h_R(\hat x)- \hat h_L(\hat x)  > 0 ) \\
&& =  \lim_{\hat w_L \to 0^+, \hat w_R \to 0^+}  
\int_{-\infty}^{+\infty} d\sigma_R [\partial_{\sigma_R} 
\hat g_{+\infty, a_{L,R}=1}(+\infty,\sigma_L,\sigma_R; \hat x) )|_{\sigma_L=\sigma_R^-} \nn\label{Hmethod} 
\eea 
The calculation of $H(-\hat x)$ is performed in Section VII, where its properties are
studied. The final formula for it is \eqref{resH2}, equivalently \eqref{resH1} in the text. 

In principle we can extract a bit more information, keeping $\sigma_1$ arbitrary we obtain the JPDF 
\bea
{\rm Prob}({\cal A}_2(-\hat x)  < \sigma_1  , \hat z_m >  - \hat x)  
= \lim_{\hat w_L \to 0^+, \hat w_R \to 0^+}  
\int_{-\infty}^{+\infty} d\sigma_R [\partial_{\sigma_R} 
\hat g_{+\infty, a_{L,R}=1}(\sigma_1,\sigma_L,\sigma_R; \hat x) )|_{\sigma_L=\sigma_R^-}
\eea
Taking a derivative w.r.t. $\sigma_1$ and dividing by $\partial_{\sigma_1} F_2(\sigma_1)$ 
it gives the probability that the longer polymer in Fig. \ref{fig:DP} passes to the right of 
the origin, given the value of the Airy process at this point (equivalently, by translational
invariance, one can shift all endpoints by $- \hat x$ in Fig. \ref{fig:DP} and ask the DP to
pass right of $- \hat x$). We do not display the resulting formula here, as
it is a simple exercise to get it from \eqref{resres1} following similar
steps as in Section VII. 1.

\item {\it Joint PDF of argmax and max, equivalently of endpoint position and free energy of a DP.}

Consider $a_L=a_R=0$ and focus on $w_{L,R} \to 0^+$. Then we have 
\bea
&& \hat h_L(\hat x) = \max_{\hat z < - \hat x}( {\cal A}_2(\hat z) - \hat z^2 ) \quad , \quad  \hat h_R(\hat x) = \max_{\hat z> - \hat x}( {\cal A}_2(\hat z) - \hat z^2 )
\eea
Let us define the value and position of the following variational problem
\bea
\hat h_m = \max_{\hat z \in \mathbb{R}}( {\cal A}_2(\hat z) - \hat z^2 ) \quad , \quad \hat z_m = {\rm argmax}_{\hat z \in \mathbb{R}}( {\cal A}_2(\hat z) - \hat z^2 )
\eea 
It can also be seen as the endpoint position $z_m$ and associated (minus) free energy $h_m$
of a directed polymer from the point $(0,0)$ to the line $(z,t)$ (with a free endpoint $z \in \mathbb{R}$, i.e. the point-to-line problem), in the
limit $t \to +\infty$, expressed in rescaled variables $\hat h_m = t^{-1/3} \hat h_m$ and 
$\hat z_m=z_m/2 t^{2/3}$. 

Then we will obtain here the joint C-PDF of the position and the value of the maximum as
\bea \label{JCPDF} 
&& {\rm Prob}( \hat z_m > - \hat x , \hat h_m) = {\rm Prob}( \hat h_R(\hat x) - \hat h_L(\hat x) >0 , \hat h_R(\hat x) = \hat h_m )
\\
&& =  \int_{-\infty}^{+\infty} d\sigma_R  \delta(\sigma_R-\hat h_m- \hat x^2) \int_{-\infty}^{\sigma_R} d \sigma_L  \partial_{\sigma_L}  \partial_{\sigma_R} 
\hat g_{+\infty, a_{L,R}=0}(+\infty,\sigma_L,\sigma_R; \hat x)
\nn \\&& = [\partial_{\sigma_R} \hat g_{+\infty, a_{L,R}=0}(+\infty,\sigma_L,\sigma_R; \hat x)
]|_{\sigma_L=\sigma_R^-, \sigma_R=\hat h_m+ \hat x^2} \nn
\eea 
using, in the integration by part, that $\hat g_{\infty}$ vanishes when any of its arguments is sent to $- \infty$. The formula for \eqref{JCPDF} obtained from the present method is given in Section XII. 2.,
formula \eqref{JCPDF2} and \eqref{JCPDF3}. 

It is interesting to note that formulae for the joint PDF of $(\hat h_m,\hat z_m)$ were obtained 
previously by very
different methods: (i) within a rigorous approach in
 \cite{QuastelEndpoint} as a formula 
 involving the Airy function and the resolvent of an associated operator, and 
 (ii) from studying non-crossing Brownian paths, 
 by Schehr in \cite{SchehrEndpoint} as a formula involving
a solution to the Lax pair for the Painlev\'e II equation. It was later shown in \cite{BaikLiechtySchehr} that these formulas are equivalent.
 
 The present derivation is much closer in spirit to the one of
 Dostenko \cite{dotsenkoEndpoint} for {\it the position} of the endpoint $\hat z_m$. 
 Dotsenko formula for the PDF of $\hat z_m$ was shown in Appendix C of
 \cite{LiechtyTail} to coincide with the predictions of \cite{QuastelEndpoint}
 and \cite{SchehrEndpoint} for the marginal distribution of of $z_m$. 
 Here however we obtain directly the joint PDF of $(h_m,z_m)$. 
 Also our integrals have a slightly different form. We will not attempt here
 to show that this formula is equivalent to the ones of \cite{QuastelEndpoint}
 and \cite{SchehrEndpoint} for the JPDF, although it is likely to be correct.
 
Note that here we obtain the more general result, i.e the triple JPDF
\bea
&& {\rm Prob}({\cal A}_2(- \hat x) < \sigma_1, \hat z_m > - \hat x , \hat h_m) =
[\partial_{\sigma_R} \hat g_{+\infty, a_{L,R}=0}(\sigma_1,\sigma_L,\sigma_R; \hat x)
]|_{\sigma_L=\sigma_R^-, \sigma_R=\hat h_m+ \hat x^2}
\eea
which contains both the above JPDF and also the JPDF of the KPZ heights for the 
droplet and flat IC in the same noise. Indeed in Section XII. 1. we will display ${\rm Prob}({\cal A}_2(- \hat x) < \sigma_1, \max_{\hat z \in \mathbb{R}}( {\cal A}_2(\hat z) - \hat z^2 ))$,
which gives the large time limit of the JPDF of the scaled KPZ heights of the flat and droplet 
solutions in the same noise, equivalently of the point-to-line and point-to-point DP scaled free energy in the same random potential.
 
\end{itemize}

\medskip

\newpage

\section{III. Calculation of the generating function}

\subsection{III.1. Moment expansion} 

For notational convenience we introduce a second set of rescaled parameters
\bea
\lambda := 2^{-2/3} t^{1/3}  \quad , \quad s = 2^{2/3} ( \sigma - \hat x^2)  \quad , \quad \tilde w_{L,R}= 
\lambda w_{L,R} = 2^{-2/3} \hat w_{L,R} \quad , \quad \tilde x = x/\lambda^2 = 2^{7/3} \hat x
\eea 
where the parameter $\lambda$ was introduced in Ref. \cite{we,dotsenko,we-flat}. With these new
parameters the generating function \eqref{genfunctLR} can be written, and expanded in series, in terms of the
joint moments, as follows
\bea \label{genfunctLR2} 
&& \hat g_t(\sigma_1,\sigma_L,\sigma_R;\hat x) = g_\lambda(s_1,s_L,s_R;\tilde  x) 
:= \overline{e^{-e^{-\lambda s_1 } Z_c(x,t) -e^{-\lambda s_L} Z_L(x,t) - e^{-\lambda s_R} Z_R(x,t)}} = \sum_{n=0}^{+\infty} \frac{(-1)^n}{n!} {\cal Z}_n \label{defgla} \\
&&  {\cal Z}_n := \overline{ (e^{-\lambda s_1 } Z_c(x,t) + e^{-\lambda s_L} Z_L(x,t) + e^{-\lambda s_R} Z_R(x,t))^n}
\eea
From the definitions \eqref{ZZdef}, upon averaging
over $\eta$ and $B_0$ (which we have chosen independent), 
the moments can be written as
\be
{\cal Z}_n=   \int dy_1..dy_n \overline{\prod_{\alpha=1}^n Z_\eta(x,t|y_\alpha,0)}^\eta 
 \langle  \prod_{\alpha=1}^{n}
(e^{-\lambda s_L+a_L B_0(y_\alpha) + w_L y_\alpha} \theta(-y_\alpha) + e^{-\lambda s_1 } \delta(y_\alpha)
+ e^{- \lambda s_R + a_R B_0(y_\alpha)-  w_R y_\alpha} \theta(y_\alpha))  \rangle_{B_0} 
\label{zn1}
\ee

\subsection{III.2. Quantum mechanics and overlap} 

As is now well known \cite{kardareplica,bb-00} the $\eta$ average in the middle of 
(\ref{zn1}) can be rewritten as 
the expectation value between initial and final states of the quantum-mechanical evolution 
operator associated to the attractive Lieb-Liniger (LL) Hamiltonian for $n$ identical particles \cite{ll}: 
\be
H_n = -\sum_{\alpha=1}^n \frac{\partial^2}{\partial {x_\alpha^2}}  - 2 \bar c \sum_{1 \leq  \alpha< \beta \leq n} \delta(x_\alpha - x_\beta) \quad , \quad \bar c=1
\label{LL}
\ee
The eigenfunctions are known from the Bethe ansatz \cite{ll}. They are parameterized by a set of rapidities
$\mu \equiv \{ \lambda_1,..\lambda_n\}$ which are solution of a set of coupled equations, the Bethe equations
(see below). The eigenfunctions are totally symmetric in the $x_\alpha$, and in the sector $x_1 \leq x_2 \leq \dots \leq x_n$,
take the (un-normalized) form 
\be \label{def1}
\Psi_\mu(x_1,..x_n) =  \sum_{P \in S_n} A_P \prod_{j=1}^n e^{i \sum_{\alpha=1}^n \lambda_{P_\alpha} x_\alpha} \, , \quad 
A_P= \prod_{1 \leq \alpha < \beta \leq n} a_{\lambda_{P_\beta},\lambda_{P_\alpha}}  \quad , \quad 
 a_{\lambda_{\beta},\lambda_{\alpha}}  = \Big(1 + \frac{i  
}{\lambda_{\beta} - \lambda_{\alpha}}\Big)\,.
\ee
They can be deduced in the other sectors from their full symmetry with respect to particle exchanges. 
The sum runs over all $n!$ permutations $P$ of the rapidities $\lambda_\alpha$. The corresponding eigenenergies are
$E_\mu=\sum_{\alpha=1}^n \lambda_\alpha^2$. One can then rewrite the moment as
a sum over eigenstates
\bea \label{sumf}
{\cal Z}_n =  \langle x \dots x | e^{-t H_n} | \Phi_0 \rangle = \sum_\mu \Psi_\mu(x,..x) \frac{e^{-t E_\mu}}{||\mu||^2}  \langle \Psi_\mu| \Phi_0 \rangle 
= \sum_\mu \Psi^*_\mu(x,..x) \frac{e^{-t E_\mu}}{||\mu||^2}  \langle \Phi_0 | \Psi_\mu \rangle
\eea 
where we have used that ${\cal Z}_n$ is real, and for convenience we will
work with the second (i.e. complex conjugate) expression. 
Here $|\Phi_0 \rangle$ is the inital state (see below) and $| x \dots x \rangle$ is the final state,
with all particles at the same point $x$.
Since this state is fully symmetric in exchanges of particles, only
symmetric eigenfunctions will contribute and we can consider particles as bosons. 
In the formula (\ref{sumf}) we first need:
\be 
\Psi^*_\mu(x,..x) = n! e^{-  i x \sum_\alpha \lambda_\alpha} \,.
\label{Psixeq}
\ee 
The wavefunction of the initial replica state is also symmetric and equal to:
\bea
&& \Phi_0(Y) = \langle y_1,..y_n | \Phi_0 \rangle = 
 \langle  \prod_{\alpha=1}^{n}
(e^{-\lambda s_L+a_L B_{0L}(-y_\alpha) + w_L y_\alpha} \theta(-y_\alpha) + e^{-\lambda s_1 }
+ e^{- \lambda s_R + a_R B_{0R}(y_\alpha)-  w_R y_\alpha} \theta(y_\alpha))  \rangle_{B_{0L},B_{0R}}
\eea 
where here and below coordinate multiplets are denoted by capital letters, e.g. $Y \equiv y_1,..y_n$. 
For convenience, we are using that $B_0(y)=\theta(-y) B_{0L}(-y) + \theta(y) B_{0R}(y)$ where 
$B_{0L,0R}$ are two independent one-sided unit Brownians (both on the positive axis). 

Taking advantage of the symmetry of the wavefunctions, we rewrite the overlap of the
initial state and any eigenstate 
\bea
&& \langle \Phi_0 | \Psi_\mu \rangle = \int dY \Psi_\mu(Y) \Phi_0(Y) =
n! \sum_{\substack{ n_1,n_L, n_R \geq 0 \\   n_1+n_L+n_R=n  }}
 \frac{1}{n_1!} e^{- \lambda n_1 s_1 - \lambda n_L s_L -  \lambda n_R s_R}
\int_{y_1< \cdots <y_{n_L}<0}  \langle e^{B_{0L}(-y_1)+..+B_{0L}(-y_{n_L})} \rangle_{B_{0L}} \nn \\
&& ~~~~~~~ \times 
\int_{0<y_{n-n_R+1}<\cdots <y_{n}} \langle e^{B_{0R}(y_{n-n_R+1})+..+B_{0R}(y_{n})} \rangle_{B_{0R}} 
\Psi_\mu(y_1,..y_{n_L},0,..0,y_{n-n_R+1},..y_{n}) 
\eea
Inserting the form \eqref{def1} of the wavefunction, we obtain the overlap as
\be \label{nLnR} 
\langle \Phi_0 | \Psi_\mu \rangle = n! \sum_{P \in S_{n}} A_P 
\sum_{\substack{ n_1,n_L, n_R \geq 0 \\   n_1+n_L+n_R=n  }} 
 \frac{1}{n_1!} e^{- \lambda n_1 s_1 - \lambda n_L s_L -  \lambda n_R s_R}
~  G^L_{n_L,w_L,a_L}[\lambda_{P_1},.., \lambda_{P_{n_L}}] G^R_{n_R,w_R,a_R}[\lambda_{P_{n-n_R+1}},.., \lambda_{P_{n}}] 
\ee 
where, as in \cite{PLDCrossover17} we define 
\begin{eqnarray} \label{Gla1} 
&& G_{p,w,a}^L[\lambda_1,..,\lambda_p] := \int_{y_1<y_2< ..<y_p<0} 
e^{\sum_{j=1}^p (w + i \lambda_j) y_j }  <e^{\sum_{j=1}^p a B_{0L}(-y_j)}>_{B_{0L}} 
\\ 
&&  G_{p,w,a}^R[\lambda_1,..,\lambda_p]  := \int_{0<y_1<y_2< ..<y_p} 
e^{\sum_{j=1}^p (- w + i \lambda_j) y_j } <e^{\sum_{j=1}^p a B_{0R}(y_j)}>_{B_{0R}}
 \nonumber 
\end{eqnarray}
with $G^L_{0,w,a}=G^R_{0,w,a}=1$. Note that for each permutation $P$ there are
three groups of rapidities, of sizes $n_L,n_1,n_R$ respectively, associated to
$y<0$, $y=0$ and $y>0$ respectively. Permutations $P \in S_n$ exchanges rapidities 
within these groups, but also between them. The above integrals can be explicitly evaluated and furthermore,
as discussed in \cite{PLDCrossover17}, see Eqs. (57-62) there, 
for the two "solvable" cases, $a=0,1$ a "miracle" occurs allowing to perform exactly the summation over the permutations, leading to a factorized form \cite{SasamotoHalfBrownReplica,we-flat} 
\bea \label{miracle1}
&& H^R_{p,w,a=1}[\{\lambda_1,..\lambda_p\}] :=  \sum_{P \in S_p} A_P G^R_{p,w,a=1}[\lambda_{P_1},..,\lambda_{P_p}] = \frac{2^p}{ \prod_{j=1}^p (2 w - 1 - 2 i \lambda_j) }  \\
&&  H^L_{p,w,a=0}[\{\lambda_1,..\lambda_p\}] := \sum_{P \in S_p} A_P G^L_{p,w,a=0}[\lambda_{P_1},..,\lambda_{P_p}] =  \frac{1}{ \prod_{\alpha=1}^{p} (w+ i \lambda_\alpha)}  \prod_{1 \leq \alpha < \beta \leq p} \frac{2 w + i \lambda_\alpha + i \lambda_\beta - 1}{ 2 w+ i \lambda_\alpha + i \lambda_\beta}  \label{miracle2}
\eea
where we have introduced two new functions which depend only 
on the set of rapidities, not on their order. These miracle identities (with $p=n$) 
allow to obtain simple expressions for the terms 
where two of the three variables $n_L,n_R,n_1$ are zero in (\ref{nLnR}) but (a priori) not for the general term,
since there are then permutations which exchange rapidities between the three groups of rapidities.

Evaluation becomes possible however when the eigenstates are strings. We now follow the strategy of  \cite{PLDCrossover17}.

\subsection{III.3. Strings and combinatorial identities} 
\label{sec:combi}

In the limit of infinite system size, the rapidities solution to the Bethe equations are the so-called strings \cite{m-65}, and the
spectrum of $H_n$ is as follows. A general eigenstate is built by partitioning the $n$ particles into a set of $1 \leq n_s \leq n$ bound states called {\it strings} 
each formed by $m_j \geq 1$ particles with $n=\sum_{j=1}^{n_s} m_j$. 
The rapidities associated to these states are written as 
\be\label{stringsol}
\lambda_{j, a}=k_j - \frac{i}2(m_j+1-2a) 
\ee 
where $k_j$ is a real momentum.
Here, $a = 1,...,m_j$ labels the rapidities within the string $j=1,\dots n_s$. We will denote
$|\mu \rangle \equiv | {\bf k}, {\bf m} \rangle$ these strings states, labelled by the set of $k_j,m_j$, $j=1,..n_s$. 
Here and below the boldface represents vectors with $n_s$ components. 
Inserting these rapidities in (\ref{def1}) leads to the Bethe eigenfunctions of the infinite system, and their corresponding
eigenenergies:
\be \label{en} 
E_\mu= \frac{1}{12} n t + \tilde E( {\bf k}, {\bf m}) \quad , \quad 
\tilde E( {\bf k}, {\bf m})  := \sum_{j=1}^{n_s} m_j k_j^2-\frac{1}{12} m_j^3
\ee 
We have separated a trivial part of the energy, which can be eliminated by defining
\bea
{\cal Z}_n  = e^{- \frac{1}{12} n t}  \tilde {\cal Z}_n  \quad , \quad Z(x,t) = e^{- \frac{1}{12} t}
\tilde Z(x,t) \quad , \quad 
h(x,t) = -  \frac{1}{12} t + \tilde h(x,t)
\eea
i.e. leading to a simple shift in the KPZ field. We will implicitly study in the remainder of the paper
$\tilde {\cal Z}_n$, $\tilde Z(x,t)$ and $\tilde h(x,t)$ but will remove the tilde in these quantities
for notational simplicity (as mentioned in the text).  The formula for the norm of the string states reads \cite{cc-07}:
\be
\frac{1}{||\mu ||^2} = \frac{1}{n!  L^{n_s} } \Phi({\bf k}, {\bf m}) \prod_{j=1}^{n_s} \frac1{m_j^{2}}  \quad , \quad \Phi({\bf k}, {\bf m}) = \prod_{1\leq i<j\leq n_s} 
\Phi_{k_i,m_i,k_j,m_j}   \, , \quad  \Phi_{k_i,m_i,k_j,m_j}=\frac{4(k_i-k_j)^2 +(m_i-m_j)^2}{4(k_i-k_j)^2 +(m_i+m_j)^2} \label{norm}
\ee
so that the formula (\ref{sumf}) for the moments becomes for $L \to +\infty$ (provided all limits exist)
\bea
 {\cal Z}_n  = 
\sum_{n_s=1}^n \frac{1}{n_s!} \sum_{(m_1,..m_{n_s})_n} \prod_{j=1}^{n_s} \int \frac{dk_j}{2 \pi m_j}  
\Phi({\bf k}, {\bf m}) e^{- t \tilde E( {\bf k} , {\bf m})} e^{ - i \sum_{j=1}^{n_s} m_j k_j x} 
 \langle \Phi_0 | {\bf k}, {\bf m} \rangle \label{momentn}
\eea 
where the second sum is over the set of partitions, denoted $(m_1,..m_{n_s})_n$, of the integer $n =\sum_{j=1}^{n_s} m_j$ into $n_s$ parts, with each $m_j \geq 1$.

It remains to calculate the overlap, formula (\ref{nLnR}). If the states are strings, the sum over permutations 
can be performed. Let us sketch the main idea, introduced by Dotsenko \cite{dotsenkoGOE}, and checked in more details in 
\cite{PLDCrossover17} (to which we refer for details). Consider a string state $ \Psi_\mu=| {\bf k} , {\bf m}   \rangle$
with rapidities given by \eqref{stringsol}. From the definition \eqref{def1}, the only permutations $P$ which have a non vanishing amplitude $A_P$, are those such that for each string the intra-string order of increasing imaginary part is
maintained. Hence if one is given a set of $3 n_s$ integers $(m^L_j,m^1_j,m^R_j)$, $j=1,\cdots,n_s$:
\bea
&& 0 \leq m_j^L \leq m_j \quad , \quad  0 \leq m_j^1 \leq m_j  \quad , \quad 0 \leq m_j^R \leq m_j \quad , \quad m_j^L+ m_j^1 +m_j^R = m_j 
\eea
which specifies how many particles in each string belongs to each of the three groups,
then one knows {\it bijectively} the three sets of rapidities which belong of each group.
For instance one knows that the first set of rapidities is:
\bea
 \Lambda_L= \{ \lambda_{j,r_j} , j=1,\cdots,n_s, r_j =1,\cdots,m_j^L \}
\eea
and that the second set is $ \Lambda_R = \{ \lambda_{j,r_j} , j=1,..n_s, r_j =m_j^L+m^1_j+1,..m_j \}$.
To treat these two sets on equal footing, it is convenient to introduce the notation:
\bea
&& \lambda_{j,r_j} = k_j - \frac{i}{2} (m_j+1 - 2 r_j)   \quad 1 \leq r_j \leq m_j^L \\
&& \bar \lambda_{j,r_j} = k_j + \frac{i}{2} (m_j+1 - 2 r_j)   \quad 1 \leq r_j \leq m_j^R
\eea 
Finally, the third set, $\Lambda_1$, is simply the complement of the first two sets. 

Consider now the overlap (\ref{nLnR}) for a given string state $|{\bf k}, {\bf m} \rangle$.
The sum over $P \in S_n$ can be made in two stages. 
In a first stage one fixes a set of $3 n_s$ integers ${\bf m}^{R,1,L}=\{m^{R,1,L}_j\}_{j=1,\cdots n_s}$. That fixes
which of the $n$ rapidities of the string belong to each of the three groups $\Lambda_L,\Lambda_1,\Lambda_R$. 
Hence there is no more permutations (i.e. with a non vanishing contribution)
exchanging rapidities between the three groups, and the only remaining permutations are permutations inside each group.
One thus performs the sum over permutations inside
each group. It then remains to sum over the variables ${\bf m}^{R,1,L}$, a sum which we
perform in a second stage.
One takes advantage that one can factor $A_P$ as (with $n=n_L+n_1+n_R$) 
\bea
&& \prod_{1 \leq \alpha < \beta \leq n} a_{\lambda_{P_\beta},\lambda_{P_\alpha}}  
= \prod_{1 \leq \alpha < \beta \leq n_L}   a_{\lambda_{P_\beta},\lambda_{P_\alpha}}
 \prod_{n-n_R+1 \leq \alpha < \beta \leq n}   a_{\lambda_{P_\beta},\lambda_{P_\alpha}} 
 \prod_{n_L+1 \leq \alpha < \beta \leq n_L+n_1}   a_{\lambda_{P_\beta},\lambda_{P_\alpha}} \times \hat G
\eea
with
\bea \label{hatG} 
&& \hat G= \prod_{\alpha =1}^{n_L} \prod_{\beta=n_L+1+n_1}^n 
a_{\lambda_{P_\beta},\lambda_{P_\alpha}} \times \prod_{\alpha =1}^{n_L} \prod_{\beta=n_L+1}^{n_L+n_1}
a_{\lambda_{P_\beta},\lambda_{P_\alpha}} \times 
\prod_{\alpha =n_L+1}^{n_L+n_1} \prod_{\beta=n_L+1+n_1}^{n}
a_{\lambda_{P_\beta},\lambda_{P_\alpha}} 
\eea 
and one defines the two fully symmetric functions of their arguments
\bea \label{defHt}
&& H_{n_L}^L[\{\lambda_1,..\lambda_{n_L}\}] = \sum_{P \in S_{n_L}} \big( \prod_{1 \leq \alpha < \beta \leq n_L}   a_{\lambda_{P_\beta},\lambda_{P_\alpha}} \big) 
G^L_{n_L,w_L,a_L}[\lambda_{P_1},.., \lambda_{P_{n_L}}] \\
&& H_{n_R}^R[\{\lambda_{n-n_R+1},..\lambda_{n}\}] = \sum_{P \in S_{n_R}} \big( \prod_{n-n_R+1 \leq \alpha < \beta \leq n}   a_{\lambda_{P_\beta},\lambda_{P_\alpha}} \big) 
G^R_{n_R,w_R,a_R}[\lambda_{P_{n-n_R+1}},.., \lambda_{P_{n}}] \nn
\eea 
One can
then evaluate $H_{n_L}^L$ on the set $\Lambda_L$ and $H_{n_R}^R$ on the set $\Lambda_R$.
One thus defines:
\bea
&& \tilde H^L[{\bf k}, {\bf m}, {\bf m}^L] = H_{n_L}^L[\lambda_{1,1},..\lambda_{1,m_1^L},..., \lambda_{n_s,1},..
\lambda_{n_s,m_{n_s}^L}] \\
&& \tilde H^R[{\bf k}, {\bf m}, {\bf m}^R] = H_{n_R}^R[\bar \lambda_{1,1},..\bar \lambda_{1,m_1^R},..., \bar \lambda_{n_s,1},..
\bar \lambda_{n_s,m_{n_s}^R}]
\eea 
Note that the functions on the left do not explicitly depend any more on the choice
$n_R,n_1,n_L$, they depend on this choice only via $n_L = \sum_j m_j^L$, $n_1 = \sum_j m_j^1$,
$n_R = \sum_j m_j^R$ and $n_L+n_R+n_1=n$,
with $m_j=m_j^L+m_j^1+m_j^R$. Note also that the sum over permutations in the middle group, corresponding
to particle with $y_\alpha=0$, yields the
simple result
\bea
\sum_{P \in S_{n_1}}  \prod_{n_L+1 \leq \alpha < \beta \leq n_L+1+n_1}   a_{\lambda_{P_\beta},\lambda_{P_\alpha}} = n_1!
\eea

So we finally arrive at the following formula for the overlap
\bea \label{overfinal} 
\langle \Phi_0 | {\bf k} , {\bf m}   \rangle = n! 
\sum_{\substack{ n_L, n_1,n_R \geq 0 \\   n_L+n_1+n_R = n  }}
\sum_{\substack{{\bf m}^L + {\bf m}^1 + {\bf m}^R = {\bf m} \\ \sum_{j=1}^{n_s} m_j^L=n_L \\ 
\sum_{j=1}^{n_s} m_j^1=n_1 \\
\sum_{j=1}^{n_s} m_j^R=n_R } }
\tilde H^L_{n_L,w_L,a_L}[{\bf k}, {\bf m}, {\bf m}^L]  \tilde H^R_{n_R,w_R,a_R}[{\bf k},{\bf m}, {\bf m}^R]  
{\cal G}[{\bf k}, {\bf m}^{ L} , {\bf m}^1, {\bf m}^{ R}] 
\eea 
Here ${\cal G}[{\bf k}, {\bf m}^{ L} , {\bf m}^1, {\bf m}^{ R}]$ is equal to the product $\hat G$ 
in \eqref{hatG}, which, for a fixed set $({\bf m}^{ L} , {\bf m}^1, {\bf m}^{ R})$ is 
independent of the permutations (since these only exchanges rapidities inside each group),
which is why formula \eqref{overfinal} holds. The explicit calculation of ${\cal G}$ 
yields a complicated product of Gamma functions (see \cite{dotsenkoGOE} and 
\cite{PLDCrossover17} for examples 
of such calculations). However we will not need
its precise form in what follows. 

Inserting of \eqref{overfinal} into \eqref{momentn} and \eqref{genfunctLR} gives an a priori exact expression for the generating function at arbitrary time. It requires evaluating the functions $\tilde H^{L,R}$ 
by injecting the string rapidities (\ref{stringsol}), into the
formula (\ref{miracle1}) and (\ref{miracle2}), according to the rule (\ref{defHt}). We find
(with the help of formula (60-63) in \cite{PLDCrossover17})
\bea
&& \tilde H^L_{n_L,w_L,a_L}[{\bf k} , {\bf m} , {\bf m}^L] = \prod_{j=1}^{n_s}  
S^{w_L,a_L}_{m_j^L,m_j,k_j} \prod_{1 \leq i < j \leq n_s} 
D^{w_L,a_L}_{m_i^L,m_i,k_i,m_j^L,m_j,k_j} \label{Htfinal}  \\
&& \tilde H^R_{n_R,w_R,a_R}[{\bf k} , {\bf m} , {\bf m}^R] = \prod_{j=1}^{n_s}  
S^{w_R,a_R}_{m_j^R,m_j,- k_j} \prod_{1 \leq i < j \leq n_s} 
D^{w_R,a_R}_{m_i^R,m_i,- k_i,m_j^R,m_j,- k_j}  \nn
\eea 
where the single string factors are:
\bea \label{single} 
&& 
S^{w,0}_{m^L,m,k} = S^{w,0,L}_{m^L,m,k} = 2^{m_L} \frac{\Gamma(2 w + 2 i k + m-m_L)}{\Gamma(2 w + 2 i k + m)}  \\
&& S^{w,1}_{m^L,m,k} = S^{w,1,L}_{m^L,m,k} = \frac{\Gamma(w + i k + \frac{m}{2} - m^L)}{
\Gamma(w + i k + \frac{m}{2} )} \nn
\eea
Note that the calculation is very similar to the one leading to Eq. (74-76) in \cite{PLDCrossover17}
with the important difference that the formula there apply only to the case $m^1_j=0$. 
Hence the notations for the $S$ and $D$ factors have slightly different arguments. 
We will not give the analogous formula to (77) for $D$ as we will not need it. 

\subsection{III.4. Large time limit and decoupling assumption: first form of the kernel}

In the large time limit, as in \cite{PLDCrossover17}, we {\it assume} that one can set the
product of factors $D$ and ${\cal G}$ to unity. 
This is of course a highly non-trivial and radical assumption,
however it is justified a posteriori by the results. It will be checked in all cases where
the solution is known by other means. This procedure follows what has been done in
other works, where it was also checked against other methods
\cite{dotsenkoGOE,PLDCrossoverDropFlat,ps-2point,dotsenko2pt,Spohn2ptnew,dotsenkoEndpoint,KPZFixedPoint,PLDCrossover17}. It is often called the "decoupling assumption" following the procedure 
introduced in 
\cite{ps-2point}. Here we use it in an a priori different form, similar to the one introduced in \cite{dotsenkoGOE}. Although these 
two implementations have been shown to coincide in some cases
\cite{dotsenko2pt,Spohn2ptnew}, to our knowledge there
is no general check of that. Nevertheless we use the name "decoupling assumption". 

Let us first obtain a closed expression once these factors are set to unity, and take
the large $\lambda$ limit in a second stage. Putting together formula
\eqref{overfinal}, \eqref{Htfinal}, \eqref{single}, with $D \to 1$ and ${\cal G} \to 1$,  
into \eqref{defgla} and \eqref{momentn} 
, we obtain
\bea
&& g_\lambda(s_1,s_L,s_R;\tilde  x)  = g^{\rm dec}_\lambda(s_1,s_L,s_R;\tilde  x)
= 1 + \sum_{n_s=1}^{+\infty} \frac{1}{n_s!} Z^{\rm dec}_\lambda(n_s,s_1,s_L,s_R;\tilde x) 
\eea
where the partition sum with fixed number of strings reads
\bea
&& Z^{\rm dec}_\lambda(n_s,s_1,s_L,s_R;\tilde x) = \prod_{j=1}^{n_s} [ \sum_{m_j=1}^\infty \int \frac{dk_j}{2 \pi m_j} 
(-1)^{m_j} e^{- i m_j k_j x } ]
 \Phi({\bf k}, {\bf m}) e^{- t \tilde E( {\bf k} , {\bf m})} \\
&& 
\times 
\sum_{{\bf m}^L + {\bf m}^1+ {\bf m}^R  = {\bf m} } 
 \prod_{j=1}^{n_s}   S^{w_L,a_L}_{m_j^L,m_j,k_j} S^{w_R,a_R}_{m_j^R,m_j,-k_j} 
 e^{- \lambda m^L_j s_L - \lambda m^1_j s_1 - \lambda m^R_j s_R } \nn
\eea 
Here the subscript "{\rm dec}" reminds that it is not the exact expression, but that
the decoupling assumption has been applied.
We have used that summing over $n$ in the generating function allows in turn to sum freely on
$n_L,n_1,n_R$ and freely on ${\bf m}$. Following the same steps as in \cite{PLDCrossover17} Section III. E.1, i.e.
using the standard determinant double-Cauchy identity:
\bea
\Phi({\bf k}, {\bf m}) = \prod_{j=1}^{n_s} (2 m_j) \det_{1 \leq i, j \leq n_s}[ \frac{1}{2 i (k_i-k_j) + m_i + m_j}] 
\eea 
performing the rescaling $k_j \to k_j/\lambda$, denoting $\tilde x=x/\lambda^2$ and
performing the Airy trick, i.e. the identity $e^{\frac{1}{3} \lambda^3 m^3} = \int dy \Ai(y) e^{\lambda m y}$ 
for $\lambda m>0$, we obtain
\bea
&& Z^{\rm dec}_\lambda(n_s,s_1,s_L,s_R;\tilde x) = 2^{n_s} \prod_{j=1}^{n_s} [ \sum_{m_j=1}^\infty \int \frac{dk_j}{2 \pi} dy_j \Ai(y_j) 
 (-1)^{m_j} e^{- i \lambda  m_j k_j \tilde x  - 4 \lambda m_j k_j^2 + \lambda m_j y_j  } ]  \\
&& 
\times \det_{1 \leq i, j \leq n_s}[ \frac{1}{2 i (k_i-k_j) + \lambda m_i + \lambda m_j}]  \times \sum_{ {\bf m}^L + {\bf m}^1 + {\bf m}^R
= {\bf m} } 
\prod_{j=1}^{n_s}  S^{w_L,a_L}_{m_j^L,m_j,\frac{k_j}{\lambda}} S^{w_R,a_R}_{m_j^R,m_j,-\frac{k_j}{\lambda}}     
e^{- \lambda m^L_j s_L - \lambda m^1_j s_1 - \lambda m^R_j s_R  } \nn
 \eea
Using standard manipulations \cite{we,dotsenko} the partition sum at fixed number of string $n_s$ 
can thus be expressed itself as a determinant:
\bea
&& Z^{\rm dec}_\lambda(n_s,s_1,s_L,s_R;\tilde x) = 
 \prod_{j=1}^{n_s} \int_{v_j>0} \;
 \det_{1 \leq i, j \leq n_s} M^\lambda_{s_1,s_L,s_R,\tilde x}(v_i,v_j) \\
\eea
with the Kernel:
\bea \label{Mss} 
 && M^\lambda_{s_1,s_L,s_R,\tilde x}(v_i,v_j) = \int \frac{dk}{2 \pi} dy \Ai(y + 4 k^2 + i k \tilde x + v_i + v_j) e^{- 2 i k (v_i-v_j)} 
 \phi_\lambda(k,y-s_L,y-s_R,y-s_1) \\
&& \phi_\lambda(k,y_L, y_R,y) =  2 \sum_{m^L, m^R, m^1 \geq 0, m=m^L+m^R + m^1 \geq 1} (-1)^{m^L+m^R+m^1} 
S^{w_L,a_L}_{m^L,m,\frac{k}{\lambda}} S^{w_R,a_R}_{m^R,m,-\frac{k}{\lambda}} 
 e^{\lambda m^L y_L +  \lambda m^R y_R + \lambda m^1 y} 
 \eea
 where the $S$ factors are given explicitly in (\ref{single}). 
The generating function thus becomes a Fredholm determinant:
\bea
g_\lambda^{\rm dec}(s_1,s_L,s_R;\tilde x) = {\rm Det}[ I + P_0 M^\lambda_{s_1,s_L,s_R,\tilde x} P_0 ]  
 \label{ggene2} 
\eea 
where $P_0(v)=\theta(v)$ is the projector on $[0,+\infty[$. Here, again, this expression is valid as soon as the factors $D$ and ${\cal G}$ are set (arbitrarily) to unity.\\

To study the large time limit, we first rewrite:
\bea
 \phi_\lambda(k,y_L, y_R,y) =  -2 + 2 \sum_{m^L \geq 0, m^R \geq 0 , m^1 \geq 0} (-1)^{m^L+m^R+m^1} 
S^{w_L,a_L}_{m^L,m^L+m^R+m^1,\frac{k}{\lambda}} S^{w_R,a_R}_{m^R,m^L+m^R+m^1,-\frac{k}{\lambda}} 
 e^{\lambda m^L y_L +  \lambda m^R y_R + \lambda m^1 y } \nn
 \eea
and use the Mellin-Barnes identity:
\bea
\sum_{m=0}^\infty (-1)^m f(m) = \frac{-1}{2 i} \int_C \frac{dz}{\sin \pi z} f(z) 
\eea 
where $C=\kappa + i \mathbb{R}$, $-1<\kappa<0$, valid provided $f(z)$ is meromorphic, with no pole 
for $z> \Re(\kappa)$, and sufficient decay at infinity. It allows to rewrite (for $2 w_{L,R}+\kappa>0$)
\be
\phi_\lambda(k,y_L, y_R,y) =  -2 + 2  (\frac{-1}{2 i})^2 \int_C \frac{dz_L}{\sin \pi z_L} \int_C \frac{dz_R}{\sin \pi z_R} 
(\frac{-1}{2 i}) \int_C \frac{dz}{\sin \pi z}
S^{w_L,a_L}_{z_L,z_R,z,\frac{k}{\lambda}} S^{w_R,a_R}_{z_R,z_L,z,-\frac{k}{\lambda}} 
 e^{\lambda z_L y_L +  \lambda z_R y_R + \lambda z y}
\ee
Here the analytic continuation $f(m) \to f(z)$ has been performed using the 
second expression in (\ref{single}), and we now define
\bea \label{single2} 
&& 
S^{w,0}_{z_L,z_R,z,k} = 2^{z_L} \frac{\Gamma(2 w + 2 i k + z_R+z)}{\Gamma(2 w + 2 i k + z_R+z_L+z)}  \\
&& S^{w,1}_{z_L,z_R,z,k} = \frac{\Gamma(w + i k + \frac{z_R+z-z_L}{2})}{
\Gamma(w + i k + \frac{z_R+z_L+z}{2} )} \nn
\eea

We now rescale $z_{L,R} \to z_{L,R}/\lambda$, and we study the large time limit $\lambda \to +\infty$. 
We first recall the definition of the rescaled drifts:
\bea
\tilde w_L = w_L \lambda \quad , \quad \tilde w_R = w_R \lambda
\eea 
and we use that for $a=0,1$:
\bea
\lim_{\lambda \to +\infty} S^{w=\tilde w/\lambda,a}_{\frac{z_L}{\lambda},\frac{z_R}{\lambda},\frac{z}{\lambda},\frac{k}{\lambda}} 
=  1 + \frac{ (1+a) z_L}{2 \tilde w + 2 i k + z_R + z - a z_L}
\eea 
as can be seen from (\ref{single2}). Thus we obtain the infinite $\lambda$ limit in the form of a multiple
contour integral:
\bea \label{phi1} 
&& \phi_{\infty}(k,y_L, y_R,y) =  -2 \\
&& - 2 \int_{C'} \frac{dz_L}{2 i \pi z_L} \int_{C'} \frac{dz_R}{2 i \pi z_R} \int_{C'} \frac{dz}{2 i \pi z} 
(1 + \frac{ (1+a_L) z_L}{2 \tilde w_L + 2 i k + z_R +z - a_L z_L})(1 + \frac{ (1+a_R) z_R}{2 \tilde w_R - 2 i k 
+ z_L + z - a_R z_R}) 
 e^{z_L y_L +  z_R y_R + z y} \nn
\eea
where $C'=0^- + i \mathbb{R}$. The calculation of this integral is performed in Section IX below. 

Let us now describe the resulting expression for the infinite time limit of the generating function. One finds
\bea \label{Mss} 
&& g_{\infty} (s_1,s_L,s_R;\tilde x) = {\rm Det}[ I + P_0 M_{s_1,s_L,s_R,\tilde x} P_0 ] \\
 && M_{s_1,s_L,s_R,\tilde x}(v_i,v_j) = \int \frac{dk}{2 \pi} dy \Ai(y + 4 k^2 + i k \tilde x + v_i + v_j) e^{- 2 i k (v_i-v_j)} 
 \phi_{\infty}(k,y-s_L,y-s_R,y-s_1) 
\eea
where from now on we are omitting from now on the "{\rm dec}" subscript, since we conjecture that it is the exact result.
From  \eqref{phitotalgen} we find
\bea \label{phitotalgen2} 
&& \frac{1}{2} \phi_{\infty}(k,y_L,y_R,y) = -1 +  \theta(-y_L) \theta(-y_R)  \theta(-y) + (1 +a_L + a_R - 3 a_R a_L) \\
&& \times \theta(y_L+ a_L y_R) \theta(y_R+ a_R y_L) \theta((1+a_R) y_L + (1+a_L) y_R-y) e^{- 2 (\tilde w_L+ i k) (y_R+ a_R y_L) - 2 (\tilde w_R - i k) (y_L+ a_L y_R)}  \nn \\
&& -  2 a_L \theta(-y_L) e^{2 y_L (\tilde w_L+ i k)} \theta(-y_L-y_R)  \theta(-y_L-y) 
- (1-a_L) \delta(y_L) \frac{e^{- 2 \max(y_R,y,0)  (\tilde w_L+ i k)} }{2 (\tilde w_L+ i k)}  \nn \\
&& -   2 a_R \theta(-y_R) e^{2 y_R (\tilde w_R - i k) } \theta(-y_R-y_L)  \theta(-y_R-y) 
- (1-a_R) \delta(y_R) \frac{e^{- 2 \max(y_L,y,0) (\tilde w_R - i k)}}{2 (\tilde w_R - i k)}   \nn \\
&& + \frac{2 a_R a_L}{\tilde w_L+\tilde w_R}
\delta(y_L + y_R) e^{ (\tilde w_L+ i k) y_L+ (\tilde w_R- i k) y_R -  (\tilde w_L+\tilde w_R)  \max(y,y_R,y_L)}  \nn
\eea 
This is a new result, and we can check that it reduces in some limits to the results of our previous work. 
Note that the limit $\lim_{y \to - \infty} \phi_{\infty}(k,y_L,y_R,y) \to  \phi_{\infty}(k,y_L,y_R)$ reproduces
correctly the function given in Eq. (B6) in \cite{PLDCrossover17}. From \eqref{Mss} we then see that the limit 
$\lim_{s_1 \to + \infty} g_{\infty} (s_1,s_L,s_R;\tilde x) = g_{\infty} (s_L,s_R;\tilde x)$ reproduces
the result given in Appendix D.1 in \cite{PLDCrossover17}. On the other hand, if we set $s_L=s_R=s$
and specialize to $s_1>s$, one can check that the present result reduces to the result, independent of $s_1$, 
obtained in \cite{PLDCrossover17} Section III. E.2. Eqs. (102)-(104). 
Although this was expected for $s_1 \to +\infty$, as discussed just above, the fact that it holds for all $s_1>s$
is a non-trivial and needed property (see below) and thus an important test for the "decoupling
assumption".

\subsection{III.5. Second form of the kernel: general triple joint CDF}

We now rewrite the kernel \eqref{Mss}-\eqref{phitotalgen2} using 
Airy function identities recalled in Appendix C of \cite{PLDCrossover17}. These manipulations
are performed in Section X below and are a generalization of similar steps 
as in Appendices C and D.2 of \cite{PLDCrossover17}. Here we only
give the final result. It is expressed as
\be
g_{\infty} (s_1,s_L,s_R;\tilde x) = \hat g_{\infty} (\sigma_1,\sigma_L,\sigma_R;\hat x) 
\ee
in terms of the variables
\be
 \sigma_{L,R,1} = 2^{-2/3} (s_{L,R,1}+ \frac{\tilde x^2}{16}) \quad , \quad \hat w = 2^{2/3} \tilde w 
\quad , \quad \hat x  = 2^{2/3} \frac{\tilde x}{8}  
\ee 
The generating function $\hat g_{\infty} $ is obtained as a Fredholm determinant
\be \label{final10} 
{\rm Prob}\left( {\cal A}_2(-\hat x)  < \sigma_1 , \hat h_L(\hat x) + \hat x^2 < \sigma_L ,  
\hat h_R(\hat x) + \hat x^2 < \sigma_R  \right) = \hat g_{\infty} (\sigma_1,\sigma_L,\sigma_R;\hat x) = {\rm Det}[ I -  P_0 K_{\sigma_1,\sigma_L,\sigma_R} P_0 ] 
\ee
where we have recalled that it is also the triple joint CDF associated to the Airy process
where $h_{L,R}(\hat x)$ are defined in \eqref{hLRdef}. The associated kernel,
written here in full generality, using the shorthand notation 
$\sigma_{m}=\min(\sigma_1,\sigma_L,\sigma_R)$ reads
\bea 
&& K_{\sigma_1,\sigma_L,\sigma_R}(v_i,v_j) = 
\int_{\sigma_{m}}^{+\infty} dy \Ai(v_i + y ) \Ai(v_j + y)  
- \frac{a_R a_L e^{(\hat w_L + \hat w_R) \sigma_{m} - (\hat w_L + \hat x) \sigma_L- (\hat w_R - \hat x) \sigma_R}  }{\hat w_L + \hat w_R}  \Ai(v_i+\sigma_L) \Ai(v_j+\sigma_R) \nn
\\
&& - (1+a_L+a_R-3 a_L a_R) e^{2 (\hat w_L + \hat x) (\sigma_R + a_R \sigma_L) + 
2 (\hat w_R - \hat x) (\sigma_L + a_L \sigma_R)} 
\int_{\max(\frac{\sigma_L+ a_L \sigma_R}{1+a_L},\frac{\sigma_R+ a_R \sigma_L}{1+a_R}, 
\frac{(1+ a_R) \sigma_L+ (1+a_L) \sigma_R- \sigma_1}{1+a_L+a_R})}^{+\infty} dy  \nn
\\
&& 
\times 
\Ai(v_i + (1-a_L+a_R) y + (1-a_R) \sigma_L - (1-a_L) \sigma_R) 
\times \Ai(v_j + (1+a_L-a_R) y- (1-a_R) \sigma_L + (1-a_L) \sigma_R ) 
\nn \\
&& \times 
 e^{- 2 y ( \hat w_L + \hat w_R + a_R (\hat w_L + \hat x) + a_L (\hat w_R - \hat x) )}  \label{K12}   \\
&& 
+  a_L 
\Ai(v_i+\sigma_L) 
 \int^{\sigma_{m}}_{-\infty} dy  \Ai(v_j   + y  )
 e^{(\hat x + \hat w_L) (y- \sigma_L)} +  a_R \Ai(v_j +\sigma_R)  
 \int^{\sigma_{m}}_{-\infty}  dy \Ai(v_i + y) 
 e^{(\hat w_R - \hat x ) (y - \sigma_R)} \nn \\
 && + (1-a_L) 
\int_{\max(\sigma_L-\sigma_R,\sigma_L-\sigma_1,0)}^{+\infty} dy \Ai(v_i + y + \sigma_L ) \Ai(v_j  - y + \sigma_L) 
e^{ - 2 y ( \hat x + \hat w_L) } \nn \\
&& + (1-a_R) 
\int_{\max(\sigma_R-\sigma_L,\sigma_R-\sigma_1,0)}^{+\infty} dy \Ai(v_i - y+\sigma_R) \Ai(v_j + y+\sigma_R) 
e^{ - 2 y (\hat w_R - \hat x ) } \nn
\eea 
We note that the kernel depends only on  
the combinations $\hat w_L + \hat x$ and $\hat w_R - \hat x$,
as required by the so-called STS symmetry (see Section XI and e.g. \cite{we-flat} or \cite{PLDCrossover17}).

We can note some desirable properties of this result \eqref{final10}, \eqref{K12} for the joint CDF, obtained via the replica calculation.
First it depends on $\sigma_1$ only when $\sigma_1 \leq \min(\sigma_L,\sigma_R)$. For $\sigma_1 > \min(\sigma_L,\sigma_R)$ all dependence in $\sigma_1$ disappears and the associated
joint PDF is zero. It can be checked considering all four cases $a_{L,R}=0,1$. 
This is a required property from the definition of this CDF: indeed, considering the point $\hat y=0$ in the quantity to be maximized in the
definition \eqref{hLRdef}, one sees that one must have ${\cal A}_2(-\hat x) \leq 
\min(\hat h_L(\hat x) + \hat x^2,\hat h_R(\hat x) + \hat x^2)$. Clearly 
this is a non-trivial check of the replica method and the decoupling assumption.

Another check of \eqref{final10}, \eqref{K12} is that for $\sigma_L,\sigma_R \to + \infty$ it should become equal to the CDF of the GUE TW distribution for 
the variable $\sigma_1$,
i.e. one must have $\hat g_{\infty} (\sigma_1,+\infty,+\infty;0) = F_2(\sigma_1) = {\rm Det}[ I -  P_{\sigma_1} K_{\Ai} P_{\sigma_1} ] $ for any value of
$a_{L,R} \in \{0,1\}^2$ and $\hat w_{L,R}$. It is easy to check that in this limit all terms in \eqref{K12},
apart from the first one, vanish. Upon the shift $v_{i,j} \to v_{i,j} -\sigma_1$ the first term
recovers exactly the Airy kernel $K_{\Ai}(v_i,v_j)=\int_{0}^{+\infty} dy \Ai(v_i + y) \Ai(v_j + y)$
and the desired property is thus correct. 

Finally, for $w_L,w_R \to +\infty$ the maximization in \eqref{hLRdef}
leads to the position of the maximum at $\hat y=0$, hence 
$\hat h_L(\hat x) + \hat x^2=\hat h_R(\hat x) + \hat x^2={\cal A}_2(-\hat x)$.
Hence in that limit one should have $\lim_{w_L,w_R \to +\infty} \hat g_{\infty} (\sigma_1,\sigma_L,\sigma_R;0)  = F_2(\min(\sigma_1,\sigma_L,\sigma_R)$,
where $F_2$ is again the CDF of the GUE-TW distribution. 
Clearly it works since again all terms apart the first one in $K_{\sigma_1,\sigma_L,\sigma_R}$
in \eqref{K12} vanish in that limit.

We will specialize below to two main cases: the Brownian case $a_L=a_R=1$ and the flat case $a_R=a_L=0$.
The mixed or crossover cases can be studied along similar lines, but we will not detail it here.

\subsection{III.6. Joint CDF for Airy minus parabola plus double-sided Brownian with drifts} 

We now specialize to $a_L=a_R=1$, i.e the double-sided Brownian case. Eq. \eqref{K12} 
simplifies into
\bea \label{K12new} 
&& K_{\sigma_1,\sigma_L,\sigma_R}(v_i,v_j) = 
K_{\Ai}(v_i + \sigma_{m}, v_j + \sigma_{m}) 
- \frac{e^{(\hat w_L + \hat w_R) \sigma_{m} - (\hat w_L + \hat x) \sigma_L- (\hat w_R - \hat x) \sigma_R}  }{\hat w_L + \hat w_R}  \Ai(v_i+\sigma_L) \Ai(v_j+\sigma_R) \nn
\\
&& +  
\Ai(v_i+\sigma_L) 
 \int^{0}_{-\infty} dy  \Ai(v_j   + y + \sigma_{m}  )
 e^{(\hat x + \hat w_L) (y + \sigma_{m}- \sigma_L)} +   \Ai(v_j +\sigma_R)  
 \int^{0}_{-\infty}  dy \Ai(v_i + y + \sigma_{m}) 
 e^{(\hat w_R - \hat x ) (y + \sigma_{m}- \sigma_R)} \nn 
 \eea
This can be written in a more compact form by defining
(as in \eqref{defBL} in the text)
\be
{\cal B}_w(v) :=e^{w^3/3-v w} - \int_0^{+\infty} dy \Ai(v+y) e^{w y}=\int_{-\infty}^0 \Ai(v+y) e^{w y} \label{defBw2} 
\ee 
where the last equality holds only for $v<0$ (when the integral is convergent),
and
\be
\Ai_{\sigma,w}(v) := \Ai(v+\sigma) e^{- w \sigma}
\ee
One can then check that one can rewrite the generating function, hence the following
triple joint CDF as
\bea \label{final10w} 
 {\rm Prob}\bigg( {\cal A}_2(-\hat x)  < \sigma_1 , &&
\max_{\hat y<0}( {\cal A}_2(\hat y-\hat x) - (\hat y - \hat x)^2 + 2 \hat w_L \hat y + 
 \sqrt{2} 
B(\hat y) ) < \sigma_L - \hat x^2 , \\
&& 
~~~~ \max_{\hat y>0}( {\cal A}_2(\hat y-\hat x) - (\hat y - \hat x)^2 -  2 \hat w_R \hat y +  
\sqrt{2} B(\hat y) ) < \sigma_R - \hat x^2 \bigg) \nn \\
& =& 
 \hat g_{\infty} (\sigma_1,\sigma_L,\sigma_R;\hat x) = {\rm Det}[ I -  P_{\sigma_{m}} 
 \hat K_{\sigma_L-\sigma_{m},\sigma_R-\sigma_{m}} P_{\sigma_{m}} ] \nn
\eea
with the kernel
\bea \label{Khat1} 
\hat K_{\sigma_L-\sigma_{m},\sigma_R-\sigma_{m}}(v_i,v_j)
 = K_{\Ai}(v_i,v_j) &+&  \Ai_{\sigma_L - \sigma_{m},\hat w_L + \hat x}(v_i) {\cal B}_{\hat w_L + \hat x}(v_j)
 +{\cal B}_{\hat w_R - \hat x}(v_i)  \Ai_{\sigma_R - \sigma_{m},\hat w_R - \hat x}(v_j) \\
&& - \frac{1}{\hat w_L + \hat w_R}  \Ai_{\sigma_L - \sigma_{m},\hat w_L + \hat x}(v_i)
\Ai_{\sigma_R - \sigma_{m},\hat w_R - \hat x}(v_j) \nn
\eea
where we recall that $\sigma_m=\min(\sigma_1,\sigma_L,\sigma_R)$. We will now study the limit where the drift go to zero $\hat w_{L,R}=0^+$ 
which relates to stationary KPZ. 

\subsection{III.7. Double-sided Brownian in limit $\hat w_{L,R}=0^+$: 
Joint CDF of Airy and Airy minus parabola plus Brownian}

We continue with the Brownian case and now perform the stationary limit $\hat w_{L,R} \to 0$. This limit
cannot be performed naively and require some manipulations. 
Since the kernel \eqref{Khat1} is singular as $\hat w_{L,R} \to 0$, to perform the limit one first rewrite it as
\bea
&& \hat K_{\sigma_L-\sigma_{m},\sigma_R-\sigma_{m}}(v_i,v_j) =
K_{\Ai}(v_i,v_j)  
+ (\hat w_L + \hat w_R) {\cal B}_{\hat w_R-\hat x}(v_i)
{\cal B}_{\hat w_L+\hat x}(v_j)  \\
&& - \big( \frac{ \Ai_{\sigma_L - \sigma_{m},\hat w_L + \hat x}(v_i)}{\sqrt{\hat w_L + \hat w_R}} 
- \sqrt{\hat w_L + \hat w_R} {\cal B}_{\hat w_R-\hat x}(v_i) \big) 
\big( \frac{\Ai_{\sigma_R - \sigma_{m},\hat w_R - \hat x}(v_j)}{\sqrt{\hat w_L + \hat w_R}} 
- \sqrt{\hat w_L + \hat w_R} {\cal B}_{\hat w_L+\hat x}(v_j) \big) \nn
\eea 
which has the form of the Airy kernel plus two projectors. 

We will now use the determinant identity, in quantum mechanical notations
\be
 {\rm Det}[ A - |U \rangle \langle V| + |R \rangle \langle S| ]  = {\rm Det} A \times \left( (1 - \langle V|A^{-1}|U \rangle)(1 + \langle S|A^{-1}|R \rangle)
+ \langle S|A^{-1}|U \rangle \langle V|A^{-1}|R \rangle \right) \label{identity2} 
\ee
where $ |U \rangle \langle V|$ stands for the projector operator $U V^T$, explicitly in coordinate
representation $(U V^T)(v_i,v_j)=U(v_i) V(v_j)$. We will use indifferently either notations, e.g.
$\langle V|A^{-1}|U \rangle \equiv {\rm Tr}[ A^{-1} U V^T]$. 
Here we apply the operator identity \eqref{identity2} to 
\bea
&& A = I - C \quad C = P_{\sigma_{m}} K_{\Ai} P_{\sigma_{m}} \quad , \quad A^{-1} = I + C (I-C)^{-1} \\
&& |U \rangle = \sqrt{\hat w_L+\hat w_R} P_{\sigma_{m}} {\cal B}_{\hat w_R-\hat x}  \quad , \quad
 |R \rangle = P_{\sigma_{m}}  \frac{ \Ai_{\sigma_L - \sigma_{m},\hat w_L + \hat x}}{\sqrt{\hat w_L + \hat w_R}} 
- \sqrt{\hat w_L + \hat w_R} P_{\sigma_{m}}{\cal B}_{\hat w_R-\hat x} \\
&& \langle V| = \sqrt{\hat w_L+\hat w_R} {\cal B}_{\hat w_L+\hat x} P_{\sigma_{m}} \quad , \quad
 \langle S| = \frac{\Ai_{\sigma_R - \sigma_{m},\hat w_R - \hat x}}{\sqrt{\hat w_L + \hat w_R}} P_{\sigma_{m}}
- \sqrt{\hat w_L + \hat w_R} {\cal B}_{\hat w_L+\hat x} P_{\sigma_{m}}
 \eea
and we use below the shorthand notation
\bea \label{shorthand2}
\Ai_\sigma(v) = \Ai_{\sigma,0}(v)  = \Ai(v+\sigma)  
\eea 

We will now set $\hat w_L=\hat w_R=\hat w$ for simplicity.
We start by estimating the small $\hat w$ expansion of 
\bea
\langle V|U \rangle = 2 \hat w {\rm Tr}[ P_{\sigma_{m}}  {\cal B}_{\hat w+\hat x} {\cal B}_{\hat w-\hat x}] 
= 1 - 2 \hat w (1+{\cal L}_{\hat x}(\sigma_m)) + O(\hat w^2) \label{scp2} 
\eea 
see e.g. formula (285) in Appendix E of \cite{deNardisPLD2timeLong}, where one shows that
\bea \label{defLx2} 
&& {\cal L}_{\hat x}(\sigma) = \sigma - 1 - \hat x^2  + \int_{\sigma}^{+\infty} dv \, (1- {\cal B}_{\hat x}(v)
{\cal B}_{-\hat x}(v)) = \sigma - 1 - \hat x^2 \\
&& + 2 \int_{\sigma}^{+\infty} du \int_{0}^{+\infty} dy 
\cosh(\frac{\hat{x}^3}{3} - (u+y) \hat x) \Ai(u+y)  - 
\int_{\sigma}^{+\infty} du \int_{0}^{+\infty} dy_1 dy_2 e^{\hat x (y_1-y_2)} \Ai(u+y_1) \Ai(u+y_2)  \nn
\eea 
This implies that the following term in \eqref{identity2} is $O(\hat w)$ as $\hat w \to 0$
\be \label{exp12}
1 - \langle V|A^{-1}|U \rangle = 1 - \langle V|U \rangle - \langle V|C (I-C)^{-1}|U \rangle = 2 \hat w (1+ {\cal L}_{\hat x}(\sigma_1) ) - 
2 \hat w \langle {\cal B}_{\hat w+\hat x} |C (I-C)^{-1}|{\cal B}_{\hat w-\hat x} \rangle
+ O(\hat w^2) 
\ee 
since the operator $C$ in the numerator of the last trace makes it convergent, due
to fast decay of Airy functions. Returning to \eqref{identity2} we can write
\bea
&& \langle S|A^{-1}|R \rangle = \frac{1}{2 \hat w} 
{\rm Tr} [P_{\sigma_m} (I-C)^{-1} \Ai_{\sigma_L-\sigma_m, \hat w+\hat x} \Ai_{\sigma_R-\sigma_m,\hat w-\hat x}^T ] -  
{\rm Tr}[  P_{\sigma_m} (I-C)^{-1} {\cal B}_{\hat w-\hat x}  \Ai_{\sigma_R-\sigma_m,\hat w-\hat x}^T  ] \\
&& -  
{\rm Tr}[  P_{\sigma_m}  (I-C)^{-1} \Ai_{\sigma_L-\sigma_m, \hat w+\hat x}  
{\cal B}_{\hat w+\hat x}^T  ]  + \langle V|U \rangle \nn
\eea 
We see from \eqref{identity2} and \eqref{exp12} that in that expression we only need 
the leading term $O(1/\hat w)$, which is the first term, since the traces here are all convergent due to the
Airy functions, and since $\langle V|U \rangle \simeq 1$ from \eqref{scp2}. We obtain
\bea
\langle S|A^{-1}|R \rangle = \frac{1}{2 \hat w} 
{\rm Tr} [P_{\sigma_m} (I-C)^{-1} \Ai_{\sigma_L-\sigma_m, \hat x} \Ai_{\sigma_R-\sigma_m,-\hat x}^T ]
+ O(1) 
\eea

Now we need to evaluate
\be
\langle V|A^{-1}|R \rangle = 
{\rm Tr}[ P_{\sigma_m}  (I-C)^{-1} \Ai_{\sigma_L-\sigma_m,\hat w+\hat x} {\cal B}_{\hat w+\hat x}^T  ] - \langle V|A^{-1}|U \rangle 
\simeq   {\rm Tr}[ P_{\sigma_m}   (I-C)^{-1} 
{\cal B}_{\hat x} \Ai_{\sigma_L-\sigma_m,\hat x}^T] - 1
\ee
Similarly one has
\be
\langle S|A^{-1}|U \rangle  \simeq 
{\rm Tr}[ P_{\sigma_m}   (I-C)^{-1} 
{\cal B}_{-\hat x} \Ai_{\sigma_R-\sigma_m,-\hat x}^T] - 1
\ee
Putting all together in the limit $\hat w \to 0$, we obtain our main result for
the triple joint CDF 
\bea  \label{defGG} 
G_{\hat x}(\sigma_1,\sigma_L,\sigma_R) := {\rm Prob}\bigg( {\cal A}_2(-\hat x)  < \sigma_1 , &&
\max_{\hat y<0}( {\cal A}_2(\hat y-\hat x) - (\hat y - \hat x)^2 + 
 \sqrt{2} 
B(\hat y) ) < \sigma_L - \hat x^2 , \\
&& 
~~~~ \max_{\hat y>0}( {\cal A}_2(\hat y-\hat x) - (\hat y - \hat x)^2  +  
\sqrt{2} B(\hat y) ) < \sigma_R - \hat x^2 \bigg) \nn
\eea
in the form, where we recall that we denote $\sigma_m:=\min(\sigma_1,\sigma_L,\sigma_R)$
\bea \label{mainres} 
&& G_{\hat x}(\sigma_1,\sigma_L,\sigma_R) =
\lim_{\hat w \to 0} \hat g_{\infty} (\sigma_1,\sigma_L,\sigma_R;\hat x) =
\lim_{\hat w \to 0}  {\rm Det}[ I -  P_{\sigma_m} \hat K_{\sigma_L-\sigma_m,\sigma_L-\sigma_m} P_{\sigma_m} ]
\\
&& = F_2(\sigma_m) Y_{\hat x}(\sigma_m) {\rm Tr} [(I- P_{\sigma_m} K_{\Ai} )^{-1}  P_{\sigma_m}  \Ai_{\sigma_L-\sigma_m} \Ai_{\sigma_R-\sigma_m}^T ] 
e^{\hat x (\sigma_R-\sigma_L)} \nn  \\
&& + F_2(\sigma_m) (e^{-  \hat x (\sigma_L-\sigma_m)} {\rm Tr}[ (I- P_{\sigma_m} K_{\Ai} )^{-1} P_{\sigma_m} \Ai_{\sigma_L-\sigma_m}
{\cal B}_{\hat x}^T] - 1)  (e^{\hat x (\sigma_R-\sigma_m)} {\rm Tr}[ (I- P_{\sigma_m} K_{\Ai} )^{-1} P_{\sigma_m} \Ai_{\sigma_R-\sigma_m}
{\cal B}_{-\hat x}^T] - 1) \nn
\eea 
We have defined, as in the main text in Eq. \eqref{defYx}, the function
\bea \label{defYx2} 
Y_{\hat x}(\sigma) = 1 + {\cal L}_{\hat x}(\sigma) -  
{\rm Tr}[ P_{\sigma} K_{\Ai} (I- P_{\sigma} K_{\Ai} )^{-1} P_{\sigma} {\cal B}_{-\hat x}
{\cal B}_{\hat x}^T]
\eea 
We recall that the vector ${\cal B}_{\hat x}(v)$ is given by \eqref{defBw2} 
and ${\cal L}_{\hat x}(\sigma)$
is given by \eqref{defLx2} (both also defined in Eq. \eqref{defBL} in the main text)
and that $\Ai_{\sigma}(u):=\Ai(u+\sigma)$.
We also recall that $F_2(\sigma) =
{\rm Det}[ I -  P_{\sigma} K_{\Ai} P_{\sigma} ]$ is the CDF of the GUE-TW distribution, and
everywhere $K_{\Ai}$ is the Airy kernel. 

Equations \eqref{defGG}, \eqref{mainres} is the master formula from 
which we will now obtain the three main results announced in the text
as particular cases, each being analyzed in details below. 

Before we do so let us indicate a useful alternative formula for $Y_{\hat x}$. Let us introduce the following notation for the two vectors
\bea \label{defBhat}
{\cal D}_{\hat x}(v) = e^{\frac{\hat x^3}{3} - \hat x v} \quad , \quad 
\hat {\cal B}_{\hat x}(v)  = {\cal D}_{\hat x}(v)  - {\cal B}_{\hat x}(v) = 
\int_0^{+\infty} dy \Ai(v+y) e^{\hat x y} 
\eea 
where $\hat {\cal B}_{\hat x}(v)$ has a fast decay at $v \to +\infty$. We see from \eqref{defLx2} that we can rewrite
\bea
&& Y_{\hat x}(\sigma) =  \sigma  - \hat x^2  + \int_{\sigma}^{+\infty} dv \, 
(\hat {\cal B}_{\hat x}(v) {\cal D}_{-\hat x}(v) + \hat {\cal B}_{-\hat x}(v) {\cal D}_{\hat x}(v)
- \hat {\cal B}_{\hat x}(v) \hat {\cal B}_{-\hat x}(v) ) -  {\rm Tr}[ P_{\sigma} K_{\Ai} (I- P_{\sigma} K_{\Ai} )^{-1} 
P_{\sigma}  {\cal D}_{-\hat x} {\cal D}_{\hat x}^T ] \nn \\
&&
- {\rm Tr}[ P_{\sigma} K_{\Ai} (I- P_{\sigma} K_{\Ai} )^{-1}  P_{\sigma}  
\hat {\cal B}_{\hat x} \hat {\cal B}_{-\hat x}^T]
+ {\rm Tr}[ P_{\sigma} K_{\Ai} (I- P_{\sigma} K_{\Ai} )^{-1} P_{\sigma} (
 {\cal D}_{\hat x}  \hat {\cal B}_{-\hat x}^T +
 {\cal D}_{-\hat x}  \hat {\cal B}_{\hat x}^T)] \nn
\eea 
Using that $K(I-K)^{-1}=(I-K)^{-1} - I$ we see that the traces involving the substraction $- I$ cancel
with the first line and it remains
\bea
&& Y_{\hat x}(\sigma) =  \sigma  - \hat x^2 -  {\rm Tr}[  (I- P_{\sigma} K_{\Ai} )^{-1} 
P_{\sigma} ( K_{\Ai} P_{\sigma}  {\cal D}_{\hat x}   {\cal D}_{-\hat x}^T  +  \hat {\cal B}_{\hat x} \hat {\cal B}_{-\hat x}^T -
   {\cal D}_{\hat x}  \hat {\cal B}_{-\hat x}^T - 
  {\cal D}_{-\hat x}  \hat {\cal B}_{\hat x}^T) ] \nn
\eea 
a form which is useful for numerical evaluations. 

\section{IV. Moments of the joint CDF and conditional averages for 2-time KPZ at large time}

\subsection{IV. 1. Joint CDF of the Airy process and the max of Airy process minus parabola plus
Brownian on the line} 

We now study
\bea  \label{defGGG} 
&& G_{\hat x}(\sigma_1,\sigma_2) := {\rm Prob}\left( {\cal A}_2(-\hat x)  < \sigma_1 , 
 \max_{\hat y}( {\cal A}_2(\hat y-\hat x) - (\hat y-\hat x)^2 + \sqrt{2} B(\hat y) ) < \sigma_2 - \hat x^2 \right) 
\eea
and its application to 2-time KPZ. Setting $\sigma_L=\sigma_R=\sigma_2$ in 
\eqref{defGG} and \eqref{mainres} we obtain the prediction
\bea \label{resFx}
&&  G_{\hat x}(\sigma_1,\sigma_2)  = \lim_{\hat w\to 0} \hat g_{\infty} (\sigma_1,\sigma_2,\sigma_2;\hat x) 
 = F_2(\sigma_1) Y_{\hat x}(\sigma_1) {\rm Tr} [(I- P_{\sigma_1} K_{\Ai} )^{-1}  P_{\sigma_1} \Ai_{\sigma_2-\sigma_1} 
\Ai_{\sigma_2-\sigma_1}^T ]  \\
&& + F_2(\sigma_1) (e^{-  \hat x (\sigma_2-\sigma_1)} {\rm Tr}[ (I- P_{\sigma_1} K_{\Ai} )^{-1} P_{\sigma_1} \Ai_{\sigma_2-\sigma_1}
{\cal B}_{\hat x}^T] - 1)  (e^{\hat x (\sigma_2-\sigma_1)} {\rm Tr}[ (I- P_{\sigma_1} K_{\Ai} )^{-1} P_{\sigma_1} \Ai_{\sigma_2-\sigma_1}
{\cal B}_{-\hat x}^T] - 1) \nn
\eea 
where the last equality is valid for $\sigma_1 \leq \sigma_2$. For $\sigma_1 \geq \sigma_2$
we have $G_{\hat x}(\sigma_1,\sigma_2) = G_{\hat x}(\sigma_2,\sigma_2)$. 
We recall that the vector ${\cal B}_{\hat x}(v)$ is given by \eqref{defBw2} 
and $Y_{\hat x}(\sigma)$
is given by \eqref{defYx2}. In the case $\hat x=0$, the formula \eqref{resFx}
reduces to the result \eqref{ResF} given in the text, where we 
denote $G(\sigma_1,\sigma_2)=G_0(\sigma_1,\sigma_2)$.
It is easy to see that the marginal CDF of $\sigma_1$, $G_{\hat x}(\sigma_1,\sigma_2=+\infty)=F_2(\sigma_1)$ is the
GUE-TW since only the last term in \eqref{resFx} survives in that limit. We now show that
the marginal CDF of $\sigma_2$ correctly coincides with the EBR distribution.

\subsection{IV. 2. Extended Baik-Rains limit}

Let us start by showing that our expression \eqref{resFx} for $G_{\hat x}(\sigma_2,\sigma_2)$ correctly
recovers the (extended) Baik-Rains distribution, $F_0(\sigma_2 - \hat x^2;\hat x)$, as required from its definition.
Indeed the stationary Airy process at a given point $\hat x$, ${\cal A}_{\rm stat}(\hat x)$, is given by
\bea
{\cal A}_{\rm stat}(\hat x)=\max_{\hat y}( {\cal A}_2(\hat y-\hat x) - (\hat y-\hat x)^2 + \sqrt{2} B(\hat y) )  
\eea 
and it is known that its one point CDF is given by 
\bea
&& {\rm Prob}({\cal A}_{\rm stat}(\hat x) < \zeta=\sigma- \hat x^2) = F_0(\zeta;\hat x) = 
 \partial_\sigma (F_2(\sigma) Y_{\hat x}(\sigma)) \nn 
 \eea
We thus need to show that Eq. \eqref{resFx} for $\sigma_1=\sigma_2=\sigma$ reduces to
\bea \label{desired} 
G_{\hat x}(\sigma,\sigma) = \partial_\sigma (F_2(\sigma) Y_{\hat x}(\sigma))
\eea 
We first use the identity
\bea
\partial_\sigma F_2(\sigma) &=& \partial_\sigma {\rm Det}[ I -  P_{0} K^\sigma_{\Ai} P_{0} ]
= - {\rm Tr} [P_{0}  (I- P_{0} K^\sigma_{\Ai})^{-1} P_0 \partial_{\sigma}  K^\sigma_{\Ai}] 
=  {\rm Tr} [P_{0}  (I- P_{0} K^\sigma_{\Ai})^{-1} P_0 \Ai_\sigma \Ai_\sigma^T ] \\
& =&  {\rm Tr} [P_{\sigma}  (I- P_{\sigma} K_{\Ai})^{-1} P_\sigma \Ai \Ai^T ] \label{F2prime} 
\eea
using that $\partial_{\sigma} K^{\sigma}_{\Ai}(v_i,v_j)=- \Ai(v_i) \Ai(v_j)$ and here and below
we often use the shorthand notation
\be
K^\sigma_{\Ai}(v) = K_\Ai(v+\sigma) 
\ee
Hence the first term in \eqref{resFx} reduces to $Y_{\hat x}(\sigma_1) \partial_{\sigma_1} F_2(\sigma_1)$. 

To show the desired result \eqref{desired}, we will now show that
\be \label{derYsigma} 
\partial_\sigma Y_{\hat x}(\sigma) = (\Tr[ (I- P_\sigma K_{\Ai} )^{-1} P_\sigma \Ai {\cal B}_{\hat x}^T] - 1)  
  (\Tr[(I- P_\sigma K_{\Ai} )^{-1} P_\sigma \Ai {\cal B}_{- \hat x}^T] - 1) 
\ee
so that the second term in \eqref{resFx} for $\sigma_2=\sigma_1$ reduces to $F_2(\sigma_1)\partial_{\sigma_1} Y_{\hat x}(\sigma_1)$. Hence if 
\eqref{derYsigma} is true, so is \eqref{desired}. 

To show \eqref{derYsigma} we note that from \eqref{defYx} we have
 \be \label{defYx} 
\partial_\sigma  Y_{\hat x}(\sigma) =  \partial_\sigma {\cal L}_{\hat x}(\sigma) -  
\partial_\sigma {\rm Tr} [ P_\sigma K_{\Ai} (I- P_\sigma K_{\Ai})^{-1} P_\sigma {\cal B}_{-\hat x} {\cal B}_{\hat x}^T]  
\ee

From the definition of $ {\cal L}_{\hat x}(\sigma)$ in \eqref{defLx2} and of ${\cal B}_{\hat x}(\sigma)$ in
\eqref{defBw2} we have
\bea \label{partialL}
&& \partial_{\sigma} {\cal L}_{\hat x}(\sigma) = {\cal B}_{\hat x}(\sigma) {\cal B}_{-\hat x}(\sigma) 
= 1 - Tr[ \Ai P_\sigma ({\cal B}_{\hat x}  + {\cal B}_{-\hat x} )] \quad , \quad 
\lim_{\sigma \to +\infty} {\cal B}_{\hat x}(\sigma) {\cal B}_{-\hat x}(\sigma)=1 
\eea
in the second equality we used the identity 
\bea
\int_0^{+\infty} \int_0^{+\infty} dy_1 dy_2 \Ai(\sigma+y_1) \Ai(\sigma+y_2) e^{x (y_1-y_2)}  =
\int_0^{+\infty} \int_0^{+\infty} du dy \Ai(\sigma+u) \Ai(\sigma+u+y) (e^{x y} + e^{-x y}) 
\eea 

Using that
\bea \label{derB2} 
&& \partial_v B_{\hat x}(v) = \Ai(v) - \hat x B_{\hat x}(v) \quad , \quad  \partial_{\sigma} B^\sigma_{\hat x}(v) = \Ai_\sigma(v) - \hat x B^\sigma_{\hat x}(v)
\eea 
we have

\bea \label{manip1} 
&&-  \partial_\sigma {\rm Tr} [ P_\sigma K_{\Ai} (I- P_\sigma K_{\Ai})^{-1} P_\sigma {\cal B}_{-\hat x} {\cal B}_{\hat x}^T]  
= - \partial_{\sigma} 
{\rm Tr} [ P_0 K^\sigma_{\Ai} (I- P_0 K^\sigma_{\Ai})^{-1} P_0 {\cal B}^\sigma_{-\hat x} ({\cal B}^\sigma_{\hat x})^T] \\
&& = - {\rm Tr} [ P_0 K^\sigma_{\Ai} (I- P_0 K^\sigma_{\Ai})^{-1} P_0 {\cal B}^\sigma_{-\hat x} (
(\Ai_\sigma - \hat x B^\sigma_{\hat x})^T]
- {\rm Tr} [ P_0 K^\sigma_{\Ai} (I- P_0 K^\sigma_{\Ai})^{-1} P_0 
(\Ai_\sigma + \hat x B^\sigma_{-\hat x}) ({\cal B}^\sigma_{\hat x})^T] \nn \\
&& + {\rm Tr} [ P_0 (I- P_0 K^\sigma_{\Ai})^{-1} 
P_0 \Ai_\sigma ({\cal B}^\sigma_{\hat x})^T ] \,
{\rm Tr} [ P_0 (I- P_0 K^\sigma_{\Ai})^{-1}  
P_0  {\cal B}^\sigma_{-\hat x} \Ai_\sigma^T ] \nn \\
&& = 
 {\rm Tr} [  \Ai_\sigma P_0 ({\cal B}^\sigma_{-\hat x} + {\cal B}^\sigma_{\hat x} ) ] \nn
\\
&& + ({\rm Tr} [ P_0 (I- P_0 K^\sigma_{\Ai})^{-1} 
P_0 \Ai_\sigma ({\cal B}^\sigma_{\hat x})^T ] -1) 
({\rm Tr} [ P_0 (I- P_0 K^\sigma_{\Ai})^{-1}  
P_0  {\cal B}^\sigma_{-\hat x} \Ai_\sigma^T ] -1) - 1 \nn
\eea
we use $C (1-C)^{-1}=(1-C)^{-1}-1$ and
\bea
&& \partial_\sigma P_0 K^\sigma_{\Ai} (I- P_0 K^\sigma_{\Ai})^{-1}
= (I- P_0 K^\sigma_{\Ai})^{-1} P_0 \partial_\sigma K^\sigma_{\Ai} (I- P_0 K^\sigma_{\Ai})^{-1} \\
&& = - (I- P_0 K^\sigma_{\Ai})^{-1} P_0 \Ai_\sigma \Ai_\sigma^T (I- P_0 K^\sigma_{\Ai})^{-1} 
\eea
using that $\partial_\sigma K^\sigma_{\Ai}  = - \Ai_\sigma \Ai_\sigma^T$ (as above).
Adding the last two lines of \eqref{manip1} to \eqref{partialL} leads to cancellations and,
upon rearranging, to \eqref{derYsigma}. This completes the proof that 
$G_{\hat x}(\sigma_1,\sigma_1)= F_0(\sigma_1-\hat x^2;\hat x)$, i.e. it equals the CDF of the 
extended Baik-Rains distribution, which is the one point CDF of the 
stationary Airy process. 

\subsection{IV. 3. Application to two-time KPZ in the large time separation limit: general framework}

Here we recall the connection between two time KPZ in the large time separation limit and the Airy process,
with little details since it was explained in Sections 7.3-7.5 in \cite{deNardisPLD2timeLong} (see also
\cite{FerrariSpohn2times}). The only difference with \cite{deNardisPLD2timeLong} is 
the way a non zero spatial shift $\hat x$ is introduced (see discussion below). Defining
$h(x,t|y,0)$ the solution of the KPZ equation \eqref{kpzeq2} with the droplet initial condition centered at $y$
(see \eqref{drop0}), we have at large times $t_1, t_2 \gg 1$ with $\Delta=(t_2-t_1)/t_1$ fixed
\bea \label{defh1h2} 
&& h_1 := \lim_{t_1 \to +\infty} t_1^{-1/3} h(0,t_1 | x, 0) = {\cal A}_2(-\hat x) - \hat x^2 \\
&& h_2 := \lim_{t_1 \to +\infty} t_1^{-1/3} h(0,t_2 | x, 0)|_{t_2=(1+\Delta) t_1} 
= \max_{\hat y \in \mathbb R} \big( {\cal A}_2(\hat y-\hat x) - (\hat y-\hat x)^2 + \Delta^{\frac{1}{3}} (\tilde {\cal A}_2(\frac{\hat y}{\Delta^{\frac{2}{3}}}) - \frac{\hat y^2}{\Delta^{\frac{4}{3}}}) \big) \nn
\eea
where $ {\cal A}_2$ and $\tilde  {\cal A}_2$ are two independent Airy processes, and we recall
that $\hat x=x/(2 t_1^{2/3})$, $\hat y=y/(2 t_1^{2/3})$. Note our unnatural choice to define $h_2$ using the scale $t_1^{1/3}$, which however is convenient for the analysis below.
In Ref. \cite{deNardisPLD2timeLong} the variable $h$ was also defined as
\bea \label{defhh} 
&& h  := \lim_{t_1 \to +\infty}  \frac{h(0,t_2 | x, 0) - h(0,t_1 | x, 0)}{(t_2-t_1)^{1/3}}|_{t_2=(1+\Delta) t_1} 
\eea
We also defined there the (unknown) exact JPDF $P_\Delta(\sigma_1,\sigma):=
\overline{\delta(h_1-\sigma_1)  \delta(h-\sigma)}$ and 
derived an approximation of it, denoted $P^{(1)}_\Delta(\sigma_1,\sigma)$, conjectured to be 
exact to leading order in large positive $\sigma_1$ at fixed $\sigma$. 
It was shown in \cite{deNardisPLDTakeuchi} to be good enough an approximation to fit 
experiments and numerics in a broad range of values 
$\sigma_1 > \langle \sigma_1 \rangle= \kappa_1^{\mbox{\tiny{GUE}}}$. Note that
the approximation which leads to $P^{(1)}_\Delta$ there is quite different from
our approach here (even if in both case the RBA was used),
in particular no decoupling assumption was necessary there.

Let us also point out an exact result for the mean heights.
From simple scaling, and convergence of one-point distributions to GUE-TW respectively at $t_1$ and
$t_2$, we know that
\bea 
&& \overline{h_1} = \kappa_1^{\mbox{\tiny{GUE}}} - \hat x^2 \label{h1exact} \quad , \quad \hat x:=\frac{x}{2 t_1^{2/3}} \\
&& \overline{h_2} = (1+ \Delta)^{1/3} \kappa_1^{\mbox{\tiny{GUE}}} - \frac{\hat x^2}{1+\Delta} \label{h2exact}
\eea 
which generalizes to non-zero $\hat x$ the results of Section 3.3.2 of \cite{deNardisPLD2timeLong}
(the last term in \eqref{h2exact} arises from the $x$-dependent shift $-x^2/(4 t_2)$ in $h(0,t_2)$ and our chosen scaling with
$t_1^{1/3}$). This implies that
\be \label{hmeanexact}
\overline{h} = \frac{\overline{h_2}-\overline{h_1}}{\Delta^{1/3}} = \frac{(1+ \Delta)^{1/3} - 1}{\Delta^{1/3}}  \kappa_1^{\mbox{\tiny{GUE}}} + \frac{\Delta^{2/3}}{1+\Delta} \hat x^2
\ee 
which is an exact result valid for any $\Delta$. \\

We focus now on the limit of large time separation $\Delta \to +\infty$. Then,
due to the property that the Airy process is locally Brownian, i.e. for fixed $\hat z$ and $a \ll 1$ 
\cite{Haag}
\bea
\tilde {\cal A}_2(\hat z + a \hat v)= \tilde {\cal A}_2(\hat z) + \sqrt{2 a} B(\hat v) + O(a) 
\eea 
where $B(\hat v)$ (a unit two sided Brownian) and $\tilde {\cal A}_2(\hat z)$ are mutually uncorrelated, one obtains
\bea
&& h_2 = \Delta^{1/3} \tilde {\cal A}_2(0) + \max_{\hat y}[ {\cal A}_2(\hat y-\hat x) - (\hat y-\hat x)^2
+ \sqrt{2} B(\hat y)  ] + O(\Delta^{-1/3}) 
\eea
We can also write, with some abuse of notations
\bea
h  = \tilde {\cal A}_2(0) + \Delta^{-1/3} (\sigma_2-\sigma_1) + O(\Delta^{-2/3}) \label{hsum2}
\eea 
where here $\sigma_1$ and $\sigma_2$ denote the random variables
\bea \label{abuse} 
&& \sigma_1 =  {\cal A}_2(-\hat x)  \\
&& \sigma_2 - \hat x^2 = \max_{\hat y}[ {\cal A}_2(\hat y-\hat x) - (\hat y-\hat x)^2
+ \sqrt{2} B(\hat y)  ] \nn
\eea 
Their JCDF is $G_{\hat x}(\sigma_1,\sigma_2)$ defined in
\eqref{defGGG}, which we obtained explicitly here in 
Eq. \eqref{resFx} for arbitrary $\hat x$, and in \eqref{ResF} in the text for $\hat x=0$. 
Note that it exhibits a non-trivial dependence in $\hat x$. In Ref. \cite{deNardisPLD2timeLong} 
instead, the dependence in the position of the final point at $t=t_2$ was studied, rather than in the position of the initial point at $t=0$ as we do here. Both dependences however can be related
using the STS symmetry (see below and Section XI).

Hence the knowledge of the JCDF $G_{\hat x}(\sigma_1,\sigma_2)$, and its associated JPDF
$p_{\hat x}(\sigma_1,\sigma_2)= \partial_{\sigma_1} \partial_{\sigma_2} G_{\hat x}(\sigma_1,\sigma_2)$,
gives some information about the two time KPZ height in the large time separation regime. 
One can ask precisely what is the extent of this information?
Since the subdominant $O(\Delta^{-2/3})$ term is actually
correlated with the (large) leading term $\Delta^{1/3} \tilde {\cal A}_2(0)$ (in an unknown way), moments of $h$ higher than one
{\it cannot} be simply obtained (see detailed discussion in Section IV. 7. 2.). Denoting here and below averages w.r.t. $p_{\hat x}(\sigma_1,\sigma_2)$ as
$\langle O(\sigma_1,\sigma_2) \rangle := \int d\sigma_1 d\sigma_2 O(\sigma_1,\sigma_2) 
p_{\hat x}(\sigma_1,\sigma_2)$, the simplest
observable which can be obtained is, by averaging \eqref{hsum2}
\bea \label{exph1} 
\overline{h} = \kappa_1^{\mbox{\tiny{GUE}}} + \Delta^{-1/3} \langle \sigma_2-\sigma_1 \rangle   + O(\Delta^{-2/3})
\eea
Since the marginals of $p_x$ are GUE-TW for $\sigma_1$, and EBR for $\sigma_2$ respectively, 
with $\langle \sigma_2 \rangle=\hat x^2$ (see the two previous subsections),
one obtains
\bea  \label{exph2} 
\overline{h} = \kappa_1^{\mbox{\tiny{GUE}}} (1 - \Delta^{-1/3})  + \hat x^2+ O(\Delta^{-2/3})
\eea
which coincides with the large $\Delta$ expansion of \eqref{hmeanexact}. Hence from the simple average we do not learn anything new, it is simply a test of the method. A more interesting two-time KPZ observable that
can be calculated at large $\Delta$ from $p_{\hat x}(\sigma_1,\sigma_2)$, 
is the two time correlation $\overline{h h_1}$ obtained from
$\langle \sigma_2 \sigma_1 \rangle$. An even more detailed information is contained in the conditional average $\langle \sigma_1-\sigma_2 \rangle_{\sigma_1}$, see
definition below. As discussed in \cite{deNardisPLDTakeuchi}, conditioning these observables to
$h_1 \ge \sigma_{1c}$ 
leads to observables which can be (and in some cases, have been)
efficiently compared to experiments and numerics. We study these observables below,
keeping close in spirit to the notations defined in \cite{deNardisPLDTakeuchi,deNardisPLD2timeLong}.
We first define them and show how they can be calculated in Sections IV. 4. and IV. 5. and in
IV. 6. we perform a numerical evaluation of the result. 

\subsection{IV. 4. Two-time KPZ universal height covariance ratio} 

The two-time height correlation, normalized to
its single time value, is defined for the droplet initial condition (centered at $x$) and in the large time limit as
\bea \label{CDelta} 
&& C_\Delta = \lim_{t_1 \to +\infty} \frac{\overline{ h(0,t_1|x,0) h(0,t_2|x,0)}^c}{\overline{h(0,t_1|x,0)^2}^c} \bigg|_{t_2=(1+\Delta) t_1} = \frac{\overline{h_1 h_2}^c}{\overline{h_1^2}^c} 
\eea 
Here we are interested in its large $\Delta$ limit which, using \eqref{defh1h2}
and \eqref{abuse},
can be obtained as
\bea \label{Cinfty22} 
&& C_{\infty} = \frac{\langle \sigma_1 \sigma_2 \rangle 
- \langle \sigma_1 \rangle \langle \sigma_2 \rangle }{\langle \sigma_1^2 \rangle^c}
= \frac{ \int d \sigma_1 d \sigma_2 \sigma_1 (\sigma_2 - \hat x^2) p_{\hat x}(\sigma_1,\sigma_2)}{\kappa_2^{\mbox{\tiny{GUE}}}} 
\eea 
using that $\langle \sigma_2 \rangle=\hat x^2$. This can be rewritten as
\bea \label{Cinfty3} 
&& C_{\infty} = \frac{\langle \sigma_1^2 \rangle + \langle \sigma_2^2 \rangle - \langle (\sigma_2-\sigma_1)^2 \rangle - 2 \hat x^2 \kappa_1^{\mbox{\tiny{GUE}} } }{2 \kappa_2^{\mbox{\tiny{GUE}} }} = \frac{1}{2} + 
\frac{(\kappa_1^{\mbox{\tiny{GUE}}} - \hat x^2)^2 + 
\kappa_2^{\mbox{\tiny{EBR}}} }
{2 \kappa_2^{\mbox{\tiny{GUE}} }} - \frac{\langle (\sigma_2-\sigma_1)^2 \rangle}{2 \kappa_2^{\mbox{\tiny{GUE}} }} 
\eea
For $\hat x=0$ we recover the formula \eqref{exact1} of the text,
and using 
\bea
&& \kappa_1^{\mbox{\tiny{GUE}}} = -1.771086807411 \\
&& \kappa_2^{\mbox{\tiny{GUE}}} = 0.81319479 \quad , \quad  \kappa_2^{\mbox{\tiny{BR}}} = 1.15039
\eea 
we obtain 
\bea
&& C_{\infty} = 3.13598 - \frac{\langle (\sigma_2-\sigma_1)^2 \rangle}{2 \kappa_2^{\mbox{\tiny{GUE}} }} \label{sumup} 
\eea

One needs to calculate $\langle (\sigma_2-\sigma_1)^2 \rangle$ numerically using the 
formula \eqref{resFx}. This is done in Section IV. 6. below, leading to the result 
\eqref{estimate1}. To perform this calculation more conveniently we first show the identity \eqref{var2}
below, valid for any $\hat x$ (which reduces to \eqref{var1} in the text for $\hat x=0$).
To this purpose, since $\lim_{\sigma_2 \to +\infty}  G_{\hat x}(\sigma_1,\sigma_2)=F_2(\sigma_1)$ with fast convergence (see previous sections), in order to deal with convergent integrals
we define 
\be \label{Gtildedef} 
\tilde G_{\hat x}(\sigma_1,\sigma_2) = G_{\hat x}(\sigma_1,\sigma_2) - F_2(\sigma_1)
\ee
Then
we have $\lim_{\sigma_2 \to \pm \infty}  \tilde G_{\hat x}(\sigma_1,\sigma_2)=0$ with fast convergence,
and $\tilde G_{\hat x}(\sigma_1 \geq \sigma_2,\sigma_2) = F_0(\sigma_2-\hat x^2;\hat x) - F_2(\sigma_1)$.
We recall that in the text we
denote $G(\sigma_1,\sigma_2)=G_{\hat x=0}(\sigma_1,\sigma_2)$.

Let us start with the identity
\bea
&& \partial_{\sigma_2} \partial_{\sigma_1} [ ({\sigma_2}-\sigma_1)^2 \tilde G_{\hat x}(\sigma_1,{\sigma_2}) ]
= 2 \partial_{\sigma_2} [(\sigma_1-{\sigma_2}) \tilde G_{\hat x}(\sigma_1,{\sigma_2})]  + \partial_{\sigma_2}  [({\sigma_2}-\sigma_1)^2 \partial_{\sigma_1} \tilde G_{\hat x}(\sigma_1,{\sigma_2})] \\
&& = 2 \partial_{\sigma_2} [(\sigma_1-{\sigma_2}) \tilde G_{\hat x}(\sigma_1,{\sigma_2}) ]
+2  ({\sigma_2}-\sigma_1)  \partial_{\sigma_1} \tilde G_{\hat x}(\sigma_1,{\sigma_2}) 
+  ({\sigma_2}-\sigma_1)^2 \partial_{\sigma_2} \partial_{\sigma_1} \tilde G_{\hat x}(\sigma_1,{\sigma_2})
\nn
\eea 
We want to integrate the last term over the sector $\sigma_1<\sigma_2$ since it vanishes for
$\sigma_1>\sigma_2$ (see previous section). We note that the integral of the three other terms are either zero, or simpler, namely
\bea
&& \int_{-\infty}^{+\infty} d{\sigma_2} \int_{-\infty}^{{\sigma_2}}  d\sigma_1  \partial_{\sigma_2} \partial_{\sigma_1} [({\sigma_2}-\sigma_1)^2 \tilde G_{\hat x}(\sigma_1,{\sigma_2})]  = \int_{-\infty}^{+\infty} d{\sigma_2} \partial_{\sigma_2} \int_{-\infty}^{{\sigma_2}}  d\sigma_1   \partial_{\sigma_1} [({\sigma_2}-\sigma_1)^2 \tilde G_{\hat x}(\sigma_1,{\sigma_2})] = 0 \\
&& 2 \int_{-\infty}^{+\infty} d\sigma_1 \int_{\sigma_1}^{+\infty}  d{\sigma_2} \partial_{\sigma_2} [ (\sigma_1-{\sigma_2}) \tilde G_{\hat x}(\sigma_1,{\sigma_2})]
=  2 \int_{-\infty}^{+\infty} d\sigma_1  [  (\sigma_1-{\sigma_2}) \tilde G_{\hat x}(\sigma_1,{\sigma_2})  ]_{\sigma_1}^{+\infty} = 0\\
&& 2 \int_{-\infty}^{+\infty} d{\sigma_2} \int_{-\infty}^{{\sigma_2}}  d\sigma_1 ({\sigma_2}-\sigma_1)  \partial_{\sigma_1} \tilde G_{\hat x}(\sigma_1,{\sigma_2}) 
=  2  \int_{-\infty}^{+\infty} d{\sigma_2} \int_{-\infty}^{{\sigma_2}}  d\sigma_1  \tilde G_{\hat x}(\sigma_1,{\sigma_2}) 
+ 2  \int_{-\infty}^{+\infty} d{\sigma_2} [ ({\sigma_2}-\sigma_1)  \tilde G_{\hat x}(\sigma_1,{\sigma_2}) ]^{\sigma_1={\sigma_2}}_{\sigma_1=-\infty} \nonumber \\
&& = 2  \int_{-\infty}^{+\infty} d{\sigma_2} \int_{-\infty}^{{\sigma_2}}  d\sigma_1  \tilde G_{\hat x}(\sigma_1,{\sigma_2}) 
\eea 
Hence we have shown that
\bea \label{var2} 
\langle ({\sigma_2}-\sigma_1)^2 \rangle := \int_{-\infty}^{+\infty} d{\sigma_2} \int_{-\infty}^{{\sigma_2}}  
d\sigma_1   ({\sigma_2}-\sigma_1)^2 \partial_{\sigma_2} \partial_{\sigma_1} \tilde G_{\hat x}(\sigma_1,{\sigma_2}) 
= - 2  \int_{-\infty}^{+\infty} d{\sigma_2} \int_{-\infty}^{{\sigma_2}}  d\sigma_1  \tilde G_{\hat x}(\sigma_1,{\sigma_2}) 
\eea 
which, for $\hat x=0$, is precisely Eq. \eqref{var1} of the text. 

It is possible to perform exactly one of the integration in that double integral, leading to
an explicit expression as a single integral.
Starting from \eqref{resFx} we can integrate over $\sigma_2$
which leads to
\bea \label{altern} 
&& \int_{-\infty}^{+\infty} d\sigma_1 \int_{\sigma_1}^{+\infty}  d\sigma_2   
[ G_{\hat x}(\sigma_1,\sigma_2) - F_2(\sigma_1)] = \int_{-\infty}^{+\infty} d\sigma_1 W_{\hat x}(\sigma_1) \\
&& W_{\hat x}(\sigma_1) = F_2(\sigma_1) 
\times \bigg( Y_{\hat x}(\sigma_1) 
\Tr [ (I- P_{\sigma_1} K_{\Ai})^{-1} P_{\sigma_1} K_{\Ai} ]  \nn
+  \Tr[  (I- P_{\sigma_1} K_{\Ai})^{-1} P_{\sigma_1} K_{\Ai}  
(I- P_{\sigma_1} K_{\Ai})^{-1} P_{\sigma_1} {\cal B}_{\hat x} {\cal B}_{- \hat x}^T] \\
&&
 -  \,
\Tr[ (I- P_{\sigma_1} K_{\Ai})^{-1} P_{\sigma_1}  (\int_{0}^{+\infty}  d\sigma e^{- \hat x \sigma} \Ai_\sigma)  {\cal B}_{\hat x}^T]  \bigg) -  \,
\Tr[ (I- P_{\sigma_1} K_{\Ai})^{-1} P_{\sigma_1}  (\int_{0}^{+\infty}  d\sigma 
e^{\hat x \sigma} \Ai_\sigma)  {\cal B}_{-\hat x}^T]  \bigg) \nn
\eea 

\subsection{IV. 5. Conditional first moment and conditional correlation}

Another observable defined in \cite{deNardisPLD2timeLong,deNardisPLDTakeuchi}
is the average of $h$ conditioned to an observed value of $h_1$, denoted $\overline{h}_{h_1}\equiv E(h|h_1)$, where here $E$ denotes expectation w.r.t. the KPZ noise. Since $h_1=\sigma_1-\hat x^2$
in loose notations, from \eqref{hsum2}
and \eqref{defh1h2} it obeys
\be
\overline{h}_{h_1=\sigma_1-\hat x^2} = \kappa_1^{\mbox{\tiny{GUE}}}  
+ \Delta^{-1/3} \langle \sigma_2-\sigma_1 \rangle_{\sigma_1} + O(\Delta^{-2/3})
\ee 
To make contact with the notations of \cite{deNardisPLD2timeLong,deNardisPLDTakeuchi}
we will define a function, $R^{\rm exact}_{1/3}$, equal to the conditional moment, defined as (see \eqref{condmom1} in the text) 
\bea
R^{\rm exact}_{1/3}(\sigma_1) := \langle \sigma_2-\sigma_1 \rangle_{\sigma_1} := \frac{1}{\partial_{\sigma_1} F_2(\sigma_1)} 
\int_{\sigma_1}^{+\infty} d\sigma_2 (\sigma_2-\sigma_1) \partial_{\sigma_2} \partial_{\sigma_1} 
\tilde G_{\hat x}(\sigma_1,\sigma_2) \label{exactR13} 
\eea 
where the index "exact" distinguish it from the approximation to this function obtained there, see below.
We also define a second function
\bea
\tilde R^{\rm exact}_{1/3}(\sigma_1) := \langle \sigma_2 - \sigma_1 \rangle_{\sigma_1} \partial_{\sigma_1} F_2(\sigma_1) \label{exactRt13} 
\eea 
which, upon integration by part of \eqref{exactR13} can be also written as
\be \label{intpart1} 
\tilde R^{\rm exact}_{1/3}(\sigma_1) = -  
\int_{\sigma_1}^{+\infty} d\sigma_2 \, \partial_{\sigma_1} \tilde G_{\hat x}(\sigma_1,\sigma_2) =  F_2(\sigma_1) - F_0(\sigma_1-\hat x^2;\hat x) -  \partial_{\sigma_1} \int_{\sigma_1}^{+\infty} d\sigma_2 \, \tilde G_{\hat x}(\sigma_1,\sigma_2) 
\ee
a form more convenient for numerical evaluation. We have used that $G_{\hat x}(\sigma_1,\sigma_1)=F_0(\sigma_1-\hat x^2;\hat x)$ and we 
recall that $\tilde G_{\hat x}(\sigma_1,\sigma_2) = G_{\hat x}(\sigma_1,\sigma_2) - F_2(\sigma_1)$ and 
that the JPDF is $p_{\hat x}(\sigma_1,\sigma_2) = \partial_{\sigma_2} \partial_{\sigma_1}G_{\hat x}(\sigma_1,\sigma_2) 
= \partial_{\sigma_2} \partial_{\sigma_1} \tilde G_{\hat x}(\sigma_1,\sigma_2)$.  \\


Another useful conditional expectation was defined 
in \cite{deNardisPLD2timeLong,deNardisPLDTakeuchi}, conditionned to
$h_1 \geq \sigma_{1c} - \hat x^2$ i.e. $\sigma_1> \sigma_{1c}$ (there for $\hat x=0$). 
One has again $
\overline{h}_{h_1 \geq \sigma_1-\hat x^2} = \kappa_1^{\mbox{\tiny{GUE}}}  
+ \Delta^{-1/3} \langle \sigma_2-\sigma_1 \rangle_{\sigma_1 \geq  \sigma_{1c}} + O(\Delta^{-2/3})$,
where such conditional averages are denoted as $\langle \ldots \rangle_{\sigma_1 \geq \sigma_{1c}}$
and for the first moment it reads
\be
\langle \sigma_2-\sigma_1\rangle_{\sigma_1 > \sigma_{1c}} 
= \frac{1}{1-F_2(\sigma_{1c})} 
\int_{\sigma_{1c}}^{+\infty} d\sigma_1 
\int_{\sigma_1}^{+\infty} d\sigma_2 (\sigma_2-\sigma_1) \partial_{\sigma_2} \partial_{\sigma_1} \tilde G_{\hat x}(\sigma_1,\sigma_2)= \frac{1}{1-F_2(\sigma_{1c})} 
\int_{\sigma_{1c}}^{+\infty} d\sigma_1   \tilde R^{\rm exact}_{1/3}(\sigma_1) \label{aga2} 
\ee
Using the form \eqref{intpart1} obtained by integration by parts, it can be rewritten as a single integral
\bea
\langle \sigma_2-\sigma_1\rangle_{\sigma_1 > \sigma_{1c}} 
= \frac{1}{1-F_2(\sigma_{1c})} 
\int_{\sigma_{1c}}^{+\infty} d\sigma (F_2(\sigma)-F_0(\sigma-\hat x^2;\hat x) +  \tilde G_{\hat x}(\sigma_{1c},\sigma) ) \label{aga1} 
\eea 
\\

Another important observable was considered in \cite{deNardisPLD2timeLong,deNardisPLDTakeuchi},
namely a conditional variant of the two-time covariance ratio \eqref{CDelta}, defined as
\bea \label{CDelta2} 
&& C_\Delta(\sigma_{1c}) := \frac{\overline{h_1 h_2}^c_{h_1> \sigma_{1c}-\hat x^2} }{\overline{h_1^2}^c_{h_1> \sigma_{1c}-\hat x^2}} 
\eea 
From the above considerations, the large $\Delta$ limit, $C_{\infty}(\sigma_{1c})  = \lim_{\Delta \to +\infty} 
C_\Delta(\sigma_{1c})$,
can be obtained as 
\bea \label{Cinfty10} 
&& C_{\infty}(\sigma_{1c}) = 1 + \frac{\langle (\sigma_2 -\sigma_1) \sigma_1 \rangle_{\sigma_1>\sigma_{1c}}
- \langle \sigma_2-\sigma_1 \rangle_{\sigma_1>\sigma_{1c}} \langle \sigma_1 \rangle_{\sigma_1>\sigma_{1c}}
}{\langle \sigma_1^2 \rangle_{\sigma_1>\sigma_{1c}}- \langle \sigma_1 \rangle_{\sigma_1>\sigma_{1c}}^2}  
\eea
a function which interpolates between $C_{\infty}(\sigma_{1c}=-\infty)= C_{\infty}$ the unconditionned 
two-time covariance ratio studied above, and $C_{\infty}(\sigma_{1c}=+\infty)=1$ (see below). It can be written as (see (20) in \cite{deNardisPLDTakeuchi})
\bea \label{Csigc} 
&& C_{\infty}(\sigma_{1c}) = 1 + \frac{
\int_{\sigma_{1c}}^{+\infty} d\sigma_1 \sigma_1 \tilde R^{\rm exact}_{1/3}(\sigma_1)  - {\cal N}(\sigma_{1c})^{-1} 
\int_{\sigma_{1c}}^{+\infty} d\sigma_1 \tilde R^{\rm exact}_{1/3}(\sigma_1) \int_{\sigma_{1c}}^{+\infty} d\sigma_1 \sigma_1 \partial_{\sigma_1} F_2(\sigma_1)  }{
\int_{\sigma_{1c}}^{+\infty} d\sigma_1 \sigma_1^2 \partial_{\sigma_1} F_2(\sigma_1) - {\cal N}(\sigma_{1c})^{-1}  
[\int_{\sigma_{1c}}^{+\infty} d\sigma_1 \sigma_1 \partial_{\sigma_1} F_2(\sigma_1) ]^2 }
\eea
where we have defined $ {\cal N}(\sigma_{1c}) = 1 - F_2(\sigma_{1c})$. In this expression, one can use \eqref{intpart1} and further integrations by parts to evaluate
\bea
\int_{\sigma_{1c}}^{+\infty} d\sigma_1 \sigma_1 \tilde R^{\rm exact}_{1/3}(\sigma_1) 
= \int_{\sigma_{1c}}^{+\infty} d\sigma_2 [ \sigma_{1c} \tilde G_{\hat x}(\sigma_{1c},\sigma_2)  
+ \sigma_2 (F_2(\sigma_2) - F_0(\sigma_2- \hat x^2, \hat x)) + 
\int_{\sigma_{1c}}^{\sigma_2} d\sigma_1  \tilde G_{\hat x}(\sigma_1,\sigma_2) ]
\eea 
remembering that $p_{\hat x}(\sigma_1,\sigma_2)=0$ for $\sigma_1>\sigma_2$. The
integral $\int_{\sigma_{1c}}^{+\infty} d\sigma_1 \tilde R^{\rm exact}_{1/3}(\sigma_1)$
was given in a simpler form above in \eqref{aga2},\eqref{aga1}.

\subsection{IV. 6. Numerical evaluation}

Here we use the general, and by now standard, method of numerical calculation of traces and determinants given in \cite{Bornemann}. It was also summarized near
Eqs (21)-(22) in \cite{deNardisPLDTakeuchi} to which we refer for details. 

To calculate $C_{\infty}$ from \eqref{Cinfty3}, we
have evaluated numerically, for $\hat x=0$, the following integral
\be
- \frac{1}{2} \langle (\sigma_2-\sigma_1)^2 \rangle = \int_{-\infty}^{+\infty} d\sigma_2 \int_{-\infty}^{\sigma_2}  d\sigma_1  (G(\sigma_1,\sigma_2)-F_2(\sigma_1)) \label{var1new}
\ee 
Using formula \eqref{ResF} for $G(\sigma_1,\sigma_2)$ we find $-2.044 \pm 0.001$ for this quantity.
As a check, we also performed an independent calculation using the formula \eqref{altern},
which led to $-2.0445 \pm 0.0005$. Using formula \eqref{sumup} we arrive at the estimate
\bea \label{estimate1} 
C_{\infty} = 0.6225 \pm 0.0015
\eea
for the infinite time-separation universal covariance ratio. 

To evaluate, for $\hat x=0$, the conditional average we define 
and calculate numerically the following integral (from \eqref{intpart1})
\bea
S(\sigma_{1c}) := \int_{\sigma_{1c}}^{+\infty} d\sigma_1 \tilde R^{\rm exact}_{1/3}(\sigma_1)
= \int_{\sigma_{1c}}^{+\infty} d\sigma (F_2(\sigma)-F_0(\sigma) + 
G(\sigma_{1c},\sigma) - F_2(\sigma_{1c}) ) \label{defSS} 
\eea 
from which we obtain the conditional average 
\bea
\langle \sigma_2-\sigma_1\rangle_{\sigma_1 > \sigma_{1c}} 
= \frac{S(\sigma_{1c})}{1-F_2(\sigma_{1c})} 
\eea 
The result is plotted in Fig. \ref{fig:condmean} of the main text,
and we refer to the caption for comments. Here we give some
numerical values
\bea
\left(
\begin{array}{ccccccccccccc}
 -4.5 & -4. & -3.5 & -3. & -2.5 & -2. & -1.5 & -1 & 0 & 1 & 2 & 3 & 4 \\
 1.771 & 1.767 & 1.752 & 1.712 & 1.639 & 1.538 & 1.423 & 1.305 & 1.094 &
   0.932 & 0.814 & 0.727 & 0.661 \\
\end{array}
\right)
\eea

To evaluate the conditional covariance ratio \eqref{Cinfty10}  we
evaluate (from \eqref{intpart1} and \eqref{defSS})
\bea
\int_{\sigma_{1c}}^{+\infty} d\sigma_1 \sigma_1 \tilde R^{\rm exact}_{1/3}(\sigma_1)
= \int_{\sigma_{1c}}^{+\infty} d\sigma_1 S(\sigma_1) 
+ \sigma_{1c} S(\sigma_{1c}) 
\eea
We then rewrite \eqref{Csigc} in a form more suitable for the numerical
evaluation
\bea \label{Csigc2} 
&& C_{\infty}(\sigma_{1c}) = 1 + \frac{ \int_{\sigma_{1c}}^{+\infty} d\sigma_1 S(\sigma_1) - 
S(\sigma_{1c}) \frac{\int_{\sigma_{1c}}^{+\infty} d\sigma_1 (1- F_2(\sigma_1))}{1- F_2(\sigma_{1c}) } }{2
\int_{\sigma_{1c}}^{+\infty} d\sigma_1 (\sigma_1-\sigma_{1c}) (1-F_2(\sigma_1)) 
- \frac{[\int_{\sigma_{1c}}^{+\infty} d\sigma_1 (1- F_2(\sigma_1)]^2}{1- F_2(\sigma_{1c}) }}
\eea
The result is plotted in Fig. \ref{fig:C} of the main text in the interval 
$\sigma_{1c} \in [-2.5,4]$, the evaluation outside this interval would require
enhanced numerical precision and was not attempted here. \\

Let us now, for sake of comparison, display the asymptotic formula
for these observables
obtained from the analysis of Section IV.7. below, which are in general
agreement with the predictions of \cite{deNardisPLD2timeLong}.
We display here the series expansions not displayed explicitly there.
For the conditional mean one finds
\bea
 \langle \sigma_2-\sigma_1\rangle_{\sigma_1 > \sigma_{1c}}^{\rm asympt} 
= \frac{S^{\rm asympt}(\sigma_{1c})}{\int_{\sigma_{1c}}^{+\infty} d\sigma K_{\Ai}(\sigma,\sigma)}
\eea
%
with 
\bea \label{Sasympt} 
S^{\rm asympt}(\sigma_{1c}) &=& \int_{\sigma_{1c}}^{+\infty} d\sigma_1
\int_{\sigma_{1}}^{+\infty} d\sigma_2 ( 2 K_{\Ai}(\sigma_1,\sigma_2) 
-  K_{\Ai}(\sigma_2,\sigma_2) ) \\
&=&
\int_{\sigma_{1c}}^{+\infty} d\sigma \big( [\int_{\sigma}^{+\infty} dy \Ai(y)]^2 
- \int_{\sigma_1}^{+\infty} d \sigma K_{\Ai}(\sigma,\sigma) \big) \nn
\eea 
We obtain the series expansion at large argument
\bea
&& S^{\rm asympt}(y) = e^{-\frac{4 y^{3/2}}{3}} 
\bigg( \frac{3}{32 \pi  y^2}-\frac{67}{256 \pi y^{7/2}}+\frac{3887}{4096
   \pi  y^5}+O\left(y^{-13/2}\right) \bigg) \\
&& \int_{y}^{+\infty} d\sigma K_{\Ai}(\sigma,\sigma) = e^{-\frac{4 y^{3/2}}{3}} 
\bigg(
\frac{1}{16 \pi y^{3/2}}-\frac{35}{384 \pi  y^3}+\frac{3745
   }{18432 \pi y^{9/2}
   }+O\left(y^{-13/2}\right)   \bigg) 
\eea
Taking the ratio 
we find
\bea
\langle \sigma_2-\sigma_1\rangle_{\sigma_1 > \sigma_{1c}}  = \frac{3}{2 \sigma_{1c}^{1/2}} - 
\frac{2}{\sigma_{1c}^2} +  \frac{473}{64 \sigma_{1c}^{7/2}} 
- \frac{4953}{128 \sigma_{1c}^5} +  O(\sigma_{1c}^{-13/2}) 
\eea 
which is found to be a good approximation for $\sigma_{1c} \geq 4$.
We also have (as also displayed in \cite{deNardisPLD2timeLong})
\bea
R^{\rm exact}_{1/3}(\sigma_1) = \langle \sigma_2-\sigma_1\rangle_{\sigma_1}  = \frac{3}{2 \sigma_{1}^{1/2}} - 
\frac{13}{8 \sigma_{1c}^2} + \frac{327}{64 \sigma_{1c}^{7/2}} 
- \frac{1513}{64 \sigma_{1c}^5} +  O(\sigma_{1c}^{-13/2}) 
\eea 


For the conditional covariance ratio, the asymptotics gives
\bea
C_{\infty}(\sigma_{1c}) = 1
-\frac{3}{4} \frac{1}{\sigma_{1c}^{3/2}}+\frac{35}{8 \sigma_{1c}^3}-\frac{4023}{128}
   \frac{1}{\sigma_{1c}^{9/2}} +O\left(\frac{1}{\sigma_{1c}^6}\right)
\eea
which, however, appears to be a good approximation only for relatively large $\sigma_{1c} \geq 4$.
Note from Fig. \ref{fig:condmean} and Fig. \ref{fig:C} that the {\it full asymptotic formula},
i.e. using the complete form \eqref{Sasympt},
is a good approximation to the exact result in a much broader region, $\sigma_{1c} \geq -0.5$ and
$\sigma_{1c} \geq 0.5$ respectively, than their power series asymptotics.

%
%


\subsection{IV. 7. Large $\sigma_1$ expansion of the joint CDF and of the conditional moments}

\subsubsection{IV. 7. 1. Leading order} 

Here we evaluate the JCDF $G_{\hat x}(\sigma_1,\sigma_2)$ and the JPDF
$p_{\hat x}(\sigma_1,\sigma_2)=\partial_{\sigma_1} \partial_{\sigma_2} G_{\hat x}(\sigma_1,\sigma_2)$,
in the limit of large positive $\sigma_1 \gg 1$ at fixed $\sigma_{21}=\sigma_2-\sigma_1$. Note that we
can restrict to $\sigma_{21} \geq 0$ since the JPDF vanishes for $\sigma_2 < \sigma_1$. 
We use the fact that $\Ai(\sigma_1) = O(\exp(- \frac{2}{3} \sigma_1^{3/2}))$, where $O(..)$ means up to a
prefactor which is an algebraic series in fractional powers of $1/\sigma_1$ which we do not write explicitly. 
One can thus organize the large $\sigma_1$ expansion as a sum of terms of the type
$O(\exp(- \frac{2 k}{3} \sigma_1^{3/2}))$, $k=1,2,.. k_{\max}$, and discard products of more than $k_{\max}$ integrals of Airy functions.

We start by focusing on $k_{\max}=2$, i.e. up to product of two Airy functions only. Instead 
of writing $\sigma_2=\sigma_1+\sigma_{21}$ explicitly, it is convenient to use the $\sigma_2$ variable, keeping in mind that it is large and of the same order as $\sigma_1$. 
Let us perform the counting of the degree $k$ for each of the building blocks of \eqref{resFx} (and, for $x=0$, of \eqref{ResF}
in the text). We note that 
\bea
&& F_2(\sigma_1) = {\rm Det}[ I -  P_{\sigma_1} K_{\Ai} P_{\sigma_1} ] \simeq_{\sigma_1 \gg 1}   1 - \Tr[P_{\sigma_1} K_{\Ai}] =
 1 - \int_{\sigma_1}^{+\infty} du K_{\Ai}(u,u) 
\eea
hence $F_2$ equals unity plus an infinite sum of even degrees $k \geq 2$. Similarly from
\eqref{defBw2}, which we rewrite as
\be
{\cal B}_{\hat x}(u) = e^{- \hat x u} (e^{\frac{\hat x^3}{3}} - \int_u^{+\infty} dy \Ai(y) e^{\hat x y} )
\ee 
i.e. we see that ${\cal B}_{\hat x}$ is the sum of a degree $k=0$ and degree $k=1$ term. From \eqref{defLx2}
we see that ${\cal L}_{\hat x}$ is a finite sum of degrees $k=0,1,2$
and from \eqref{defYx2}, that $Y_{\hat x}$ equals $1+ {\cal L}_{\hat x}$ 
plus an infinite sum of degrees $k \geq 2$.

The first line in 
\eqref{resFx} contains already two explicit Airy functions ($k=2$), hence we only need $F_2(\sigma_1) Y_{\hat x}(\sigma_1)$ to order $k=0$. This means that in the first line we can replace $F_2(\sigma_1) \to 1$ and, from
\eqref{defYx2} and \eqref{defLx2}, $Y_{\hat x}(\sigma_1) \to 1 + {\cal L}_{\hat x}(\sigma_1) \to \sigma_1-\hat x^2$. Hence
\bea
\text{First line in \eqref{resFx}} \simeq (\sigma_1-\hat x^2) {\rm Tr} [  P_{\sigma_1}  \Ai_{\sigma_2-\sigma_1} \Ai_{\sigma_2-\sigma_1}^T ]
=  (\sigma_1-\hat x^2) K_{\Ai}(\sigma_2,\sigma_2) \to \sigma_1 K_{\Ai}(\sigma_2,\sigma_2) 
\eea 
in the last step we have discarded terms which do not depend on {\it both} $\sigma_1$ and
$\sigma_2$, since they do not contribute to $p_{\hat x}(\sigma_1,\sigma_2)$.

The second line of \eqref{resFx}, upon expanding, contain four terms. The term $F_2(\sigma_1)$
alone can be discarded (it depends on only one of the two variables). The term product of two traces
contains already two Airy functions ($k=2$) hence we can replace in that term $F_2(\sigma_1) \to 1$ 
and $(I- P_{\sigma_1} K_{\Ai} )^{-1} \to I$ in the traces leading to
\bea \label{equiv11} 
 {\rm Tr}[ P_{\sigma_1} \Ai_{\sigma_2-\sigma_1}
{\cal B}_{\hat x}^T]  {\rm Tr}[  P_{\sigma_1} \Ai_{\sigma_2-\sigma_1}
{\cal B}_{-\hat x}^T] \simeq (\int_0^{+\infty} \Ai(v+\sigma_2) e^{- \hat x v})
 (\int_0^{+\infty} \Ai(v+\sigma_2) e^{ \hat x v})
\to 0
\eea
In the 
first equivalence in \eqref{equiv11} we have replaced ${\cal B}_{\pm \hat x}(v)$ 
by its $k=0$ piece, $e^{\pm \frac{\hat x^3}{3} \mp \hat x v}$, and shifted the
integrals by $\sigma_1$ and in the last step
we observed that the result does not depend on $\sigma_1$, hence can be discarded. In the cross terms
we see that we can again set $F_2(\sigma_1) \to 1$ 
and $(I- P_{\sigma_1} K_{\Ai} )^{-1} \to I$ since their leading degree is already $k=1$ and
the Airy kernel increases the degree by $k \to k+2$. The cross term becomes
\bea
&& - F_2(\sigma_1) (e^{-  \hat x (\sigma_2-\sigma_1)} {\rm Tr}[ (I- P_{\sigma_1} K_{\Ai} )^{-1} P_{\sigma_1} \Ai_{\sigma_2-\sigma_1}
{\cal B}_{\hat x}^T] - e^{\hat x (\sigma_2-\sigma_1)} {\rm Tr}[ (I- P_{\sigma_1} K_{\Ai} )^{-1} P_{\sigma_1} \Ai_{\sigma_2-\sigma_1} {\cal B}_{-\hat x}^T]) \\
&& \simeq - e^{-  \hat x (\sigma_2-\sigma_1)} {\rm Tr}[ P_{\sigma_1} \Ai_{\sigma_2-\sigma_1}
{\cal B}_{\hat x}^T] - e^{\hat x (\sigma_2-\sigma_1)} {\rm Tr}[ P_{\sigma_1} \Ai_{\sigma_2-\sigma_1} {\cal B}_{-\hat x}^T]  \nn
\eea 
Now we can calculate
\bea
- e^{-  \hat x (\sigma_2-\sigma_1)} {\rm Tr}[ P_{\sigma_1} \Ai_{\sigma_2-\sigma_1}
{\cal B}_{\hat x}^T] &=& - e^{-  \hat x (\sigma_2-\sigma_1)} \int_{\sigma_1}^{+\infty} du \Ai(u+\sigma_2-\sigma_1)
e^{- \hat x u} (e^{\frac{\hat x^3}{3}} - \int_u^{+\infty} dy \Ai(y) e^{\hat x y} ) \\
& =& -  \int_{\sigma_2}^{+\infty} du \Ai(u)
e^{- \hat x u} (e^{\frac{\hat x^3}{3}} - \int_{u+\sigma_1-\sigma_2}^{+\infty} dy \Ai(y) e^{\hat x y} ) \nn
\eea
where we have shifted $u \to u+\sigma_1-\sigma_2$ in the second line: this shows that the first
term does not depend on $\sigma_1$, hence we can discard it.

Putting together and taking the derivatives, we finally obtain the leading behavior of the JPDF for $\sigma_{21} \geq 0$ fixed and large positive $\sigma_1 \gg 1$, which we call $p_{\hat x}^{(1)}$, as
\bea
p_{\hat x}(\sigma_1,\sigma_2) &= & \partial_{\sigma_2} \partial_{\sigma_1} G_{\hat x}(\sigma_1,\sigma_2)   \simeq 
p_{\hat x}^{(1)}(\sigma_1,\sigma_2)  + O(\exp(- 2 \sigma_1^{3/2})) \\
p_{\hat x}^{(1)}(\sigma_1,\sigma_2) & = &
\partial_{\sigma_2} [ K_{\Ai}(\sigma_2,\sigma_2) - 2 \cosh(\hat x(\sigma_1-\sigma_2) )
K_{\Ai}(\sigma_1,\sigma_2) ] \\
&=& - \Ai(\sigma_2)^2 - 2 \partial_{\sigma_2} [ \cosh(\hat x(\sigma_1-\sigma_2) ) K_{\Ai}(\sigma_1,\sigma_2) ]  \nn
\eea
which is $O(\exp(- \frac{4}{3} \sigma_1^{3/2}))$ and we have neglected 
all terms with $k > k_{\max}=2$ i.e. of order $O(\exp(- 2 \sigma_1^{3/2}))$. One can 
check that to this order 
\bea
\int_{\sigma_1}^{+\infty} d\sigma_2 p_{\hat x}^{(1)}(\sigma_1,\sigma_2) = K_{\Ai}(\sigma_1,\sigma_1) = \partial_{\sigma_1} F_2^{(1)}(\sigma_1) 
\eea 
as required, since the exact sum rule is
\bea \label{sumruleexact} 
\int_{\sigma_1}^{+\infty} d\sigma_2 p_{\hat x}(\sigma_1,\sigma_2)  = \partial_{\sigma_1} F_2(\sigma_1) 
\eea 
and $F^{(1)}_2$ is the leading tail approximation of the CDF of the GUE-TW distribution
of the same order accuracy $O(\exp(- \frac{4}{3} \sigma_1^{3/2}))$ ($k=2$). \\

Let us now examine the behaviour of the PDF approximant $p_{\hat x}^{(1)}$ for $\sigma_{21} \to 0$. We recall that $p_{\hat x}(\sigma_1,\sigma_2)=0$ for
$\sigma_{21}=\sigma_2-\sigma_1<0$ and one finds
\be \label{exppx} 
p^{(1)}_{\hat x}(\sigma_1,\sigma_2) = \frac{2}{3} \left(\sigma _1 (\sigma_1+ 3 \hat x^2) \Ai \left(\sigma _1\right){}^2- (\sigma _1 + 3 \hat x^2)
   \Ai'\left(\sigma _1\right){}^2-2 \Ai \left(\sigma _1\right)
   \Ai'\left(\sigma _1\right)\right) \sigma_{21} +O\left(\sigma_{21}^2\right) 
\ee
hence the PDF exhibits a linear cusp singularity at $\sigma_{21}=0$, at least to this order of
approximation. We note that the function $p_{\hat x}^{(1)}(\sigma_1,\sigma_2)$ is positive only for $\hat x^2 < \hat x^*(\sigma_1)^2$ where
\be
 [\hat x^*(\sigma_1)]^2 := - \frac{\sigma_1}{3} + \frac{2 \Ai(\sigma_1) \Ai'(\sigma_1) }{3 \sigma_1 \Ai(\sigma_1)^2 - 3 \Ai'(\sigma_1)^2} \simeq_{\sigma_1 \gg 1} \sigma_1 + \frac{1}{\sigma_1^{1/2}} - \frac{3}{4 \sigma_1^2} + O(\frac{1}{\sigma_1^3}) 
\ee 
is the value of $\hat x^2$ at which the $O(\sigma_{21})$ term in \eqref{exppx} changes sign. The function
$\hat x^*(\sigma_1)^2$ has a mininum at 
$\sigma_1= -1.41801$ where $(\hat x^*)^2=0.267978$. This change of sign, which does not occur for
$\hat x=0$, is not a contradiction, but simply means that $p_{\hat x}^{(1)}(\sigma_1,\sigma_2)$ is not a uniformly good approximation in $\hat x$ to $p_{\hat x}^{(1)}(\sigma_1,\sigma_2)$, i.e. the values of $\sigma_1$ needed for it to be accurate increase with $\hat x^2$. The large $\sigma_1$ asymptotics, at fixed $\sigma_{21}>0$ reads
\be
p^{(1)}_{\hat x}(\sigma_1,\sigma_2) \simeq  
e^{- \frac{2}{3} \sigma_2^{3/2} - \frac{2}{3} \sigma_1^{3/2} }
\left( \frac{\cosh(\sigma_{21} \hat x)}{4 \pi \sigma_1^{1/2}} - \frac{\hat x \sinh(\sigma_{21} \hat x)}{4 \pi \sigma_1} + 
O(\frac{f(\sigma_{21} \hat x)}{\sigma_1^2}) \right) 
+ e^{- \frac{4}{3} \sigma_2^{3/2} } \left( \frac{-1}{4 \pi \sigma_1^{1/2}} + O(\frac{\sigma_{21}}{\sigma_1^{3/2}},\frac{1}{\sigma_1^2}) \right)
\ee 
which is, as required, positive in all cases. \\

We can now calculate the conditional first moment, $\langle \sigma_2-\sigma_1 \rangle_{\sigma_1}$,
defined in \eqref{exactR13},
to this (leading) order of the large $\sigma_1$ expansion. Specifically we compute
the approximation to $\tilde R^{\rm exact}_{1/3}(\sigma_1)$ defined in \eqref{exactRt13}. That defines a function,
which we call $\tilde R_{1/3}(\sigma_1)$ to stick with the notations of \cite{deNardisPLD2timeLong,deNardisPLDTakeuchi}. We find
\bea \label{rtilde} 
\tilde R_{1/3}(\sigma_1) &:= & \int_{\sigma_1}^{+\infty} d\sigma_2 (\sigma_2-\sigma_1) p^{(1)}(\sigma_1,\sigma_2) 
= \int_{\sigma_1}^{+\infty} d\sigma_2 (\sigma_2-\sigma_1)
\partial_{\sigma_2} [ K_{\Ai}(\sigma_2,\sigma_2) - 2 \cosh(\hat x(\sigma_1-\sigma_2) ) K_{\Ai}(\sigma_1,\sigma_2) ] \nn \\
& =&   \int_{\sigma_1}^{+\infty} d\sigma_2 \left( 2 \cosh(\hat x(\sigma_1-\sigma_2) )  K_{\Ai}(\sigma_1,\sigma_2) - K_{\Ai}(\sigma_2,\sigma_2) \right)
\eea
since the boundary term in the integration by part vanishes. For $\hat x=0$, we can use the identity
\bea
2 \int_{\sigma_1}^{+\infty} d\sigma K_{\Ai}(\sigma_1,\sigma)  =  \left(\int_{\sigma_1}^{+\infty} dy \Ai(y) \right)^2 
\eea 
to obtain the result \eqref{firstmomcond} in the text. As noted there, remarkably, the function 
$\tilde R_{1/3}(\sigma_1)$ exactly coincide with the one obtained in \cite{deNardisPLD2timeLong} (and similarly for $R_{1/3}(\sigma_1)=
\tilde R_{1/3}(\sigma_1)/K_{\Ai}(\sigma_1,\sigma_1)$) formulae (173) there.

This agreement extends to finite $\hat x$. Indeed the result of \cite{deNardisPLD2timeLong}, see formula (169) there, can be rewritten as
\bea \label{aga} 
&& \tilde R_{1/3}(\sigma_1) = - \partial_{\sigma_1} \int_{\sigma_1}^{\infty} dy_1 
\int_{\sigma_1}^{\infty}  dy_2 K_{\Ai}(y_1,y_2) \cosh( \bar X(y_1-y_2)) - \int_{\sigma_1}^{\infty} 
dy K_{\Ai}(y,y) \\
&&= 2  \int_{\sigma_1}^{\infty}  dy_2 \cosh( \bar X(\sigma_1-y_2)) K_{\Ai}(\sigma_1,y_2) - \int_{\sigma_1}^{\infty} 
dy K_{\Ai}(y,y) \nn
\eea 
which is exactly the same formula as \eqref{rtilde} is we identify $\bar X$ there with $\hat x$ here. 
There, the variable $\bar X$ was defined as $\bar X \simeq x_2/(2 t_2 t_1^{-1/3})$ for large $\Delta$
(see formula (164) there) where $x_2$ is the position at time $t_2$. The STS allows to 
relate this to shifting the position at time $t=0$, see Eq. \eqref{shift2} in Section XI below and the discussion around it
(there $\hat x$ is denoted
$\hat X_0$).

One can now replace $\tilde R^{\rm exact}_{1/3}(\sigma_1)$ by its approximation to any desired
order (here the leading one for large $\sigma_1$) in the formula for the two-time conditional covariance ratio \eqref{Csigc}, and obtain the corresponding approximation for this observable,
see Section IV. 6. where it is displayed. We thus fully confirm the
correctness of the leading order prediction for $C_{\infty}(\sigma_{1c})$ obtained in \cite{deNardisPLD2timeLong}, which was successfully tested in the experiments
\cite{deNardisPLDTakeuchi} and found accurate even beyond its naive range of validity
(i.e. $\sigma_1$ large positive).
Below, we obtain the next order corrections, which go beyond the
method of \cite{deNardisPLD2timeLong}. Before doing so let us discuss higher moments.

\subsubsection{IV. 7. 2. Higher moments}

The higher moments of $p^{(1)}_{\hat x}$ can also be calculated. Let us restrict to $\hat x=0$ here.
One has
\bea
\int_{\sigma_1}^{+\infty} d\sigma_2 (\sigma_2-\sigma_1)^2 p^{(1)}(\sigma_1,\sigma_2) 
= - 2 \int_{\sigma_1}^{+\infty} d\sigma_2 (\sigma_2-\sigma_1) [ K_{\Ai}(\sigma_2,\sigma_2) - 2 K_{\Ai}(\sigma_1,\sigma_2) ] 
\eea 
which is found {\it not} equal to $2 \tilde R_{2/3}(\sigma_1)$ defined in \cite{deNardisPLD2timeLong}. 
This is in agreement with the discussion
at the end of Section 7.4 in \cite{deNardisPLD2timeLong}. Indeed in the infinite times limit at fixed $\Delta=(t_2-t_1)/t_1$, we have, for large $\Delta$
\be
h  \simeq \tilde {\cal A}_2(0) + \Delta^{-\frac{1}{3}} ( \max_{\hat y \in \mathbb R} \big( {\cal A}_2(\hat y) - \hat y^2
+ \sqrt{2} B(\hat y) \big) - {\cal A}_2(0) ) + {\cal C} \Delta^{-\frac{2}{3}} + ..
\ee
denoting $\sigma$ the random variable $h$ as in the text, we rewrite in shorthand notations
\be
\sigma = \chi_2 + \Delta^{-\frac{1}{3}} (\sigma_2 - \sigma_1) + {\cal C} \Delta^{-\frac{2}{3}} + ..
\ee
where the JPDF of the variables $\sigma_1,\sigma_2$ 
is $p(\sigma_1,\sigma_2)$ obtained here exactly, the first term is the standard GUE-TW random variable, 
{\it uncorrelated} from $\sigma_1,\sigma_2$. Very little is known however about the random variable ${\cal C}$. In particular 
it is likely to be correlated with the first two terms. What information about the PDF of $h$ (or the JPDF of $h$ and 
$h_1$ can we then obtain from our exact knowledge of $p(\sigma_1,\sigma_2)$ ? Writing the cumulant generating
function
\bea
\overline{e^{ i \lambda \sigma}} = \overline{e^{ i \lambda \chi_2 + i \lambda  \Delta^{-\frac{1}{3}} (\sigma_2 - \sigma_1) + \lambda \Delta^{-\frac{2}{3}} {\cal C}} }
\eea
shows, by expansion in $\lambda$, that mixed correlations of $\chi_2$ and ${\cal C}$ prevent to determine the cumulants higher than the second one solely from the knowledge of $p(\sigma_1,\sigma_2)$. One can also write the JPDF defined
in \cite{deNardisPLD2timeLong} and recalled in the main text as
\bea
&& P_{\Delta}(\sigma_1,\sigma) = \overline{\delta(h_1-\sigma_1)  \delta(h-\sigma)}
= \overline{\delta(h_1-\sigma_1)  \delta(\chi_2 + \Delta^{-\frac{1}{3}} (\sigma_2 - \sigma_1) + {\cal C} \Delta^{-\frac{2}{3}} + ... -\sigma)} \nn \\
&& = \partial_{\sigma} F_2(\sigma) \partial_{\sigma_1} F_2(\sigma_1)
+ \Delta^{-\frac{1}{3}} \partial^2_{\sigma} F_2(\sigma)   \langle \sigma_2 - \sigma_1 \rangle_{\sigma_1} \partial_{\sigma_1} F_2(\sigma_1) + O( \Delta^{-\frac{2}{3}}) \label{expanP} 
\eea 
by expanding the delta function in powers of $\Delta^{-\frac{1}{3}}$. This result, for large $\sigma_1$, 
agrees with the first two term 
of the expansion of $P_{\Delta}^{(1)}(\sigma_1,\sigma)$ in Eq. (168) Section 6.6 of \cite{deNardisPLD2timeLong},
using the result \eqref{firstmomcond}. The present result \eqref{expanP} however is exact for all $\sigma_1$. 
Conversely, the result of \cite{deNardisPLD2timeLong} is conjectured to be exact only for large $\sigma_1$
but to all orders in $\Delta^{-\frac{1}{3}}$, a conjecture which we are unable to check here beyond the
order $O(\Delta^{-\frac{1}{3}})$, since it contains information about the term ${\cal C}$ and subleading ones. 

\subsubsection{IV. 7. 3. Next order}

To show that the expansion can be carried further we obtain now the JPDF to the next order $k=3$ (three Airy functions)
which we denote as $p_{\hat x}^{(2)}$. Let us collect the terms. The first line in \eqref{resFx} gives
\bea
&& p_{\hat x}^{(2)}(\sigma_1,\sigma_2)|_{
\text{First line in \eqref{resFx}}} \simeq 2 \partial_{\sigma_1}  \partial_{\sigma_2} 
 \int_{\sigma_1}^{+\infty} du \int_{0}^{+\infty} dy 
\cosh(\frac{\hat{x}^3}{3} - (u+y) \hat x) \Ai(u+y)
K_{\Ai}(\sigma_2,\sigma_2) \\
&& = 2 \Ai(\sigma_2)^2 \int_{0}^{+\infty} dy 
\cosh(\frac{\hat{x}^3}{3} - (\sigma_1+y) \hat x) \Ai(\sigma_1+y) \nn
\eea 
The term in the second line of \eqref{resFx} which is a product of two traces gives (in loose notations)
\bea \label{equiv1} 
&&  {\rm Tr}[ P_{\sigma_1} \Ai_{\sigma_2-\sigma_1}
{\cal B}_{\hat x}^T]  {\rm Tr}[  P_{\sigma_1} \Ai_{\sigma_2-\sigma_1}
{\cal B}_{-\hat x}^T] |_{k=3} = - {\rm sym}_{\hat x}
{\rm Tr}[ P_{\sigma_1} \Ai_{\sigma_2-\sigma_1} \int_0^{+\infty} dy \Ai(y+u) e^{\hat x y} ]
e^{-\frac{\hat x^3}{3}} 
{\rm Tr}[ P_{\sigma_1} \Ai_{\sigma_2-\sigma_1} e^{ \hat x v}] \nn \\
&& = - {\rm sym}_{\hat x} 
\int_{\sigma_2}^{+\infty} du \Ai(u) \int_{\sigma_1-\sigma_2}^{\infty} dy \Ai(y+u) e^{\hat x y} 
\times \int_{\sigma_2}^{+\infty} dv \Ai(v) e^{\hat x v} \nn
\eea
where we denote ${\rm sym}_{\hat x} f(\hat x) = f(\hat x)+f(-{\hat x})$.
Hence 
\be
 p_{\hat x}^{(2)}(\sigma_1,\sigma_2)|_{
\text{Second line in \eqref{resFx},TrTr}} = {\rm sym}_{\hat x} e^{-\frac{\hat x^3}{3}} \partial_{\sigma_2} [
  e^{\hat x (\sigma_1-\sigma_2)} K_{\Ai}(\sigma_1,\sigma_2)
\times \int_{\sigma_2}^{+\infty} dv \Ai(v) e^{\hat x v} ]
\ee
The sum of the two cross terms becomes
\bea
&& - {\rm sym}_{\hat x}  e^{\frac{\hat x^3}{3}} F_2(\sigma_1) e^{-  \hat x (\sigma_2-\sigma_1)}
 {\rm Tr}[ (I + P_{\sigma_1} K_{\Ai} ) P_{\sigma_1} \Ai_{\sigma_2-\sigma_1} e^{- \hat x u}] 
\eea
which gives two terms. The first one
\bea
&&  {\rm sym}_{\hat x}  e^{\frac{\hat x^3}{3}} \int_{\sigma_1}^{+\infty} K_{\Ai}(y,y)
 \int_{\sigma_2}^{+\infty} \Ai(u) e^{- \hat x u}
\eea
leads to
\bea
 p_{\hat x}^{(2)}(\sigma_1,\sigma_2)|_{\rm first} = {\rm sym}_{\hat x}  e^{\frac{\hat x^3}{3}} 
 K_{\Ai}(\sigma_1,\sigma_1) \Ai(\sigma_2) e^{- \hat x \sigma_2} 
\eea 
The second one is
\bea
&& - {\rm sym}_{\hat x}  e^{\frac{\hat x^3}{3}} e^{-  \hat x (\sigma_2-\sigma_1)}
 {\rm Tr}[ P_{\sigma_1} K_{\Ai}  P_{\sigma_1} \Ai_{\sigma_2-\sigma_1} e^{- \hat x u}] 
 = - {\rm sym}_{\hat x}  e^{\frac{\hat x^3}{3}} e^{-  \hat x (\sigma_2-\sigma_1)} 
 \int_{\sigma_1}^{\infty} du  \int_{\sigma_1}^{\infty}  dv 
 K_{\Ai}(u,v) \Ai(v+\sigma_2-\sigma_1) e^{- \hat x u} \nn \\
 && = - {\rm sym}_{\hat x}  e^{\frac{\hat x^3}{3}} 
 \int_{\sigma_2}^{\infty} du  \int_{\sigma_2}^{\infty}  dv 
 K_{\Ai}(u+\sigma_1-\sigma_2,v+\sigma_1-\sigma_2) \Ai(v) e^{- \hat x u} 
 \eea
Taking a derivative w.r.t. $\sigma_1$ gives
\bea
&& p_{\hat x}^{(2)}(\sigma_1,\sigma_2)|_{
\text{last} }=  {\rm sym}_{\hat x}  e^{\frac{\hat x^3}{3}} \partial_{\sigma_2} [
 \int_{\sigma_2}^{\infty} du  \int_{\sigma_2}^{\infty}  dv 
 \Ai(u+\sigma_1-\sigma_2) \Ai(v+\sigma_1-\sigma_2) \Ai(v) e^{- \hat x u} ] \\
 && =  {\rm sym}_{\hat x}  e^{\frac{\hat x^3}{3}} \partial_{\sigma_2}
[ K_{\Ai}(\sigma_1,\sigma_2) ( e^{- \hat x (\sigma_2-\sigma_1)}  \int_{\sigma_1}^{\infty} du 
 \Ai(u) e^{- \hat x u} )  ] \nn
\eea 

Putting together all terms, 
%
we find that the total can be rewritten as a derivative
\bea
&& p_{\hat x}^{(2)}(\sigma_1,\sigma_2) = \partial_{\sigma_2} Q(\sigma_1,\sigma_2) \\
&& Q(\sigma_1,\sigma_2) = {\rm sym}_{\hat x}  e^{\frac{\hat x^3}{3}} 
\bigg[ (K_{\Ai}(\sigma_1,\sigma_2) e^{- \hat x (\sigma_2-\sigma_1)} - 
K_{\Ai}(\sigma_2,\sigma_2) ) \int_{\sigma_1}^{+\infty} du \Ai(u) e^{- u \hat x} 
\\
&& + (K_{\Ai}(\sigma_1,\sigma_2) e^{- \hat x (\sigma_1-\sigma_2)} - 
K_{\Ai}(\sigma_1,\sigma_1) ) \int_{\sigma_2}^{+\infty} du \Ai(u) e^{- u \hat x} ) \bigg] \nn
\eea
which immediately implies that 
\bea
&& \int_{\sigma_1}^{+\infty} d\sigma_2 p_{\hat x}^{(2)}(\sigma_1,\sigma_2) = 0
\eea
as required by \eqref{sumruleexact}, since the GUE-TW CDF has no term
with three Airy function ($k=3$). It is easy to see, using the symmetry of $Q(\sigma_1,\sigma_2)$ in its arguments, that $Q$ vanishes quadratically and that the probability vanishes
linearly, at coinciding points, i.e. $p_{\hat x}^{(2)}(\sigma_1,\sigma_1) = 0$, again with a cusp
singularity. Finally note that the correction to the conditional first moment 
can be expressed using $Q$ by integration by part, as
\bea
\tilde R^{\rm exact}_{1/3}(\sigma_1) = \tilde R_{1/3}(\sigma_1) -
\int_{\sigma_1}^{+\infty} d\sigma_2 Q(\sigma_1,\sigma_2) 
+ O(e^{-\frac{8}{3} \sigma_1^{3/2}}) 
\eea 

\section{V. Half axis results} 

\subsection{V.1. joint PDF of Airy and maximum of Airy minus parabola plus Brownian on the half-axis}

Here we concentrate on the simplified case $\sigma_L=+\infty$. In that case, from 
\eqref{final10w}-\eqref{Khat1} it is easy to see that the dependence in $\hat x, \hat w_R$ is only
on the variable $\hat x- \hat w_R$ (from the STS symmetry). Hence we can set directly $\hat w_R=0$ 
with no loss of information and we obtain the JPDF of
Airy and the maximum of Airy minus parabola plus Brownian on the half-axis defined as 
\bea \label{final10w2} 
&& G^R_{\hat x}(\sigma_1,\sigma_R) := {\rm Prob}\bigg( {\cal A}_2(-\hat x)  < \sigma_1, ~~ \max_{\hat y>0}( {\cal A}_2(\hat y-\hat x) - (\hat y - \hat x)^2  +  
\sqrt{2} B(\hat y) ) < \sigma_R - \hat x^2 \bigg) \\
&& = 
 \hat g_{\infty} (\sigma_1,+\infty,\sigma_R;\hat x) = {\rm Det}[ I -  P_{\sigma_{m}} 
 \hat K_{\sigma_R-\sigma_{m}} P_{\sigma_{m}} ] \nn
\eea
with $\sigma_m=\min(\sigma_1,\sigma_R)$ and the kernel
\bea \label{Khat12} 
\hat K_{\sigma_R-\sigma_{m}}(v_i,v_j)
 = K_{\Ai}(v_i,v_j)  +{\cal B}_{ - \hat x}(v_i)  \Ai(v_j+  \sigma_R - \sigma_{m}) e^{\hat x (\sigma_R 
 - \sigma_{m})}
\eea
Using that the second term in the kernel is a projector, 
one can rewrite
\bea \label{final10w3} 
&& G^R_{\hat x}(\sigma_1,\sigma_R) = F_2(\sigma_m) (1 - e^{\hat x (\sigma_R-\sigma_m)} 
{\rm Tr}[ (I- P_{\sigma_m} K_{\Ai} )^{-1} P_{\sigma_m} \Ai_{\sigma_R-\sigma_m}
{\cal B}_{-\hat x}^T] )
\eea
where we recall that $\Ai_{\sigma}(v)=\Ai(\sigma+v)$, which agrees with the limit of formula \eqref{mainres} for $\sigma_L \to +\infty$. The marginal of $\sigma_1$ is the GUE-TW 
distribution $F_2(\sigma_1)$, and
the marginal of $\sigma_R$ is the CDF of the one point distribution for the
transition process ${\cal A}_{2 \to {\rm stat}}$
\bea \label{final10w4} 
&& {\rm Prob}( {\cal A}_{2 \to {\rm stat}} < \sigma_R - \hat x^2) 
= G^R_{\hat x}(+\infty,\sigma_R) = G^R_{\hat x}(\sigma_1 \geq \sigma_R,\sigma_R) \\
&& = F_2(\sigma_R) (1 - 
{\rm Tr}[ (I- P_{\sigma_R} K_{\Ai} )^{-1} P_{\sigma_R} \Ai \,
{\cal B}_{-\hat x}^T] ) = {\rm Det}[ I - P_{\sigma_R} K_{\Ai} - P_{\sigma_R} \Ai \, {\cal B}_{-\hat x}^T] \nn
\eea
At point $\hat x=0$ this kernel is also the one of the BPP transition (also called GUE1) which can also be written as $F_1^2$, i.e. GOE$^2$, see e.g. \cite{wang}.

\subsection{V.2. Two-time persistent correlations with half axis constraint}

Consider now the two-time problem, or the two directed polymer (DP) problem, where the first (shorter) DP goes from 
$(\hat x,0)$ to $(0,t_1)$ and the second DP goes from $(\hat x,0)$ to $(0,t_2)$
with the constraint on its path, denoted $x(t)$, that $y=x(t_1)>0$ (we denote $y$ its position at $t_1$). 
Following similar arguments as in the previous section, with the same definitions as in \eqref{defh1h2}-\eqref{defhh} 
of the rescaled heights one has now
\bea
&& h_1 = {\cal A}_2(-\hat x) - \hat x^2 \\
&& h_2 = \Delta^{1/3} \tilde {\cal A}_2(0) + \max_{\hat y>0}[ {\cal A}_2(\hat y-\hat x) - (\hat y-\hat x)^2
+ \sqrt{2} B(\hat y)  ] + O(\Delta^{-1/3}) \\
&& h  = \tilde {\cal A}_2(0) + \Delta^{-1/3} (\sigma_R-\sigma_1) + O(\Delta^{-2/3}) \label{hsum}
\eea 
where $\sigma_1$ and $\sigma_R$ are the random variables
\bea \label{abuse2} 
&& \sigma_1 =  {\cal A}_2(-\hat x)  \\
&& \sigma_R - \hat x^2 = \max_{\hat y>0}[ {\cal A}_2(\hat y-\hat x) - (\hat y-\hat x)^2
+ \sqrt{2} B(\hat y)  ] 
\eea 
We can write again the persistent correlation ratio defined as the $\Delta \to +\infty$ 
limit of \eqref{CDelta} as
\bea \label{Cinfty23} 
&& C_{\infty} = \frac{\langle \sigma_1 \sigma_R \rangle 
- \langle \sigma_1 \rangle \langle \sigma_R \rangle }{\langle \sigma_1^2 \rangle^c}
= \frac{\langle \sigma_1^2 \rangle + \langle \sigma_R^2 \rangle - \langle (\sigma_R-\sigma_1)^2 \rangle - 2 \langle \sigma_R \rangle \kappa_1^{\mbox{\tiny{GUE}} } }{2 \kappa_2^{\mbox{\tiny{GUE}} }} \\
&& = \frac{1}{2} + 
\frac{ (\kappa_1^{\mbox{\tiny{GUE}}})^2 + \langle \sigma_R^2 \rangle - 
2 \langle \sigma_R \rangle \kappa_1^{\mbox{\tiny{GUE}}}}
{2 \kappa_2^{\mbox{\tiny{GUE}} }} - \frac{\langle (\sigma_R-\sigma_1)^2 \rangle}{2 \kappa_2^{\mbox{\tiny{GUE}} }} \nn
\eea 

We calculate, 
\bea \label{var2R} 
\langle ({\sigma_R}-\sigma_1)^2 \rangle := \int_{-\infty}^{+\infty} d{\sigma_R} \int_{-\infty}^{{\sigma_R}}  
d\sigma_1   ({\sigma_R}-\sigma_1)^2 \partial_{\sigma_R} \partial_{\sigma_1} G^R_{\hat x}(\sigma_1,{\sigma_R}) 
= - 2  \int_{-\infty}^{+\infty} d{\sigma_R} \int_{-\infty}^{{\sigma_R}}  d\sigma_1  
(G^R_{\hat x}(\sigma_1,{\sigma_R}) - F_2(\sigma_1)) \nn
\eea 
One integration can be performed exactly. Indeed, we can define for $\sigma_1<\sigma_R$
\bea
 W_{\hat x}^R(\sigma_1) := \int_{\sigma_1}^{+\infty} d{\sigma_R} [ G^R_{\hat x}(\sigma_1,{\sigma_R}) - F_2(\sigma_1)] &=&
-  \int_{\sigma_1}^{+\infty} d{\sigma_R} F_2(\sigma_1)
e^{\hat x (\sigma_R-\sigma_1)} 
{\rm Tr}[ (I- P_{\sigma_1} K_{\Ai} )^{-1} P_{\sigma_1} \Ai_{\sigma_R-\sigma_1}
{\cal B}_{-\hat x}^T] ) \nn
\\
& = & -  F_2(\sigma_1) {\rm Tr}[ (I- P_{\sigma_1} K_{\Ai} )^{-1} P_{\sigma_1} \hat {\cal B}_{\hat x}
{\cal B}_{-\hat x}^T]
\eea 
where ${\cal B}_{\hat x}$ and $\hat {\cal B}_{\hat x}$ are two vectors defined in \eqref{defBhat}.
Then we have
\bea
\langle ({\sigma_R}-\sigma_1)^2 \rangle = - 2  \int_{-\infty}^{+\infty} d{\sigma_1} W_{\hat x}^R(\sigma_1) 
\eea 
which is thus the simpler version of \eqref{altern}. 
Let us denote $F_{\hat x}^{\rm half}(\sigma)={\rm Det}[ I - P_{\sigma} K_{\Ai} - P_{\sigma} \Ai \, {\cal B}_{-\hat x}^T]$ then for $p=1,2$
\bea
\langle \sigma_R^p \rangle = p \int_0^{+\infty} d\sigma \sigma^{p-1} (1-F_{\hat x}^{\rm half}(\sigma)) - 
p \int_{-\infty}^{0} d\sigma \sigma^{p-1} F_{\hat x}^{\rm half}(\sigma) 
\eea 
we find for $\hat x=0$ the numerical estimate
\bea
\langle \sigma_R \rangle \approx -0.49368  \quad , \quad \langle \sigma_R^2 \rangle \approx 1.47525
\eea 
and we find
\bea
C_{\infty} \approx 0.6925 \label{reshalf} 
\eea 
which is slightly larger than the result in the full space. It means that the persistent correlation, i.e. the memory effect is increased by the constraint. Most likely the finite overlap between the two corresponding optimal paths, which is responsible for this correlation, increases (the longer path midpoint $y$ 
tending to get closer to the origin). 

Finally, note that one could further restrict the longer DP to pass through $y$, in
which case the result for $C_{\infty}$
would be simply equal to two 
point correlation of the Airy process at separation $\hat y$, normalized 
to its value at $\hat y=0$. Note also that the probability that the unconstrained DP (starting
at $\hat x$) passes 
right or left of $\hat y=0$ is related to the function $H(\hat x)$ (see text).

\section{VI. Extended Baik-Rains distribution and its moments}

Here we recall the definition and the exact expression for the CDF of the 
extended Baik-Rains distribution (EBR), and provide simpler expressions
for its low moments using integrations by parts. 

By definition the EBR distribution is the 
one-point distribution of the stationary Airy process \cite{BaikFerrariPeche2010,BaikLiuAiry,QR14} 
(defined at the one point level in 
\eqref{hstat}) with associated CDF 
\be
 {\rm Prob}({\cal A}_{\rm stat}(\hat x) < \sigma) =: F_0(\sigma-\hat x^2,\hat x) = 
 \partial_\sigma (F_2(\sigma) Y_{\hat x}(\sigma)) \label{EBR2}
 \ee
where $F_0(\sigma-\hat x^2,\hat x) = H(\sigma,\frac{\hat x}{2},- \frac{\hat x}{2})$
where the function $H$ was defined in definition 3 in \cite{png} (see also Sec. 2.4. and formula (63) in 
\cite{CorwinLiuWang}). The explicit form \eqref{EBR2} was
obtained by the RBA method in \cite{SasamotoStationary}, and we recall the definition of the function 
\be \label{defYx} 
  Y_{\hat x}(\sigma) :=   \sigma  - \hat x^2 
+  \int_{\sigma}^{+\infty} dv (1- {\cal B}_{\hat x}(v) {\cal B}_{-\hat x}(v))
  -  
{\rm Tr} [ P_\sigma K_{\Ai} (I- P_\sigma K_{\Ai})^{-1} P_\sigma {\cal B}_{-\hat x} {\cal B}_{\hat x}^T]  
\ee
and of the auxiliary function
\bea
&& {\cal B}_w(v) = e^{\frac{1}{3} w^3 - v w} - \int_0^{+\infty} dy \Ai(v+y) e^{w y} \label{defBL9} 
\eea
For $\hat x=0$, the function $F_0(\sigma,0) = F_0(\sigma) =
 \partial_\sigma (F_2(\sigma) Y_{0}(\sigma))$ is the CDF of the
 standard Baik-Rains (BR) distribution $F_0$.\\

Integration by parts allow to simplify the expressions for the moments. Let us 
first recall the calculation of the first moment of the EBR distribution, which is simple
\bea \label{int0}
&&  \kappa_1^{EBR} = \langle \sigma \rangle_{F_0,\hat x} = \int d\sigma \sigma \partial_\sigma^2 [ F_2(\sigma) Y_{\hat x}(\sigma) ] 
= - \int_{-\infty}^0  d\sigma  \partial_\sigma [ F_2(\sigma) Y_{\hat x}(\sigma) ] 
- \int^{+\infty}_0 \partial_\sigma [ F_2(\sigma) Y_{\hat x}(\sigma) - (\sigma - \hat x^2)] \\
&& = - [ F_2(\sigma) Y_{\hat x}(\sigma) ]_{-\infty}^0  - [ F_2(\sigma) Y_{\hat x}(\sigma) - (\sigma - \hat x^2)]^{+\infty}_0 = \hat x^2 \nn
\eea 
and $\kappa_1^{BR}=0$. We have used that for large positive $\sigma$, $Y_{\hat x}(\sigma) \simeq
F_2(\sigma) Y_{\hat x}(\sigma) \simeq \sigma - \hat x^2$ up to fast decaying terms.

The second moment of EBR distribution can be written as
\bea
&& \mu_2^{EBR}= \langle \sigma^2 \rangle_{F_0,\hat x} =
\int  d\sigma \sigma^2 \partial_\sigma^2 [ F_2(\sigma)  Y_{\hat x}(\sigma) ] =
\int_{-\infty}^0 d\sigma \sigma^2 \partial_\sigma^2 [ F_2(\sigma)  Y_{\hat x}(\sigma) ] 
+ \int^{+\infty}_0 d\sigma \sigma^2 \partial_\sigma^2 [ F_2(\sigma)  Y_{\hat x}(\sigma) - (\sigma - \hat x^2)] \nn
\\
&& 
= - 2 \int_{-\infty}^0 d\sigma \sigma \partial_\sigma [ F_2(\sigma) Y_{\hat x}(\sigma) ] - 2
\int^{+\infty}_0 d\sigma \sigma \partial_\sigma [ F_2(\sigma) Y_{\hat x}(\sigma) - (\sigma - \hat x^2)] \nn \\
&& = 2 \int_{-\infty}^0 d\sigma F_2(\sigma) Y_{\hat x}(\sigma) + 2 \int^{+\infty}_0 d\sigma [ F_2(\sigma) Y_{\hat x}(\sigma) - (\sigma - \hat x^2)] \label{mu2BR} 
\eea 
and one can check that the boundary terms vanish in each equality. This formula is useful for
numerical evaluations.

\section{VII. PDF of Argmax of Airy minus parabola plus Brownian}
\label{argmax1} 

\subsection{VII. 1. CDF of Argmax $H(\hat x)$} 

Here we calculate the CDF
\bea
&& H(- \hat x) := {\rm Prob}(\hat z_m >  - \hat x) \quad , \quad 
\hat z_m={\rm argmax}_{\hat z \in \mathbb{R}} \left( {\cal A}_2(\hat z) - \hat z^2 + \sqrt{2} B(\hat z) \right) \label{maxgen3} 
\eea
From Eq. \eqref{Hmethod} we first need to calculate $\hat g_{+\infty, a_{L,R}=1}(\sigma_1=+\infty,\sigma_L,\sigma_R; \hat x)$
in the limit $\hat w_L \to 0^+, \hat w_R \to 0^+$. 
We can thus use the more general result \eqref{mainres} in the particular case $\sigma_1=+\infty$. We only need the case $\sigma_L \leq \sigma_R$ (see below), in which case $\sigma_m=\min(\sigma_1,\sigma_L,\sigma_R)=\sigma_L$ and we obtain
\bea
&& \lim_{\hat w_R=\hat w_L=\hat w \to 0} \hat g_{+\infty, a_{L,R}=1}(\sigma_1=+\infty,\sigma_L,\sigma_R; \hat x)  
= 
{\rm Det}[ I -  P_0 K_{\Ai}^{\sigma_L} P_0 ] \label{resres1} 
\\
&& 
\times \bigg(Y_{\hat x}(\sigma_L) 
e^{\hat x (\sigma_R-\sigma_L)} {\rm Tr} [(I- P_0 K_{\Ai}^{\sigma_L})^{-1} P_0 \Ai_{\sigma_R} \Ai_{\sigma_L}^T ] \nn
\\
&& + ({\rm Tr}[  (I- P_0 K_{\Ai}^{\sigma_L} )^{-1} P_0 \Ai_{\sigma_L} ({\cal B}^{\sigma_L}_{\hat x})^T ] - 1)
(e^{\hat x (\sigma_R-\sigma_L)}  {\rm Tr}[ (I- P_0 K_{\Ai}^{\sigma_L} )^{-1} P_0 \Ai_{\sigma_R} ({\cal B}^{\sigma_L}_{-\hat x})^T  ] - 1) \nn
 \bigg)
\eea 
where we use the shorthand notations
\bea \label{shorthand1}
K_{\Ai}^{\sigma}(v_i,v_j)=K_{\Ai}(v_i+\sigma,v_j+\sigma) \quad , \quad 
{\cal B}^{\sigma}_{\hat w}(v)={\cal B}_{\hat w}(v+\sigma) \quad , \quad 
\Ai_\sigma(v) = \Ai(v+\sigma)  
\eea 
where ${\cal B}_{\hat x}(v)$ is given in \eqref{defBw2} (Eq. \eqref{defBL} in the main text), 
and $Y_{\hat x}(\sigma)$ was defined in
\eqref{defYx2} (Eq. \eqref{defYx} in the main text). In \eqref{resres1} we have shifted the
arguments of the kernel by $\sigma_m=\sigma_L$ to use projectors $P_0$, which simplifies the evaluation of 
derivatives w.r.t. $\hat x$ and $\hat \sigma_L$ (see below). One can check (as above) using the identities \eqref{F2prime}, \eqref{derYsigma} 
that for $\sigma_R=\sigma_L$ this becomes the CDF of the 
extended BR distribution
\bea
\lim_{\hat w_R=\hat w_L=\hat w \to 0} \hat g_{+\infty, a_{L,R}=1}(\sigma_1=+\infty,\sigma_L,\sigma_L; \hat x) 
 = F_0(\sigma_L-\hat x^2,\hat x) =
\partial_{\sigma_L} [ F_2(\sigma_L) Y_{\hat x}(\sigma_L) ]
\eea \\

Let us now calculate
\bea
  H(- \hat x) = \lim_{\hat w_L \to 0^+, \hat w_R \to 0^+}  
\int_{-\infty}^{+\infty} d\sigma_R [\partial_{\sigma_R} 
\hat g_{+\infty, a_{L,R}=1}(+\infty,\sigma_L,\sigma_R; \hat x) )|_{\sigma_L=\sigma_R^-} 
\eea 
Taking the derivative w.r.t. $\sigma_R$ in \eqref{resres1} using \eqref{derB2}
and rearranging we obtain the main result given in the text in \eqref{resH1} namely
\bea \label{resH2} 
&& H(- \hat x) = \int  d\sigma  F_2(\sigma) \times \bigg( Y_{\hat x}(\sigma)
{\rm Tr} [(I- P_\sigma K_{\Ai})^{-1} P_\sigma (\Ai' + \hat x \Ai)  \Ai^T ] \nn
\\
&& + ({\rm Tr}[ (I- P_\sigma K_{\Ai} )^{-1} P_\sigma \Ai {\cal B}_{\hat x}^T] - 1)  \, {\rm Tr}[(I- P_\sigma K_{\Ai} )^{-1} 
P_\sigma (\Ai' + \hat x \Ai) {\cal B}_{-\hat x}^T] 
 \bigg)
\eea 

\subsection{VII. 2. Proof that PDF of Argmax coincides with $f_{\rm KPZ}$} 

We will now show that
\bea
H(\hat x) = \frac{1}{2} - \frac{1}{4} \partial_{\hat x} g(\hat x)  \label{Hg}
\eea 
where
\bea
g(\hat x) = \langle \sigma^2 \rangle_{F_0,\hat x} - \langle \sigma \rangle^2_{F_0,\hat x} 
=  \int  d\sigma \sigma^2 \partial_\sigma^2 [ F_2(\sigma) Y_{\hat x}(\sigma) ]
- 
( \int  d\sigma \sigma \partial_\sigma^2 [ F_2(\sigma) Y_{\hat x}(\sigma) ] )^2
\eea 
is the second cumulant of the EBR distribution (note that although we denote 
it by $g(\hat x)$ it has nothing to do with the generating function $g(\sigma_1,\sigma_L,\sigma_R;\hat x)$). 
Since $Y_{- \hat x}(\sigma)=Y_{\hat x}(\sigma)$ the function $g(\hat x)$ is even in $\hat x$
hence $ \partial_{\hat x} g(\hat x)$ is an odd function of $\hat x$.

To show \eqref{Hg} we proceed in two steps. First we show that the even part of $H(\hat x)$ is constant and equal to $1/2$,
i.e. that $H(- \hat x) + H(\hat x)  = 1$. We see that in the first line of \eqref{resH2}, since $Y_{\hat x}(\sigma)$ is even the
term proportional to $\hat x$ cancels in the sum and what remains in the integrand can be simply rewritten
as $Y_{\hat x}(\sigma) \partial_\sigma^2 F_2(\sigma)$ using the following identity obtained by derivation of \eqref{F2prime}, see Eq. (327) in Appendix G2
of \cite{deNardisPLD2timeLong} for details 
\bea
 F_2(\sigma) {\rm Tr} [(1-P_\sigma K_{\Ai})^{-1} P_\sigma \Ai' \Ai^T] = \frac{1}{2} \partial_\sigma^2 F_2(\sigma)
\eea 
To deal with the symmetrized version of the second line in \eqref{resH2} we need some further
identities. We first recall the identity \eqref{derYsigma}
\be \label{derYsigma2} 
\partial_\sigma Y_{\hat x}(\sigma) = (\Tr[ (I- P_\sigma K_{\Ai} )^{-1} P_\sigma \Ai {\cal B}_{\hat x}^T] - 1)  
  (\Tr[(I- P_\sigma K_{\Ai} )^{-1} P_\sigma \Ai {\cal B}_{- \hat x}^T] - 1) 
\ee
We will need to take a derivative of this expression. An intermediate formula is 
obtained, using \eqref{derB2}, as
\bea
&& \partial_\sigma {\rm Tr}[ (I- P_\sigma K_{\Ai} )^{-1} P_\sigma \Ai {\cal B}_{\hat x}^T]  =
\partial_\sigma {\rm Tr}[ (I- P_0 K_{\Ai}^\sigma )^{-1} P_0 \Ai_{\sigma} ({\cal B}^\sigma_{\hat x})^T] \\
&& =
{\rm Tr}[ (I- P_0 K_{\Ai}^\sigma )^{-1} P_0 \Ai'_{\sigma} ({\cal B}^\sigma_{\hat x})^T]
-  
{\rm Tr}[ (I- P_0 K_{\Ai}^\sigma )^{-1} P_0 \Ai_{\sigma} ({\cal B}^\sigma_{\hat x})^T] 
{\rm Tr}[ (I- P_0 K_{\Ai}^\sigma )^{-1} P_0 \Ai_{\sigma} \Ai_\sigma^T] \nn
\\
&& +
{\rm Tr}[ (I- P_0 K_{\Ai}^\sigma )^{-1} P_0 \Ai_\sigma (\Ai_\sigma - \hat x {\cal B}^\sigma_{\hat x})^T] \nn \\
&& =
{\rm Tr}[ (I- P_\sigma K_{\Ai} )^{-1} P_\sigma \Ai' ({\cal B}_{\hat x})^T]
-  
({\rm Tr}[ (I- P_\sigma K_{\Ai} )^{-1} P_\sigma \Ai ({\cal B}_{\hat x})^T] -1)
{\rm Tr}[ (I- P_\sigma K_{\Ai} )^{-1} P_\sigma \Ai  \Ai^T] 
\\
&& - \hat x
{\rm Tr}[ (I- P_\sigma K_{\Ai} )^{-1} P_\sigma \Ai {\cal B}_{\hat x}^T] \nn
\eea 
Hence taking the derivative of \eqref{derYsigma2} we find that
\bea \label{Yxderder} 
&& F_2(\sigma) \partial_\sigma^2 Y_{\hat x}(\sigma)  = - 2 \partial_\sigma F_2(\sigma) \partial_\sigma Y_{\hat x}(\sigma) \\
&& + (\Tr[ (I- P_\sigma K_{\Ai} )^{-1} P_\sigma \Ai {\cal B}_{\hat x}^T] - 1) 
( {\rm Tr}[ (I- P_\sigma K_{\Ai} )^{-1} P_\sigma \Ai' ({\cal B}_{-\hat x})^T] 
+ \hat x
{\rm Tr}[ (I- P_\sigma K_{\Ai} )^{-1} P_\sigma \Ai {\cal B}_{-\hat x}^T] ) \nn \\
&& 
+  (\Tr[(I- P_\sigma K_{\Ai} )^{-1} P_\sigma \Ai {\cal B}_{- \hat x}^T] - 1) 
( {\rm Tr}[ (I- P_\sigma K_{\Ai} )^{-1} P_\sigma \Ai' ({\cal B}_{\hat x})^T] 
- \hat x
{\rm Tr}[ (I- P_\sigma K_{\Ai} )^{-1} P_\sigma \Ai {\cal B}_{\hat x}^T] ) \nn
\eea 
where to obtain the first term we used \eqref{F2prime} and \eqref{derYsigma2}.
Now we can check that the sum of the last two lines in \eqref{Yxderder} is precisely the 
integrand which appears from the symmetrized version of the second line in \eqref{resH2}.
This finally leads to 
\bea
&& H(- \hat x) + H(\hat x)  =  \int  d\sigma  [ Y_{\hat x} \partial_\sigma^2 F_2(\sigma) 
+ F_2(\sigma) \partial_\sigma^2 Y_{\hat x}(\sigma)  + 2 \partial_\sigma F_2(\sigma) \partial_\sigma Y_{\hat x}(\sigma)] 
 =  \int  d\sigma  \partial_\sigma F_0(\sigma-\hat x^2,\hat x)  = 1
\eea
i.e. a total derivative, where we have used the definition \eqref{EBR1} of the EBR distribution. 
This is the announced result for the symmetrized part.

%
%
%
%
%

We will now complete the proof of \eqref{Hg} by showing that the antisymmetric part can be written as
\bea
&& \frac{1}{2} (H(- \hat x) - H(\hat x)) = \frac{1}{2} \int  d\sigma \frac{\sigma^2}{2} \partial_\sigma^2 [ F_2(\sigma) \partial_{\hat x} Y_{\hat x}(\sigma) ]
- \frac{1}{4} \partial_{\hat x}
( \int  d\sigma \sigma \partial_\sigma^2 [ F_2(\sigma) Y_{\hat x}(\sigma) ] )^2
= \frac{1}{4} \partial_{\hat x} g_{\infty}(\hat x)
\eea 

We will first obtain the following expression for the following derivative w.r.t. $\hat x$
\bea
&&  \partial_{\hat x} Y_{\hat x}(\sigma) = - 2 \hat x  \label{derxY}
  +  {\rm Tr}[(I- P_\sigma K_{\Ai} )^{-1} P_\sigma    ( {\cal B}_{\hat x} -  {\cal B}_{-\hat x}) (\Ai')^T]  + \hat x 
 {\rm Tr}[(I- P_\sigma K_{\Ai} )^{-1} P_\sigma  ( {\cal B}_{-\hat x} +  {\cal B}_{\hat x}) \Ai^T] \\
&& - {\rm Tr}[(I- P_\sigma K_{\Ai} )^{-1} P_\sigma \Ai' {\cal B}_{\hat x}^T] {\rm Tr}[(I- P_\sigma K_{\Ai} )^{-1} P_\sigma
 \Ai {\cal B}_{-\hat x}^T ]
 + {\rm Tr}[(I- P_\sigma K_{\Ai} )^{-1} P_\sigma \Ai' {\cal B}_{-\hat x}^T] {\rm Tr}[(I- P_\sigma K_{\Ai} )^{-1} P_\sigma
 \Ai {\cal B}_{\hat x}^T ] \nn 
\eea

Let us recall the definition
 \be \label{defYx3} 
  Y_{\hat x}(\sigma) := 1+ {\cal L}_{\hat x}(\sigma) -  
{\rm Tr} [ P_\sigma K_{\Ai} (I- P_\sigma K_{\Ai})^{-1} P_\sigma {\cal B}_{-\hat x} {\cal B}_{\hat x}^T]  
\ee
From the definition of \eqref{defBw2} of $B_{\hat x}(v)$ one first obtain the derivative formulae
\bea
&& \partial_{\hat x} B_{\hat x}(v) = (\hat x^2-v) B_{\hat x}(v) + \Ai'(v)- \hat x \Ai(v) \label{derxB1} \\
&& (\partial_{\hat x} - (v_2-v_1)) {\cal B}_{\hat x}(v_1) {\cal B}_{-\hat x}(v_2)
= (\Ai'(v_1) - \hat x \Ai(v_1) ) {\cal B}_{-\hat x}(v_2) -  {\cal B}_{\hat x}(v_1) (\Ai'(v_2) + \hat x \Ai(v_2)) \label{derxB2}
\eea 
From the definition \eqref{defLx2} of ${\cal L}_{\hat x}(\sigma)$ as well as \eqref{derxB2} we thus obtain
\bea \label{derxL}
&& 
\partial_{\hat x} {\cal L}_{\hat x}(\sigma) = - 2 \hat x - \int_{\sigma}^{+\infty} du
\partial_{\hat x} ({\cal B}_{\hat x}(u) {\cal B}_{-\hat x}(u))
\\
&& = - 2 \hat x - \int_{\sigma}^{+\infty} du [ \Ai'(u) ({\cal B}_{-\hat x}(u) - {\cal B}_{\hat x}(u)) 
- \hat x \Ai(u) ({\cal B}_{-\hat x}(u) + {\cal B}_{\hat x}(u)) ] \\
&& = - 2 \hat x +  \hat x {\rm Tr}[P_\sigma ({\cal B}_{-\hat x} + {\cal B}_{\hat x}) \Ai^T ] 
-  {\rm Tr}[ P_\sigma ({\cal B}_{-\hat x} - {\cal B}_{\hat x}) (\Ai')^T] 
\eea 
Now from \eqref{derxB2} the derivative of the last term in \eqref{defYx3} reads
\bea \label{lastterm}
&& -  {\rm Tr}[ K_{\Ai} (I- P_\sigma K_{\Ai} )^{-1}   P_\sigma \partial_{\hat x} ( {\cal B}_{\hat x}   {\cal B}^T_{-\hat x})]  = -  {\rm Tr}[ K_{\Ai} (I- P_\sigma K_{\Ai} )^{-1}  P_\sigma D ] \\
&& -  {\rm Tr}[ K_{\Ai}  (I- P_\sigma K_{\Ai} )^{-1} P_\sigma    ( {\cal B}_{-\hat x} -  {\cal B}_{\hat x}) (\Ai')^T]  + \hat x 
 {\rm Tr}[K_{\Ai}   (I- P_\sigma K_{\Ai} )^{-1} P_\sigma  ( {\cal B}_{-\hat x} +  {\cal B}_{\hat x}) \Ai^T] \nn
\eea 
where we have denoted $D(v_1,v_2)=(v_2-v_1) {\cal B}_{\hat x}(v_1) {\cal B}_{-\hat x}(v_2)$.
Let us now calculate ${\rm Tr}[ K_{\Ai} (I- K_{\Ai} )^{-1}  D ]$. To this aim let us first calculate
for any integer $p \geq 2$
\bea
&& {\rm Tr}[ (P_\sigma K_{\Ai})^{p-1} D ] = \int_{v_1,..v_p>\sigma} K_{\Ai}(v_1,v_2) \ldots K_{\Ai}(v_{p-1},v_p) (v_1-v_p) 
{\cal B}_{\hat x}(v_p) {\cal B}_{-\hat x}(v_1) \\
&& = \sum_{k=0}^{p-2} {\rm Tr}[ (P_\sigma K_{\Ai})^{k} P_\sigma (\Ai (\Ai')^T - \Ai' (\Ai)^T)  (P_\sigma K_{\Ai})^{p-2-k}  P_\sigma
{\cal B}_{\hat x}   {\cal B}^T_{-\hat x}]] \label{derKK}
\eea 
we have rewritten the factor in the first line $v_1-v_p=v_1-v_2 + v_{2} - v_{3} + \ldots v_{p-1} - v_p$
and used that 
\be
K_{\Ai}(v_i,v_j)=\frac{\Ai(v_i) \Ai'(v_j) - \Ai'(v_i) \Ai(v_j) }{v_i-v_j}
\ee
In summary $D$ acts as an operator derivative, replacing one of the $K_{\Ai}$ by the antisymmetric 
combination $\Ai (\Ai')^T - \Ai' (\Ai)^T$. Hence we have
\bea \label{manip10}
&&   {\rm Tr}[ (I- P_\sigma K_{\Ai} )^{-1}  P_\sigma K_{\Ai} P_\sigma D]  =   {\rm Tr}[(I- P_\sigma K_{\Ai} )^{-1} P_\sigma (\Ai (\Ai')^T - \Ai' \Ai^T)  (I- P_\sigma K_{\Ai} )^{-1} P_\sigma {\cal B}_{\hat x} {\cal B}_{-\hat x}^T] \\
&& = {\rm Tr}[(I- P_\sigma K_{\Ai} )^{-1} P_\sigma \Ai' {\cal B}_{\hat x}^T] {\rm Tr}[(I- P_\sigma K_{\Ai} )^{-1} P_\sigma
 \Ai {\cal B}_{-\hat x}^T ]
 - {\rm Tr}[(I- P_\sigma K_{\Ai} )^{-1} P_\sigma \Ai' {\cal B}_{-\hat x}^T] {\rm Tr}[(I- P_\sigma K_{\Ai} )^{-1} P_\sigma
 \Ai {\cal B}_{\hat x}^T ] \nn
\eea 
using that $(I-K)^{-1} K = (I-K)^{-1} -1$. Putting together \eqref{defYx3}, \eqref{derxL},\eqref{lastterm} 
and \eqref{manip10} we finally obtain the wanted formula \eqref{derxY}.

Let us now write explicitly from \eqref{resH2} the antisymmetric combination
\bea \label{anti1} 
&& \frac{1}{2} (H(- \hat x) - H(\hat x)) = \int  d\sigma  \bigg(  \hat x ( \partial_\sigma F_2(\sigma) Y_{\hat x}(\sigma) + F_2(\sigma) \partial_\sigma Y_{\hat x}(\sigma) - F_2(\sigma)) \\
&& + F_2(\sigma)  \hat x \Tr [ (1-P_\sigma K_\Ai)^{-1} P_\sigma \frac{{\cal B}_x+{\cal B}_{-x}}{2}  \Ai^T] 
+ F_2(\sigma)  \Tr [ (1- P_\sigma K_\Ai)^{-1} P_\sigma \frac{{\cal B}_x-{\cal B}_{-x}}{2}  (\Ai')^T] \nn
\\
&& + \frac{1}{2} F_2(\sigma)  \Tr[ (1- P_\sigma K_\Ai)^{-1} P_\sigma {\cal B}_{\hat x} \Ai^T] 
\Tr[(1- P_\sigma K_\Ai)^{-1} P_\sigma {\cal B}_{-\hat x} (\Ai')^T ] \nn \\
&&
- \frac{1}{2} F_2(\sigma)  \Tr[(1- P_\sigma K_\Ai)^{-1} P_\sigma {\cal B}_{-\hat x}  \Ai^T] \Tr[(1- P_\sigma K_\Ai)^{-1} P_\sigma {\cal B}_{\hat x} (\Ai')^T ] \nn
\eea 
where the term $\hat x  \partial_\sigma F_2(\sigma) Y_{\hat x}(\sigma)$ is the contribution of the 
first line in \eqref{resH2}, using \eqref{F2prime}. We have rearranged terms so as to make
appear $\partial_\sigma Y_{\hat x}(\sigma)$ which is given by \eqref{derYsigma2}.

Comparison of \eqref{anti1} and of \eqref{derxY} immediately shows that
\bea
&& \frac{1}{2} (H(- \hat x) - H(\hat x)) = \int  d\sigma  \bigg(  \hat x \partial_\sigma( F_2(\sigma)  Y_{\hat x}(\sigma) ) 
+ \frac{1}{2} F_2(\sigma) \partial_{\hat x} Y_{\hat x}(\sigma) \bigg) \\
&& = \int  d\sigma  \hat x ( \partial_\sigma( F_2(\sigma)  Y_{\hat x}(\sigma))  - \theta(\sigma) ) +
\int d\sigma  (\frac{1}{2} F_2(\sigma) \partial_{\hat x} Y_{\hat x}(\sigma) + \theta(\sigma) \hat x) \label{Hantfin} 
\eea
where in the second line we have added and substracted a Heaviside function so as to
make each integral convergent. We have used that for large positive $\sigma$, $Y_{\hat x}(\sigma) \simeq
F_2(\sigma) Y_{\hat x}(\sigma) \simeq \sigma - \hat x^2$ up to fast decaying terms.
Now we can check that the second term is $\frac{1}{4} \partial_{\hat x}$ applied to the last line
of Eq. \eqref{mu2BR}, which defines the second moment of the EBR distribution. Hence we have
\bea \label{id10} 
&& \int d\sigma (\frac{1}{2} F_2(\sigma) \partial_{\hat x} Y_{\hat x}(\sigma) + \theta(\sigma) \hat x) = 
\frac{1}{2} \partial_{\hat x} \int  d\sigma \frac{\sigma^2}{2} \partial_\sigma^2 [ F_2(\sigma)  Y_{\hat x}(\sigma) ] = \frac{1}{4} \partial_{\hat x}  \langle \sigma^2 \rangle_{F_0,\hat x} 
\eea 

Let us now recall the calculation of the first moment of the EBR distribution
in \eqref{int0}.
The first term in \eqref{Hantfin} is calculated in exactly the same way
\bea \label{int2} 
 \hat x  \int  d\sigma ( \partial_\sigma( F_2(\sigma)  Y_{\hat x}(\sigma))  - \theta(\sigma) )  
 =  \hat x  \int  d\sigma ( \partial_\sigma( F_2(\sigma)  Y_{\hat x}(\sigma))  - \theta(\sigma) \partial_\sigma (\sigma - \hat x^2) )
= - \hat x^3 =  - \frac{1}{4} \partial_{\hat x}  \langle \sigma \rangle^2_{F_0,\hat x} 
\eea
since the second integral in \eqref{int2} is exactly minus the sum of the two last terms in the first line of \eqref{int0}. Putting together
\eqref{Hantfin} \eqref{id10} and \eqref{int2}, we have shown
\bea
&& \frac{1}{2} (H(- \hat x) - H(\hat x)) = \frac{1}{4} \partial_{\hat x} 
( \langle \sigma^2 \rangle_{F_0,\hat x}  - \langle \sigma \rangle^2_{F_0,\hat x} )
= \frac{1}{4} \partial_{\hat x} g(\hat x)
\eea 
Since $\partial_{\hat x} g(\hat x)$ is odd this finally implies our desired result \eqref{Hg}. 

Note that the function
\bea
f_{\rm KPZ}(y) = \frac{1}{4} \partial^2_{\hat x} g(y)
\eea 
was introduced by Prahofer and Spohn \cite{PrahoferSpohn2004}
in the context of the PNG and TASEP models,
and is known to be a probability distribution, i.e. positive
with $\int dy f_{\rm KPZ}(y) =1$. 
Some values for $f_{\rm KPZ}(y)$ (see e.g. \cite{MaesThiery}) are $f_{\rm KPZ}(0)=0.54$, and
the second and fourth moments are $0.714$ and
$0.733$ respectively, with a kurtosis of $2.812 - 3$.
It is interesting to have recovered this
function here by a completely different calculation, on a different observable
(the argmax of Airy plus Brownian), as the connection between the
two observables was pointed out only very recently by Maes and Thiery \cite{MaesThiery}. 
They obtained this connection from fluctuation dissipation relations exploiting the stationarity of
the Brownian initial condition, and using the Burgers equation.

\section{VIII. Stationary KPZ in presence of a step}  
\label{step-BB} 

Here we provide the derivation of the result \eqref{GH} in the text for the CDF 
\be
G_{\hat H}(\sigma_L) ={\rm Prob}( \hat h_{\rm step}(\hat x)- \hat H + \hat x^2  < \sigma_L) 
\ee
where (with $\hat H>0$ with no loss of generality)
\bea \label{hhatstep} 
 \hat h_{\rm step}(\hat x) := \max_{\hat y} \left( {\cal A}_2(\hat x-\hat y) - (\hat x-\hat y)^2 + \sqrt{2} B(y) - \hat H {\rm sgn}(\hat y) \right) \nn
\eea
This is relevant for the KPZ class when the initial condition is stationary far on each side of $x=0$ 
with a mismatch of height around $x=0$. More precisely, whenever the rescaled initial
condition has the form ${\sf h}_0(\hat y) = \sqrt{2} B(\hat y) - \hat H {\rm sgn}(\hat y)$.
One example is the solution $h(x,t)$ of the continuum KPZ equation \eqref{kpzeq} with initial conditions (in our units
$\nu=1$, $\lambda_0=D=2$) $h(x,t=0)= B_0(x) - H {\rm sgn}(x)$, i.e. equal to a two sided unit Brownian
(with $B_0(0)=0$) with a downward step of size $2 H$ at $x=0$, which is scaled as $H = \hat H t^{1/3}$, with fixed 
$\hat H$,
so as to remain relevant in the large time limit which we study here. Other examples are discussed
in \cite{PLDCrossover17} (see Eq. (2) and discussion there) where the step is smooth but scales 
appropriately. In all these cases, at large time $t$ 
one has $\lim_{t \to +\infty} t^{- \frac{1}{3}} h(x=2 t^{\frac{2}{3}} \hat x,t) = \hat h_{\rm step}(\hat x)$, where 
$\hat h_{\rm step}(\hat x)$ is 
defined in \eqref{hhatstep}, a definition which holds at a given point $\hat x$, not as a process in $\hat x$
(the latter would require replacing ${\cal A}_2(\hat x-\hat y)$ by the so-called Airy sheet). 

Using Eq. \eqref{gcontain} and the definitions \eqref{hLRdef} we see that the desired CDF is
given by our generating function as follows
\bea
&& G_{\hat H}(\sigma_L) = {\rm Prob} (\hat h_L(\hat x) + \hat x^2 <\sigma_L, \hat h_R(\hat x) + \hat x^2 < \sigma_R=\sigma_L +2 \hat H ) \\
&& = \lim_{\hat w_L,\hat w_R \to 0^+} \hat g_{+\infty, a_L=a_R=1}(\sigma_1=+\infty,\sigma_L,\sigma_R=\sigma_L +2 \hat H;\hat x) \nn
\eea
As we stressed in Section, we already calculated this generating function for $\hat w_{L,R}>0$ in 
\cite{PLDCrossover17} (which is also a particular case of the more general calculation performed here). 
It lead to the more general result for Brownian IC plus a wedge plus a step (see Section II C.2 item 4 there). 
However the limit $\hat w_{L,R}=0^+$ is quite non-trivial, and not obtained there. It is performed
here in Section III. 7. Let us translate the result in the present setting. From
\eqref{resres1} we obtain, after some rearrangement
\bea
&& G_{\hat H}(\sigma_L) = 
{\rm Det}[ I -  P_{\sigma_L} K_{\Ai} P_{\sigma_L} ] \label{resres2} 
\\
&& 
\times \bigg(Y_{\hat x}(\sigma_L) 
e^{2 \hat x \hat H} {\rm Tr} [(I- P_{\sigma_L} K_{\Ai})^{-1} P_{\sigma_L} \Ai_{2 \hat H} \Ai^T ] \nn
\\
&& + ({\rm Tr}[  (I- P_{\sigma_L} K_{\Ai} )^{-1} P_{\sigma_L} \Ai ({\cal B}_{\hat x})^T ] - 1)
(e^{2 \hat x \hat H}  {\rm Tr}[ (I- P_{\sigma_L} K_{\Ai} )^{-1} P_{\sigma_L} \Ai_{2 \hat H} ({\cal B}_{-\hat x})^T  ] - 1) \nn
 \bigg)
\eea 
which leads to the result \eqref{GH} in the text.

\section{IX. Calculation of the auxiliary function $\phi_{\infty}(k,y_L,y_R,y)$}
\label{app:1} 

Here we compute the integral defined in \eqref{phi1}. Let us denote $A_{L,R}=2 \tilde w_{L,R} \pm 2 i k$ and
assume ${\rm Re}(A_{L,R}) >0$. We recall that $\{a_L,a_R\} \in \{0,1\}^2$. 
We expand the product in \eqref{phi1}, leading to four terms. 
We use the elementary integrals
\bea
&& (\frac{-1}{2 \pi i}) \int_{C'} \frac{dz}{z} e^{z y  } = \theta(-y)  \\
&&  (\frac{-1}{2 \pi i}) \int_{C'} \frac{dz}{z} \frac{1}{A +z}  e^{z y  } =
\int_0^{+\infty} dv e^{- A v} \theta(-y+v) =  \frac{1}{A} ( \theta(-y) + \theta(y) e^{- A y})  \quad , \quad {\rm Re}(A)>0 
\eea

Let us write
\bea
\frac{e^{ z_L y_L+z_R y_R + z y  }}{A_R + z_L + z - a_R z_R}  = \int_0^{+\infty} dv 
e^{- v A_R + z_L (y_L-v) + z_R (y_R + a_R v) + z (y-v)} 
\eea 
Hence
\bea
(\frac{-1}{2 \pi i})^3 \int_{C'} \frac{dz_L}{z_L}\int_{C'} 
\frac{dz_R}{z_R} \int_{C'} \frac{dz}{z} \frac{e^{ z_L y_L+z_R y_R + z y  }}{A_R + z_L + z - a_R z_R} =
 \int_0^{+\infty} dv 
e^{- v A_R} \theta(v-y_L) \theta(- a_R v - y_R) \theta(v-y)
\eea 
Taking a derivative w.r.t. $y_R$ one obtains:
\bea
&& (\frac{-1}{2 \pi i})^3 \int_{C'} \frac{dz_L}{z_L}\int_{C'} 
\frac{dz_R}{z_R} \int_{C'} \frac{dz}{z} \frac{z_R e^{ z_L y_L+z_R y_R + z y  }}{A_R + z_L + z - a_R z_R} =
-  \int_0^{+\infty} dv \delta(a_R v + y_R)
e^{- v A_R} \theta(v-y_L)  \theta(v-y) \\
&& = - \delta_{a_R,1} \theta(-y_R) e^{y_R A_R} \theta(-y_R-y_L)  \theta(-y_R-y) 
- \delta_{a_R,0} \delta(y_R) \frac{1}{A_R} e^{- \max(y_L,y,0) A_R}
\eea 
which allows to evaluate the two cross-terms in (\ref{phi1}). We also need:
\bea
&& \frac{e^{ z_L y_L+z_R y_R + z y }}{(A_L+ z_R+z- a_L z_L)(A_R + z_L+z-a_R z_R)} 
=  \int_{v_1>0,v_2>0} e^{-A_L v_1- A_R v_2} 
e^{z_L (y_L - v_2+a_L v_1) + z_R(y_R -v_1+a_R v_2) + z (y -v_1-v_2)} \nn
\eea
which similarly leads to
\bea
&& (\frac{-1}{2 \pi i})^3 \int_{C'} \frac{dz_L}{z_L}\int_{C'} 
\frac{dz_R}{z_R} \int_{C'} \frac{dz}{z} \frac{e^{ z_L y_L+z_R y_R + z y }}{(A_L+ z_R+z- a_L z_L)(A_R + z_L+z-a_R z_R)} \\
&& 
= \int_{v_1>0,v_2>0} e^{-A_L v_1- A_R v_2} \theta(v_1-y_R-a_R v_2) \theta(v_2-y_L-a_L v_1) 
\theta(v_1+v_2-y) \nn
\eea
and taking two derivatives we obtain:
\bea
&&  (\frac{-1}{2 \pi i})^3 \int_{C'} \frac{dz_L}{z_L}\int_{C'} 
\frac{dz_R}{z_R} \int_{C'} \frac{dz}{z} \frac{z_L z_R e^{ z_L y_L+z_R y_R + z y }}{(A_L+ z_R+z- a_L z_L)(A_R + z_L+z-a_R z_R)}
\\
&& 
= \int dv_1 dv_2 e^{-A_L v_1- A_R v_2} \delta(v_1-y_R-a_R v_2) \delta(v_2-y_L-a_L v_1) 
\theta(v_1+v_2-y) \theta(v_1) \theta(v_2) \nn \\
&& = \theta(y_L+ a_L y_R) \theta(y_R+ a_R y_L) \theta((1+a_R) y_L + (1+a_L) y_R-y)
e^{- A_L (y_R+ a_R y_L) - A_R (y_L+ a_L y_R)}  (1 - a_R a_L) \nn \\
&& + \frac{a_R a_L}{A_L+A_R}
\delta(y_L + y_R) e^{A_L y_L} 
 e^{- (A_L + A_R) \max(0,y_L,\frac{y+y_L}{2})}    \nn
\eea
upon enumeration of all four cases $a_{L,R}=0,1$. Putting all together we finally obtain
the general result
\bea \label{phitotalgen} 
&& \frac{1}{2} \phi_{\infty}(k,y_L,y_R,y) = -1 +  \theta(-y_L) \theta(-y_R)  \theta(-y)
\\
&& -  2 a_L \theta(-y_L) e^{y_L A_L} \theta(-y_L-y_R)  \theta(-y_L-y) 
- (1-a_L) \delta(y_L) \frac{1}{A_L} e^{- \max(y_R,y,0) A_L} ) \nn \\
&& -   2 a_R \theta(-y_R) e^{y_R A_R} \theta(-y_R-y_L)  \theta(-y_R-y) 
- (1-a_R) \delta(y_R) \frac{1}{A_R} e^{- \max(y_L,y,0) A_R} ) \nn \\
&& +   (1 +a_L + a_R - 3 a_R a_L) \theta(y_L+ a_L y_R) \theta(y_R+ a_R y_L) \theta((1+a_R) y_L + (1+a_L) y_R-y)
e^{- A_L (y_R+ a_R y_L) - A_R (y_L+ a_L y_R)}  \nn \\
&& + \frac{4 a_R a_L}{A_L+A_R}
\delta(y_L + y_R) e^{\frac{1}{2} (A_L y_L+ A_R y_R) - \frac{1}{2} (A_L + A_R)  \max(y,y_R,y_L)}  \nn
\eea 
which leads to \eqref{phitotalgen2} in Section III. 4.


\section{X. Derivation of the second form of the kernel \eqref{K12}}

We now rewrite the kernel \eqref{Mss}-\eqref{phitotalgen2} by integrating over $k$, using 
Airy function identities recalled in Appendix C of \cite{PLDCrossover17} and following similar
steps as in Appendix D there.
This gives $M_{s_1,s_L,s_R,\tilde x}(v_i,v_j) = e^{\frac{\tilde x}{4}(v_j-v_i)} \tilde M_{s_1,s_L,s_R,\tilde x}(v_i,v_j)$ where
\bea
&& \tilde M_{s_1,s_L,s_R,\tilde x}(v_i,v_j) = - \int dy \bigg( (1- \theta(\min(s_L,s_R,s_1)-y))
2^{-1/3} \Ai(2^{1/3} ( v_i + \frac{y}{2} + \frac{\tilde x^2}{32}) ) \Ai(2^{1/3} (v_j+ \frac{y}{2} + \frac{\tilde x^2}{32}))   \label{secondform2} \\
&& -  (1 +a_L + a_R - 3 a_R a_L) 2^{-1/3} \Ai(2^{1/3} (v_i +\frac{s_L(1-a_R)-s_R(1-a_L)+(1+a_R-a_L) y}{2} + \frac{\tilde x^2}{32}) )  \theta(y-\frac{s_L+a_L s_R}{1+a_L}) \nn \\
&& \times 
\Ai(2^{1/3} (v_j +\frac{s_R(1-a_L)-s_L(1-a_R)+(1+a_L-a_R) y}{2} + \frac{\tilde x^2}{32}) ) \theta(y-\frac{s_R+a_R s_L}{1+a_R})
 e^{\frac{\tilde x}{4} ( y(a_L-a_R) + s_R (1-a_L) - s_L (1-a_R))} \nn \\
&& \times 
\theta(y-\frac{(1+a_R) s_L + (1+a_L) s_R -s_1}{1+a_R+a_L})
 e^{- 2 (\tilde w_L  (1+ a_R) + \tilde w_R (1+ a_L)) y + 2 \tilde w_L(s_R+a_R s_L) + 2 \tilde w_R(s_L+a_L s_R) }  \nn  \\
&& + 2 a_L \theta(s_L-y) \theta(s_L+s_R-2 y) \theta(s_L+s_1-2y)  2^{-1/3}  \Ai(2^{1/3} (v_i + \frac{s_L}{2} + \frac{\tilde x^2}{32}) ) \Ai(2^{1/3} (v_j+ y - \frac{s_L}{2} + \frac{\tilde x^2}{32})) e^{(2 \tilde w_L+\frac{\tilde x}{4}) (y-s_L)} \nn \\
&& + 2 a_R \theta(s_R-y) \theta(s_L+s_R-2 y) \theta(s_R+s_1-2y)  2^{-1/3} 
\Ai(2^{1/3} (v_i+ y - \frac{s_R}{2} + \frac{\tilde x^2}{32})) \Ai(2^{1/3} (v_j + \frac{s_R}{2} + \frac{\tilde x^2}{32}) )  e^{(2 \tilde w_R - \frac{\tilde x}{4} ) (y-s_R)} \nn \\
&& + \frac{1}{2} (1-a_L) \delta(y-s_L) 
e^{-  (2 \tilde w_L + \frac{\tilde x}{4}) \max(s_L-s_R,s_L-s_1,0)} \int_{0}^{+\infty} dr e^{ - r ( \frac{\tilde x}{8} + \tilde w_L) } \nn  \\
&& \times 
2^{-1/3} \Ai(2^{1/3} (v_i + \frac{s_L+\max(s_L-s_R,s_L-s_1,0)}{2}  + \frac{r}{4} +  \frac{\tilde x^2}{32}) ) 
\Ai(2^{1/3} (v_j + \frac{s_L-\max(s_L-s_R,s_L-s_1,0)}{2}  - \frac{r}{4} + \frac{\tilde x^2}{32}))  \nn \\
&& + \frac{1}{2} (1-a_R) \delta(y-s_R) 
e^{-  (2 \tilde w_R - \frac{\tilde x}{4}) \max(s_R-s_L,s_R-s_1,0)} \int_{0}^{+\infty} dr e^{ - r ( - \frac{\tilde x}{8} + \tilde w_R) }  \nn \\
&& \times 
2^{-1/3} \Ai(2^{1/3} (v_i + \frac{s_R-\max(s_R-s_L,s_R-s_1,0)}{2}  - \frac{r}{4} + \frac{\tilde x^2}{32})) 
\Ai(2^{1/3} (v_j + \frac{s_R+\max(s_R-s_L,s_R-s_1,0)}{2}  + \frac{r}{4} +  \frac{\tilde x^2}{32}) ) 
 \nn \\
&& - \delta(y- \frac{s_L+s_R}{2}) \frac{a_R a_L }{\tilde w_L + \tilde w_R} 
e^{-  \tilde w_L s_L - \tilde w_R s_R + (\tilde w_L+\tilde w_R) \min(s_1,s_L,s_R)} 
2^{-1/3} \Ai(2^{1/3} ( v_i + \frac{s_L}{2} + \frac{\tilde x^2}{32}) ) \Ai(2^{1/3} (v_j+ \frac{s_R}{2} + \frac{\tilde x^2}{32}))   \bigg)
\nn
\eea
We have written the terms almost in the same order as they appear in \eqref{phitotalgen2}. 
We used (C1) of \cite{PLDCrossover17} for the first term in \eqref{secondform2} with $a=v_i+y/2$, $b=v_j+y/2$, for the second term with $a=v_i + y \frac{1+a_R-a_L}{2} + \frac{s_L-s_R+a_L s_R - a_R s_L}{2}$
and $b=v_j + y \frac{1+a_L-a_R}{2} - \frac{s_L-s_R+a_L s_R - a_R s_L}{2}$, for
the third term with $a=v_i + s_L/2$ and $b=v_j+y - s_L/2$, for the fourth term 
$a=v_i+y-s_R/2$ and $b=v_j+s_R/2$, for the last (eighth) term 
$a=v_i+s_L/2$ and $b=v_j+s_R/2$ (using that $y=(s_L+s_R)/2$). For the fifth term in \eqref{secondform2}
we used (C2) of \cite{PLDCrossover17} with $a=v_i + \frac{1}{2}(y + \max(s_L-s_R,s_L-s_1,0)$
and $b=v_j + \frac{1}{2}(y - \max(s_L-s_R,s_L-s_1,0)$ with $y=s_L$, and for the
sixth term in \eqref{secondform2}
we used (C3) of \cite{PLDCrossover17} with $a=v_i + \frac{1}{2}(y - \max(s_R-s_L,s_R-s_1,0)$
and $b=v_j + \frac{1}{2}(y + \max(s_R-s_L,s_R-s_1,0)$ with $y=s_R$.

We now rescale $y \to 2^{2/3} y$ in the first and second term,
$y \to 2^{-1/3} y$ in the third and fourth term, and substitute $r \to 2^{5/3} y$ (after integration over $y$)
in the fifth and sixth terms. Switching to the variables
\be
 \sigma_{L,R,1} = 2^{-2/3} (s_{L,R,1}+ \frac{\tilde x^2}{16}) \quad , \quad \hat w = 2^{2/3} \tilde w 
\quad , \quad \hat x  = 2^{2/3} \frac{\tilde x}{8}  
\ee
we define a new kernel $K_{\sigma_1,\sigma_L,\sigma_R}$ related to the old one 
by a similarity transformation, i.e.
$\tilde M_{s_1,s_L,s_R,\tilde x}(v_1,v_2) = - 2^{1/3} K_{\sigma_1,\sigma_L,\sigma_R}(2^{1/3} v_1, 2^{1/3} v_2)$. This then leads, after some rearrangements (such as shifts in the $y$ integrals) to
formula \eqref{final10} and the expression \eqref{K12} for the kernel $K_{\sigma_1,\sigma_L,\sigma_R}$.
Note that in the final Fredholm determinant the common factor $e^{\frac{\tilde x}{4}(v_j-v_i)}$ 
can be discarded, since ${\rm Det}[I + P_0 M_{s_1,s_L,s_R;\tilde x} P_0 ] = {\rm Det}[I + P_0 
\tilde M_{s_1,s_L,s_R;\tilde x}  P_0]$.

\section{XI. STS symmetry and varying endpoint positions}

Let us recall the "statistical tilt symmetry" (STS) symmetry. If $h(x,t)$ is the solution of the KPZ equation \eqref{kpzeq2} 
with initial condition $h(x,0)=h_0(x)$ and white noise $\eta$,
then, for any $u$, $\tilde h(x,t)= h(x- u t,t) - x u/2 + u^2 t/4$ is also solution
with a tilted white noise $\tilde \eta$, which has the same correlation as $\eta$,
and initial condition $\tilde h(x,0)=h_0(x) - x u/2$. 
If the initial condition of $h$ is droplet at $x=x_0=0$, then the initial condition
of $\tilde h$ is also droplet at $x=x_0=0$. Hence one has the joint equivalence in law
\bea
\{ h(x_1,t_1 | 0, 0) , h(x_2,t_2|0,0) \}  \equiv \{ h(x_1- u t_1,t_1 | 0, 0) - \frac{u x_1}{2} + \frac{u^2 t_1}{4},
h(x_2- u t_2,t_2 | 0, 0) - \frac{u x_2}{2} + \frac{u^2 t_2}{4} \} \label{STS1} 
\eea
valid for any $u$. Let us now define $H_j = h(x_j,t_j | x_0, 0)$, $j=1,2$, and denote $P_{x_0,x_1,x_2}(H_1,H_2)$ the associated
JPDF. First one has the translational invariance property
\bea
P_{x_0,x_1,x_2}(H_1,H_2) = P_{x_0+y , x_1+y,x_2+y}(H_1,H_2) 
\eea 
for any $y$. This can be combined with the STS symmetry \eqref{STS1}, and we obtain, for any $u,y$:
\bea
P_{x_0,x_1,x_2}(H_1,H_2) = P_{x_0+y,x_1+y- u t_1, x_2+y-u t_2}(H_1+ \frac{u(x_1-x_0)}{2} - \frac{u^2 t_1}{4} ,
H_2+ \frac{u(x_2-x_0)}{2} - \frac{u^2 t_2}{4}) 
\eea 

Choosing $u=(x_1-x_0)/t_1$ and $y=-x_0$ we can express the JPDF with 
general endpoints in terms of the one where only
the last point varies, as
\bea
P_{x_0,x_1,x_2}(H_1,H_2) = P_{0,0,X_2=x_2- \frac{t_2}{t_1} x_1 - (1- \frac{t_2}{t_1}) x_0}( H_1+ \frac{(x_1-x_0)^2}{4 t_1}, H_2 + \frac{(x_2-x_0)^2-X_2^2}{4 t_2} )
\eea

Choosing $u=\frac{x_2-x_1}{t_2-t_1}$ and $y=\frac{t_1 x_2 - t_2 x_1}{t_2-t_1}$ we can express the JPDF with 
general endpoints in terms of the one where only
the first point varies, as
\bea
&& P_{x_0,x_1,x_2}(H_1,H_2) \\
&& = P_{X_0=\frac{t_1(x_2-x_0)-t_2(x_1-x_0)}{t_2-t_1},0,0}
\left( H_1+ \frac{(x_1-x_0)(x_2-x_1)}{2 (t_2-t_1)} - 
\frac{t_1 (x_2-x_1)^2}{4 (t_2-t_1)^2}, H_2 +  \frac{(x_2-x_0)(x_2-x_1)}{2 (t_2-t_1)} - 
\frac{t_2 (x_2-x_1)^2}{4 (t_2-t_1)^2} \right) \nn
\eea 

It is clear from this expression that, in the limit of large times $t_1,t_2$ and large time separation
$t_2/t_1 \gg 1$, to be in the regime where
$\hat X_0= X_0/(2 t_1^{2/3})$ is fixed, one needs
to take $(x_0,x_1,x_2)=(0,0,x_2 \simeq 2 \hat X_0 t_1^{-1/3} t_2)$, namely
\be \label{shift2} 
P_{0,0,x_2 = 2 \hat X_0 t_1^{-1/3} t_2}(H_1,H_2) \simeq
P_{X_0=2 t_1^{2/3} \hat X_0,0,0}(H_1 - t_1^{1/3} \hat X_0^2, H_2- \frac{x_2^2}{4 t_2})
\ee 
This is the regime where the variable defined in \cite{deNardisPLD2timeLong} $\bar X=x_2/(2 (t_2-t_1) t_1^{-1/3}) \simeq \hat X_0$
is fixed. It can thus be compared with our results in the present paper where we vary
instead the first endpoint position, see discussion in Section IV. 7. 1. below Eq. \eqref{aga}.

\section{XII. Flat limit: max and argmax of Airy$_2$ minus parabola, and Airy$_2$ at a point, jointly}

We now specify the result \eqref{final10}, \eqref{K12}, to the case $a_L=a_R=0$
(wedge) in the limit $\hat w_R=\hat w_L=0^+$ (flat). In that case we denote
\bea
&& \hat h_L(\hat x) = \max_{\hat z < - \hat x}( {\cal A}_2(\hat z) - \hat z^2 ) \quad , \quad  \hat h_R(\hat x) = \max_{\hat z> - \hat x}( {\cal A}_2(\hat z) - \hat z^2 )
\eea
and we have an exact expression for the following triple joint CDF as a Fredholm determinant
\be \label{final10f} 
{\rm Prob}\left( {\cal A}_2(-\hat x)  < \sigma_1 , \hat h_L(\hat x) + \hat x^2 < \sigma_L ,  
\hat h_R(\hat x) + \hat x^2 < \sigma_R  \right) = \hat g_{\infty} (\sigma_1,\sigma_L,\sigma_R;\hat x) = {\rm Det}[ I -  P_0 K_{\sigma_1,\sigma_L,\sigma_R} P_0 ] 
\ee
in terms of the kernel 
\bea \label{K12f} 
&& K_{\sigma_1,\sigma_L,\sigma_R}(v_i,v_j) = \int_{\sigma_{m}}^{+\infty} dy \Ai(v_i + y ) \Ai(v_j + y)  
  \\
  && - e^{2  \hat x (\sigma_R  - \sigma_L)} 
\int_{\max(\sigma_L,\sigma_R,\sigma_L+\sigma_R- \sigma_1)}^{+\infty} dy  
\Ai(v_i +  y +  \sigma_L - \sigma_R)  \Ai(v_j + y-  \sigma_L +  \sigma_R ) 
 \nn \\
 && +  
\int_{\max(\sigma_L-\sigma_R,\sigma_L-\sigma_1,0)}^{+\infty} dy \Ai(v_i + y + \sigma_L ) \Ai(v_j  - y + \sigma_L) 
e^{ - 2 y  \hat x  }   \\
&& +  
\int_{\max(\sigma_R-\sigma_L,\sigma_R-\sigma_1,0)}^{+\infty} dy \Ai(v_i - y+\sigma_R) \Ai(v_j + y+\sigma_R) 
e^{ + 2 y  \hat x  } \nn
\eea 
where $\sigma_{m}=\min(\sigma_1,\sigma_L,\sigma_R)$. We now consider two
applications of this formula.

\subsection{XII.1. 
JPDF of the max of Airy$_2$ minus parabola and of Airy$_2$ at a point, i.e. JPDF of flat and droplet
KPZ heights, i.e. point-to-point and point-to-line DP free energies}

%

Let us further specialize to the case $\sigma_L=\sigma_R=\sigma$. 
We obtain
\bea \label{JCDFAiAi} 
&& {\rm Prob}( {\cal A}_2(-\hat x) < \sigma_1,  \max_y ({\cal A}_2(y) - y^2) < \sigma - \hat x^2) 
= {\rm Det}[ I -  P_\sigma K_{\max(\sigma-\sigma_1,0)} P_\sigma ]  \\
&& K_{\sigma-\sigma_1}(v_i,v_j) =
\int_{\sigma_1-\sigma}^{\sigma-\sigma_1} dy \Ai(v_i + y) \Ai(v_j + y)  - 
\int_{\sigma_1-\sigma}^{\sigma-\sigma_1} dy \Ai(v_i + y) \Ai(v_j - y) e^{-2 y \hat x} \nn
 \\
 && ~~~~~~~~~~~~~~~~~~ + \int_{-\infty}^{+\infty} dy \Ai(v_i + y) \Ai(v_j  - y) e^{-2 y \hat x}  \nn
\eea 
Note that for $\sigma \leq \sigma_1$, using the identity
\bea \label{ident3} 
K_0(v_i,v_j) = \int_{-\infty}^{+\infty} dy \Ai(v_i + y) \Ai(v_j  - y) e^{-2 y \hat x}  
= 2^{-1/3} \Ai(2^{-1/3} (v_i+v_j- 2 \hat x^2)) e^{\hat x(v_i-v_j)} 
\eea 
and defining (with $\sigma'=\sigma - \hat x^2 = 2^{-2/3} s$)
\bea
\tilde K_\sigma(v_i,v_j)=K_0(v_i+\sigma,v_j+\sigma) = 2^{-1/3} \Ai(2^{-1/3} (v_i+v_j) + s) e^{\hat x(v_i-v_j)} 
\eea 
one recovers, using a similarity transformation of the kernel, for $\sigma_1 \geq \sigma$
\bea
&& {\rm Prob}( {\cal A}_2(-\hat x) < \sigma_1,  \max_y ({\cal A}_2(y) - y^2) < \sigma - \hat x^2) 
= {\rm Prob}(\max_y ({\cal A}_2(y) - y^2) < \sigma') \\
&& = 
{\rm Det}[ I -  P_\sigma K_{0} P_\sigma ]
= {\rm Det}[ I -  P_0 \tilde K_{\sigma} P_0 ] 
=  {\rm Det}[ I -  P_0 K_{s}^{\rm GOE} P_0] = F_1(s=2^{2/3} \sigma')
\eea 
where $K_s^{\rm GOE}(v_i,v_j)=\Ai(v_i+v_j+s)$ is the GOE kernel. Hence we
correctly recover the well known result that the maximum of Airy$_2$ minus a parabola is distributed
by the (scaled) GOE TW distribution. Here in addition we have obtained in \eqref{JCDFAiAi} the
JCDF of this maximum and the value of the Airy$_2$ process at some
arbitrary point $-\hat x$. Hence, Eq.  \eqref{JCDFAiAi} 
also gives the large time limit of the joint CDF of the (properly centered and scaled)
KPZ height fields with respectively flat and droplet
initial conditions {\it in the same realization of the noise}, i.e. of the couple
$(h_{\rm flat}(0,t),h(0,t|x,0))$, equivalently of the point-to-point and point-to-line
DP free energies {\it in the same random potential}.

\subsection{XII.2. 
JPDF of max and argmax, i.e. of endpoint position and free energy of a point-to-line DP}

Let us now specialize to $\sigma_1=+\infty$ and $\sigma_L < \sigma_R$. Then we obtain from  
\eqref{final10f}-\eqref{K12f}
\bea
&& \hat g_{\infty} (+\infty,\sigma_L,\sigma_R;\hat x) = {\rm Det}[ I -  P_{\sigma_L} 
K_{\sigma_R-\sigma_L} P_{\sigma_L} ] \\
&& K_{\sigma_R-\sigma_L}(v_i,v_j) = 
\int_{0}^{+\infty} dy \Ai(v_i + y  ) \Ai(v_j + y )   - e^{2  \hat x (\sigma_R  - \sigma_L)} 
\int_{0}^{+\infty} dy  
\Ai(v_i +  y  )  \Ai(v_j + y+  2 (\sigma_R-\sigma_L) ) 
   \\
 &&
+ \int_{0}^{+\infty} dy \Ai(v_i + y  ) \Ai(v_j  - y ) 
e^{ - 2 y  \hat x  }  +  
e^{  2 \hat x ( \sigma_R-\sigma_L )   } 
\int_{0}^{+\infty} dy \Ai(v_i - y) \Ai(v_j + y+2 (\sigma_R-\sigma_L) ) 
e^{ 2 y  \hat x  } \nn
\eea 
For $\sigma_L=\sigma_R^-$ we obtain the same kernel as in \eqref{ident3} 
\bea
&& K_{0}(v_i,v_j) =  
\int_{-\infty}^{+\infty} dy \Ai(v_i + y  ) \Ai(v_j  - y ) 
e^{ - 2 y  \hat x  }  = 2^{-1/3} \Ai(2^{-1/3} (v_i+v_j- 2 \hat x^2)) e^{\hat x(v_i-v_j)} 
\eea 
which, as noted above, is identical to the GOE kernel by a similarity transformation.\\


We can now obtain an explicit formula for the JCPDF of max and argmax, defined as
\bea
\hat h_m = \max_{\hat z \in \mathbb{R}}( {\cal A}_2(\hat z) - \hat z^2 ) \quad , \quad \hat z_m = {\rm argmax}_{\hat z \in \mathbb{R}}( {\cal A}_2(\hat z) - \hat z^2 )
\eea 
Recalling Eq. \eqref{JCPDF} and, taking a derivative, we obtain (we will denote $\sigma_R=\sigma$ for notational simplicity below)
\bea \label{JCPDF2} 
&& {\rm Prob}( \hat z_m > - \hat x , \hat h_m)  = [\partial_{\sigma_R} \hat g_{+\infty, a_{L,R}=0}(+\infty,\sigma_L,\sigma_R; \hat x)
]|_{\sigma_L=\sigma_R^-, \sigma_R=\sigma=\hat h_m+ \hat x^2} \nn
\\
&& = {\rm Det}[ I -  P_{\sigma} K_{0}  ]
{\rm Tr}[ (I- P_{\sigma} K_{0} )^{-1} P_{\sigma} K^{(1)} ]|_{\sigma=\hat h_m+ \hat x^2}
\eea
where we have defined the derivative kernel $K^{(1)}(v_i,v_j):
= - \partial_{\sigma_R} K_{\sigma_R-\sigma_L}(v_i,v_j)|_{\sigma_L=\sigma_R^{-}, \sigma_R=\sigma}$,
which reads, explicitly
\be
 K^{(1)}(v_i,v_j)= 
 2 \int_{0}^{+\infty} dy  
\Ai(v_i +  y  )  ( \Ai'(v_j + y) + \hat x \Ai(v_j + y) ) 
 -  2 
\int_{0}^{+\infty} dy \Ai(v_i - y)
e^{ 2 y  \hat x  }  ( \Ai'(v_j + y) + \hat x \Ai(v_j + y) ) 
\ee


We can rewrite this result in a slightly different form
\bea \label{JCPDF3} 
&& {\rm Prob}( \hat z_m > - \hat x , \hat h_m)  
= {\rm Det}[ I -  P_0 \tilde K_{\sigma}  ]
{\rm Tr}[ (I- P_0 \tilde K_{\sigma}  )^{-1} 
P_0 \tilde K^{(1)}_{\sigma} ]|_{\sigma=\hat h_m+ \hat x^2}
\eea
with
\bea
&& \tilde K_\sigma(v_i,v_j)=K_0(v_i+\sigma,v_j+\sigma) \quad , \quad 
K^{(1)}_\sigma(v_i,v_j)= K^{(1)}_\sigma(v_i+\sigma ,v_j+\sigma)
\eea

%

Let us recall the result of \cite{QuastelEndpoint} (which was proved
equivalent in \cite{BaikLiechtySchehr} to the one of \cite{SchehrEndpoint}) in the
present notations.
First the marginal distribution of the maximum is
\bea
&& {\rm Prob}(\hat h_m \leq \sigma') = F_1(2^{2/3} \sigma')   \quad , \quad F_1(s) = {\rm Det}[I - P_0 K^{\rm GOE}_s P_0] \quad , \quad K^{\rm GOE}_s(v_1,v_2) = \Ai(v_1+v_2+s) 
\eea 
i.e. it is the GOE TW distribution (as also found here, see previous subsection). Note that
the GOE kernel can also be rewritten in terms of $K_0$ as
\bea
\Ai(v_i+v_j + 2^{2/3}(\sigma - \hat x^2))  = 2^{1/3} K_0(2^{1/3} v_i + \sigma , 2^{1/3} v_j + \sigma) e^{- 2^{1/3} (v_i-v_j) \hat x} 
\eea 
Using this relation we can rewrite the formula in Theorem 2 of \cite{QuastelEndpoint} as follows
(after a similarity transformation). The formula for the JPDF of $\hat h_m$ and $\hat z_m$ (more precisely its density) is then
\bea
&& P(\hat h_m,\hat z_m)=  {\rm Det}[ I - P_0 \tilde K_\sigma P_0 + P_0 \tilde \psi_{\hat z_m,\hat h_m}
\tilde \psi_{-\hat z_m,\hat h_m}^T P_0] - F_1(2^{2/3} \hat h_m) \\
&& 
\tilde \psi_{\hat z_m,\hat h_m}(v) = 2 \hat z_m \Ai(v + \hat h_m + \hat z_m^2) + 2 \Ai'(v + \hat h_m + \hat z_m^2) ) = 2 \hat z_m \Ai_\sigma(v) + 2 \Ai'_\sigma(v) |_{\sigma = \hat h_m + \hat z_m^2}
\eea 

It remains to check that our result agrees with the one of \cite{QuastelEndpoint}, that is
\bea
 \partial_{\hat x} {\rm Prob}( \hat z_m > - \hat x , \hat h_m)  = P(\hat h_m,-\hat x)
\eea 
To show the agreement is thus equivalent to show that 
\bea \label{check11} 
&& \partial_{\hat x}  
 {\rm Det}[ I -  P_0 \tilde K_{\sigma}  ]
{\rm Tr}[ (I- P_0 \tilde K_{\sigma}  )^{-1} 
P_0 \tilde K^{(1)}_{\sigma} ]|_{\sigma=\hat h_m+ \hat x^2} \\
&& = 
{\rm Det}[ I - P_0 \tilde K_\sigma P_0 + P_0 (-2 \hat x \Ai_\sigma + 2 \Ai'_\sigma)
(2 \hat x \Ai_\sigma + 2 \Ai'_\sigma) P_0] - {\rm Det}[ I - P_0 \tilde K_\sigma P_0] \nn
\eea 
where we recall that $\tilde K_\sigma$ and $\tilde K^{(1)}_{\sigma}$ also depend explicitly
on $\hat x$ (not indicated for notational simplicity).

We will not attempt here to show that \eqref{check11} is correct (preliminary investigations
show that it may not be trivial), hence it is left for the future.

\end{widetext}

\end{document}